\documentclass[12pt]{article}

\usepackage[margin=1in]{geometry}
\usepackage[utf8]{inputenc}
\usepackage[T1]{fontenc}
\usepackage{amsmath,amssymb,amsthm}
\usepackage{mathtools}
\usepackage{bm}
\usepackage{graphicx}
\usepackage{float}
\usepackage{booktabs}
\usepackage{array}
\usepackage{xcolor}
\usepackage{hyperref}
\usepackage{algorithm}
\usepackage{algorithmic}
\usepackage{multirow}
\usepackage{subcaption}
\usepackage{tikz}
\usetikzlibrary{calc,patterns,decorations.markings,arrows.meta}

\theoremstyle{definition}
\newtheorem{remark}{Remark}[section]

\DeclareMathOperator{\rank}{rank}
\DeclareMathOperator{\diag}{diag}

\newcommand{\bx}{\bm{x}}
\newcommand{\bv}{\bm{v}}

\newcommand{\ba}{\bm{a}}

\newcommand{\bd}{\bm{d}}

\newcommand{\bff}{\bm{f}}

\newcommand{\bC}{\bm{C}}
\newcommand{\bM}{\bm{M}}
\newcommand{\bK}{\bm{K}}
\newcommand{\bA}{\bm{A}}
\newcommand{\bJ}{\bm{J}}

\newcommand{\bB}{\bm{B}}

\newcommand{\bg}{\bm{g}}
\newcommand{\bb}{\bm{b}}
\newcommand{\bc}{\bm{c}}
\newcommand{\blambda}{\bm{\lambda}}

\newcommand{\Real}{\mathbb{R}}
\newcommand{\pd}[2]{\frac{\partial #1}{\partial #2}}
\newcommand{\Null}{\operatorname{Null}}
\newcommand{\tr}{\operatorname{tr}}

\begin{document}

\title{A finite element formulation for incompressible viscous flow
based on the principle of minimum pressure gradient}

\author{Julian J. Rimoli\\[4pt]
\small Department of Mechanical and Aerospace Engineering,
University of California, Irvine, CA 92697, USA\\
\small\texttt{jrimoli@uci.edu}}

\date{}

\maketitle

\begin{abstract}
We present a finite element formulation for incompressible viscous flow based on
the principle of minimum pressure gradient (PMPG). This variational principle,
recently established by Taha, Gonzalez \& Shorbagy (\textit{Phys.\ Fluids},
vol.~35, 2023), states that the Navier--Stokes equations are equivalent to
determining the rate of change $\partial_t \bv$ at each instant by minimizing
the $L^2$ norm of the implied pressure gradient, subject to incompressibility
and boundary conditions. We discretize the PMPG functional directly using Q9
biquadratic finite elements and minimize over the nodal velocity rates (Rayleigh--Ritz). No pressure degrees of freedom appear;
incompressibility and boundary conditions are enforced as linear
equality constraints through a monolithic saddle-point system, whose Lagrange
multipliers provide wall forces without pressure reconstruction. We verify the
formulation against exact Poiseuille flow (machine-precision recovery) and the
Kovasznay solution (convergence rate $\sim 3.3$), and validate it against
published benchmarks for the lid-driven cavity, the backward-facing step, and
flow past a circular cylinder. The formulation produces smooth, oscillation-free
solutions on coarse meshes in the convection-dominated regime without
stabilization. We further show that the element-wise PMPG
functional density serves as a built-in error indicator for adaptive mesh
refinement, and that the stationarity condition can be read backwards to estimate
the kinematic viscosity directly from velocity field measurements.
\end{abstract}


\section{Introduction}
\label{sec:introduction}

\subsection{The principle of minimum pressure gradient}

The numerical simulation of incompressible viscous flow has been a central topic
in computational mechanics for over half a century. The prevailing approaches
(finite element, finite volume, and spectral methods) discretize the
Navier--Stokes equations directly, treating the pressure either as a primary
unknown (mixed methods) or through projection/fractional-step
algorithms~\cite{chorin1968,chang2002analysis, guermond2006}. In mixed formulations, the
velocity--pressure pair must satisfy the Babu\v{s}ka--Brezzi inf-sup
condition~\cite{babuska1973,brezzi1974} to ensure stability, which constrains
the choice of interpolation spaces. Projection methods decouple velocity and
pressure through operator splitting, but can introduce a splitting error that
compromises accuracy at steady state~\cite{guermond2006}.

A fundamentally different perspective was recently established by Taha, Gonzalez
\& Shorbagy~\cite{taha2023minimization}, who proved that the incompressible
Navier--Stokes equations are equivalent to a constrained optimization problem:
at each instant in time, the rate of change $\partial_t \bv$ is determined by
minimizing the $L^2$ norm of the implied pressure gradient over the domain,
subject to incompressibility and boundary conditions. This is the
\emph{principle of minimum pressure gradient} (PMPG). The principle has deep
roots in classical analytical mechanics (Gauss's principle of least
constraint~\cite{gauss1829} and Appell's formulation~\cite{appell1900}), first
extended to ideal flows by Gonzalez \&
Taha~\cite{gonzalez2022variational} and subsequently generalized to viscous
flows~\cite{taha2023minimization}.

The PMPG principle recasts the Navier--Stokes equations as a constrained
minimization rather than a boundary-value problem, calling for a fundamentally
different discretization strategy: rather than writing the weak form and applying
Galerkin projection, one can substitute a finite-dimensional velocity
approximation directly into the PMPG functional and minimize over the nodal
velocity rates (the classical Rayleigh--Ritz approach), with
constraints enforced exactly through Lagrange multipliers. Demonstrations of this principle
in~\cite{taha2023minimization,gonzalez2022variational}, however, relied on
analytical or semi-analytical velocity representations, limiting their
applicability to simple geometries. A general-purpose implementation requires a
finite element discretization capable of handling arbitrary domains and boundary
conditions; this is the subject of the present work.

\subsection{Objectives and contributions}

The goal of this work is to develop a general-purpose finite element
implementation of the PMPG principle for viscous incompressible flow. We make
five contributions:

\begin{enumerate}
  \item \textbf{First FEM implementation of PMPG for viscous flow.}
    We apply the Rayleigh--Ritz method to the PMPG functional: the Q9
    biquadratic velocity interpolation is substituted into the variational
    principle, converting it into a finite-dimensional constrained minimization
    over the nodal velocity rates. The resulting solver handles
    general two-dimensional geometries with no-slip walls, prescribed inlet
    profiles, and free-outflow boundaries.

  \item \textbf{Velocity(rate)-only primal formulation.}
    No pressure FE space is introduced and no inf--sup velocity--pressure pairing is required. Pressure-related quantities enter through the dual variables (Lagrange multipliers) associated with the imposed incompressibility and boundary constraints, enabling force extraction without pressure reconstruction.

  \item \textbf{Monolithic constraint-based architecture.}
    Divergence-free flow and all boundary conditions are encoded as linear
    equality constraints and enforced through a monolithic saddle-point system
    at each time step. Unlike projection methods, which introduce splitting
    errors between the viscous and incompressibility sub-problems, the
    monolithic approach enforces all constraints simultaneously and converges to
    the exact steady state at machine precision for problems whose solutions lie
    in the finite element space. Because convection enters only through the
    explicit right-hand side, the system matrix is symmetric positive definite
    regardless of the Reynolds number, and the formulation produces
    oscillation-free solutions without stabilization even on coarse meshes in
    the convection-dominated regime.

  \item \textbf{Built-in error indicator and adaptive refinement.}
    The element-wise PMPG functional density measures the local momentum
    residual and serves as a zero-cost error indicator: no adjoint problem,
    reference mesh, or additional solve is needed.  We demonstrate its use
    for adaptive mesh refinement on the backward-facing step.

  \item \textbf{Inverse viscosity estimation from velocity data.}
    We show that the PMPG stationarity condition can be read backwards to
    recover the kinematic viscosity directly from velocity field measurements
    (e.g., PIV data).  The constraint forces are eliminated by projecting
    onto the divergence-free subspace, and the viscosity is obtained from a
    single ratio of inner products, requiring no forward solve, pressure
    reconstruction, or iterative optimization.
\end{enumerate}

\subsection{Outline}

The remainder of this paper is organized as follows.
Section~\ref{sec:formulation} develops the PMPG formulation: the Rayleigh--Ritz
discretization of the variational principle, the Q9 finite element
interpolation, the divergence-free constraint, boundary conditions, the
monolithic saddle-point solver, adaptive Courant--Friedrichs--Lewy (CFL) time stepping, and the extraction
of wall forces from Lagrange multipliers.
Section~\ref{sec:verification} presents verification against the exact Poiseuille
and Kovasznay solutions. Section~\ref{sec:demonstrations} validates the method
against published benchmark data for the lid-driven cavity, the
backward-facing step, and flow past a circular cylinder.
Section~\ref{sec:error_estimation} shows that the element-wise PMPG functional
provides a natural error indicator and demonstrates its use for adaptive mesh
refinement.
Section~\ref{sec:inverse} develops the inverse viscosity estimation method
and demonstrates it on synthetic PIV data.
Section~\ref{sec:conclusions} summarizes the contributions and
discusses future directions.

\section{Formulation}
\label{sec:formulation}

\subsection{The PMPG variational principle}
\label{sec:pmpg_principle}

Consider incompressible viscous flow of a Newtonian fluid with constant density
$\rho = 1$ and kinematic viscosity $\nu$ in a domain $\Omega \subset \Real^2$.
The velocity field $\bv(\bx, t)$ satisfies the Navier--Stokes equations:
\begin{align}
  \pd{\bv}{t} + (\bv \cdot \nabla)\bv &= -\nabla p + \nu \nabla^2 \bv
  && \text{in } \Omega, \label{eq:ns} \\
  \nabla \cdot \bv &= 0
  && \text{in } \Omega, \label{eq:incomp}
\end{align}
with boundary conditions on $\partial\Omega$.

Taha et al.~\cite{taha2023minimization} proved that the solution
of~\eqref{eq:ns}--\eqref{eq:incomp} can be equivalently characterized through
a minimization principle. The PMPG principle states that, at each instant, the
time derivative $\partial_t \bv$ is the one that minimizes the $L^2$ norm of
the implied pressure gradient:
\begin{equation}
  \min_{\partial_t \bv} \; \mathcal{J}(\partial_t \bv) = \frac{1}{2}
  \int_\Omega \left| \pd{\bv}{t}
  + (\bv \cdot \nabla)\bv - \nu \nabla^2 \bv \right|^2 d\Omega,
  \label{eq:pmpg_continuous}
\end{equation}
for all velocity fields that satisfy $\nabla \cdot \bv = 0$ and Dirichlet boundary conditions. In this formulation,
the current velocity $\bv$ and its spatial derivatives are known at a given instant of time; the unknown is
$\partial_t \bv$, which appears as the first term in the integrand. The functional is
therefore quadratic in $\partial_t \bv$, and its minimization subject to the aforementioned constraints.

The quadratic structure of the PMPG functional is the key property that makes it
attractive to finite element discretization. Just as the minimum potential energy
principle in linear elasticity (quadratic in displacements) leads via
Rayleigh--Ritz to symmetric linear systems that form the basis of structural
finite element methods, the PMPG functional (quadratic in $\partial_t \bv$)
should yield a symmetric linear system when discretized with finite element shape
functions. The next section develops this discretization.

\subsection{Rayleigh--Ritz discretization}
\label{sec:rayleigh_ritz}

We discretize the velocity field using finite element shape functions
$N_a(\bx)$, writing $\bv(\bx, t) = \sum_a N_a(\bx) \, \bd_a(t)$, where
$\bd \in \Real^{n_{\text{dofs}}}$ collects the nodal velocity coefficients. The
time derivative (the field with respect to which the minimization is carried out ) is interpolated with the same basis:
$\partial_t \bv = \sum_a N_a \, \dot{\bd}_a$. The specific choice of element
and shape functions is deferred to Section~\ref{sec:q9_element}; the
development here is independent of that choice.

Substituting into the PMPG functional~\eqref{eq:pmpg_continuous} converts
the infinite-dimensional minimization into a finite-dimensional quadratic
program over $\dot{\bd}$. Denoting the residual vector
\begin{equation}
  \bm{R}(\dot{\bd}) = \pd{\bv}{t} + (\bv \cdot \nabla)\bv
  - \nu \nabla^2 \bv
  = \sum_a N_a \dot{\bd}_a + \bg,
  \label{eq:residual}
\end{equation}
where $\bg = (\bv \cdot \nabla)\bv - \nu \nabla^2 \bv$ collects the terms
that depend on the known velocity field, the functional becomes
$\mathcal{J} = \frac{1}{2} \int_\Omega |\bm{R}|^2 \, d\Omega$, which is
quadratic in $\dot{\bd}$ with a unique minimum. Differentiating with respect
to $\dot{d}_i$ (the DOF associated with component $k$ at node $a$) gives
\begin{equation}
  \pd{\mathcal{J}}{\dot{d}_i}
  = \int_\Omega \bm{R} \cdot \pd{\bm{R}}{\dot{d}_i} \, d\Omega
  = \int_\Omega R_k \, N_a \, d\Omega = 0.
  \label{eq:stationarity}
\end{equation}
Expanding $R_k = \sum_b N_b \dot{d}_{k,b} + g_k$ and separating the unknown
$\dot{\bd}$ from the known terms:
\begin{equation}
  \sum_b \left( \int_\Omega N_a \, N_b \, d\Omega \right) \dot{d}_{k,b}
  + \int_\Omega N_a \left[ (\bv \cdot \nabla) v_k
  - \nu \nabla^2 v_k \right] d\Omega = 0.
  \label{eq:stationarity_expanded}
\end{equation}
Applying integration by parts to the viscous term
($\int_\Omega N_a \nabla^2 v_k \, d\Omega = -\int_\Omega \nabla N_a \cdot
\nabla v_k \, d\Omega + \int_{\partial\Omega} N_a \, \partial_n v_k \,
d\Gamma$; the boundary integral vanishes on Dirichlet boundaries where the
velocity is prescribed by constraints, and provides the natural zero-stress
condition on free-outflow boundaries)
and collecting the coefficients, we obtain the semi-discrete system
\begin{equation}
  \bM \, \dot{\bd} = -\bm{r}_{\text{conv}}(\bd) - \nu \bK \, \bd,
  \label{eq:pmpg_stationarity}
\end{equation}
with the global operators
\begin{alignat}{2}
  M_{ij} &= \int_\Omega N_i \, N_j \, d\Omega
  &\qquad& \text{(mass matrix)}, \label{eq:mass} \\
  K_{ij} &= \int_\Omega \nabla N_i \cdot \nabla N_j \, d\Omega
  && \text{(stiffness matrix)}, \label{eq:stiffness} \\
  r_{\text{conv},i} &= \int_\Omega N_i \left[ (\bv \cdot \nabla) \bv \right]
  d\Omega
  && \text{(convective residual)}. \label{eq:r_conv}
\end{alignat}
In practice, the global integrals decompose as sums of element contributions,
e.g.\ $M_{ij} = \sum_e \int_{\Omega_e} N_i \, N_j \, d\Omega_e$, evaluated
on the reference element with the appropriate Gauss quadrature ruleCa.
Both $\bM$ and $\bK$ have a block-diagonal
structure $\diag(\bM_s, \bM_s)$ and $\diag(\bK_s, \bK_s)$, reflecting the
independent interpolation of $v_x$ and $v_y$. The convective residual is
quadratic in $\bd$ and must be reassembled at each time step.

The variational structure of the PMPG formulation endows these operators with
favorable properties. The mass matrix $\bM$ is symmetric positive definite,
and the stiffness matrix $\bK$ is symmetric positive semi-definite, since
$\bd^T \bK \bd = \int_\Omega |\nabla \bv|^2 \, d\Omega \geq 0$. It follows
that any linear combination $\bM + \alpha \bK$ with $\alpha \geq 0$ is
symmetric positive definite, a property that guarantees well-posedness of the
time-stepping system derived below.

\paragraph{Constrained instantaneous problem.}
Equation~\eqref{eq:pmpg_stationarity} is an instantaneous condition: at each
time $t$, the velocity field $\bd$ is known, and the velocity rate $\dot{\bd}$
is the unknown to be determined. The solution must additionally satisfy the
incompressibility constraint and boundary conditions. These constraints are all
linear in $\bd$ (trivially so for Dirichlet boundary conditions, and as a
consequence of the subparametric geometry mapping for the divergence constraint;
see Section~\ref{sec:subparametric}) and can be collected as
\begin{equation}
  \bC \, \bd = \bb,
  \label{eq:constraints_abstract}
\end{equation}
where $\bC$ is the global constraint matrix and $\bb$ the prescribed values;
the specific construction of $\bC$ and $\bb$ is detailed in
Sections~\ref{sec:div_constraint} and~\ref{sec:boundary_conditions}. Since the
current state satisfies $\bC \bd = \bb$, the constraint on the rate of change
is $\bC \dot{\bd} = \bm{0}$, i.e., the velocity rate must itself be
divergence-free and must not alter the prescribed boundary values.

The procedure is then, in principle: solve~\eqref{eq:pmpg_stationarity} for
$\dot{\bd}$ subject to $\bC \dot{\bd} = \bm{0}$, update
$\bd^{n+1} = \bd^n + \Delta t \, \dot{\bd}$, and repeat.

\paragraph{Time discretization.}
A key feature of the present formulation is that the forward Euler
discretization $\dot{\bd} \approx (\bd^{n+1} - \bd^n) / \Delta t$ transforms
the semi-discrete equation~\eqref{eq:pmpg_stationarity} into a system for
$\bd^{n+1}$ directly. This is significant because $\bd^{n+1}$ is the quantity
to which we apply the constraints: replacing $\bC \dot{\bd} = \bm{0}$ with
$\bC \bd^{n+1} = \bb$ is exact, and the constrained system can be solved in a
single monolithic step (Section~\ref{sec:monolithic}).

Substituting the forward Euler discretization
into~\eqref{eq:pmpg_stationarity} yields a fully explicit time step:
$\bM \, \bd^{n+1} = \bM \, \bd^n - \Delta t \, \bm{r}_{\text{conv}}
- \nu \Delta t \, \bK \, \bd^n$.
For numerical stability, we replace the explicit viscous term
$-\nu \bK \bd^n$ with the implicit evaluation $-\nu \bK \bd^{n+1}$,
yielding the semi-implicit system
\begin{equation}
  (\bM + \nu \Delta t \, \bK) \, \bd^{n+1}
  = \bM \, \bd^n - \Delta t \, \bm{r}_{\text{conv}}(\bd^n).
  \label{eq:semi_implicit}
\end{equation}
This removes the diffusive stability constraint
$\Delta t \sim h^2 / \nu$, leaving only the convective CFL condition
$\Delta t \sim h / |\bv|$. The modification does not change the steady-state
solution (where $\bd^{n+1} = \bd^n$) and is essential for practical
computations at moderate Reynolds numbers.

\begin{remark}\label{rem:stability}
The semi-implicit system~\eqref{eq:semi_implicit} has a structural property
that distinguishes it from standard Galerkin discretizations. The system matrix
$\bA = \bM + \nu \Delta t \, \bK$ is symmetric positive definite for any
positive viscosity and time step, regardless of the Reynolds number: the mass
matrix $\bM$ is SPD, the stiffness matrix $\bK$ is positive semi-definite, and
convection appears only in the explicit right-hand side
$\bm{r}_{\text{conv}}(\bd^n)$. Each time step is therefore a constrained
minimization of a convex quadratic functional, and the solution is uniquely
defined.  In Galerkin finite element methods, by contrast, the convection
operator contributes an asymmetric term to the system matrix that can violate
discrete coercivity when the element P\'{e}clet number
$\text{Pe}_h = |\bv| h / (2\nu) \gg 1$, producing spurious
oscillations~\cite{brooks1982}. The PMPG formulation avoids this mechanism
entirely: the CFL condition restricts the time step for temporal accuracy, not
for preventing spatial oscillations. Numerical evidence for this property at
$Re = 1000$ on coarse meshes with $\text{Pe}_h$ up to~42 is presented in
Section~\ref{sec:cavity_re1000}.
\end{remark}

The development above is independent of the specific element type or problem dimension.
Next we will detail the specifics of our implementation: the Q9
biquadratic element used for velocity interpolation in 2D problems
(Section~\ref{sec:q9_element}), the pointwise divergence-free constraint
construction (Section~\ref{sec:div_constraint}), boundary conditions
(Section~\ref{sec:boundary_conditions}), and the monolithic saddle-point
solver that integrates all components (Section~\ref{sec:monolithic}).

\subsection{Q9 finite element discretization}
\label{sec:q9_element}

The velocity and velocity rate fields are discretized using 9-node biquadratic (Q9) Lagrangian
finite elements. Each element carries 9 nodes, and the field is interpolated
using the full set of biquadratic shape functions. The coordinate mapping from
the reference to the physical element uses a \emph{subparametric} approach
detailed in Section~\ref{sec:subparametric}.

\subsubsection{Reference element}

\begin{figure}[htbp]
\centering
\begin{tikzpicture}[scale=1.2, every node/.style={font=\small}]
  \begin{scope}[shift={(-2.8,0)}]
    \draw[thick] (-1,-1) -- (1,-1) -- (1,1) -- (-1,1) -- cycle;
    \foreach \x/\y/\lab in {-1/-1/1, 1/-1/2, 1/1/3, -1/1/4}{
      \fill (\x,\y) circle (2.5pt);
      }
    \node[below left]  at (-1,-1) {$1$};
    \node[below right] at ( 1,-1) {$2$};
    \node[above right] at ( 1, 1) {$3$};
    \node[above left]  at (-1, 1) {$4$};
    \foreach \x/\y in {0/-1, 1/0, 0/1, -1/0}{
      \draw[fill=white, thick] (\x,\y) circle (2.5pt);
    }
    \node[below right] at ( 0,-1) {$5$};
    \node[above right] at ( 1, 0) {$6$};
    \node[above right] at ( 0, 1) {$7$};
    \node[above left]  at (-1, 0) {$8$};
    \draw[fill=white, thick] (0,0) circle (2.5pt);
    \node[above right] at (0,0) {$9$};
    \draw[->, thick, gray] (-1.6,0) -- (1.7,0) node[right] {$\xi$};
    \draw[->, thick, gray] (0,-1.6) -- (0,1.7) node[above] {$\eta$};
    \node[below left, gray, font=\scriptsize] at (-1,0) {$-1$};
    \node[below right, gray, font=\scriptsize] at (1,0) {$1$};
    \node[below left, gray, font=\scriptsize] at (0,-1) {$-1$};
    \node[above left, gray, font=\scriptsize] at (0,1) {$1$};
    \node[below] at (0,-1.7) {Reference element};
  \end{scope}

  \draw[->, very thick, >=stealth] (-0.2,0) -- (1.0,0)
    node[midway, above, font=\footnotesize] {$\bx(\xi,\eta)$};

  \begin{scope}[shift={(3.2,0)}]
    \coordinate (P1) at (-1.3,-0.9);
    \coordinate (P2) at ( 1.1,-1.1);
    \coordinate (P3) at ( 1.4, 0.8);
    \coordinate (P4) at (-0.8, 1.2);
    \coordinate (P5) at ($0.5*(P1)+0.5*(P2)$);
    \coordinate (P6) at ($0.5*(P2)+0.5*(P3)$);
    \coordinate (P7) at ($0.5*(P3)+0.5*(P4)$);
    \coordinate (P8) at ($0.5*(P4)+0.5*(P1)$);
    \coordinate (P9) at ($0.25*(P1)+0.25*(P2)+0.25*(P3)+0.25*(P4)$);
    \draw[thick] (P1) -- (P2) -- (P3) -- (P4) -- cycle;
    \foreach \p in {P1,P2,P3,P4}{
      \fill (\p) circle (2.5pt);
    }
    \node[below left]  at (P1) {$1$};
    \node[below right] at (P2) {$2$};
    \node[above right] at (P3) {$3$};
    \node[above left]  at (P4) {$4$};
    \foreach \p in {P5,P6,P7,P8}{
      \draw[fill=white, thick] (\p) circle (2.5pt);
    }
    \node[below]       at (P5) {$5$};
    \node[right]       at (P6) {$6$};
    \node[above]       at (P7) {$7$};
    \node[left]        at (P8) {$8$};
    \draw[fill=white, thick] (P9) circle (2.5pt);
    \node[above right] at (P9) {$9$};
    \draw[->, thick, gray] (-1.7,-1.6) -- (1.8,-1.6) node[right] {$x$};
    \draw[->, thick, gray] (-1.7,-1.6) -- (-1.7,1.6) node[above] {$y$};
    \node[below] at (0.1,-1.9) {Physical element};
  \end{scope}
\end{tikzpicture}
\caption{Q9 reference element (left) and a representative physical element
  (right) under the subparametric mapping $\bx(\xi,\eta)$.
  Filled circles (\protect\tikz\protect\fill circle (2pt);) denote the 4 corner
  nodes that define the bilinear geometry mapping; open circles
  (\protect\tikz\protect\draw[fill=white,thick] circle (2pt);) denote the 5
  midside and center nodes used only for the biquadratic field
  interpolation. In the physical element, the midside nodes lie at the
  midpoints of the (straight) edges, and the center node lies at the
  intersection of the bimedians, as determined by the bilinear map.}
\label{fig:q9_element}
\end{figure}
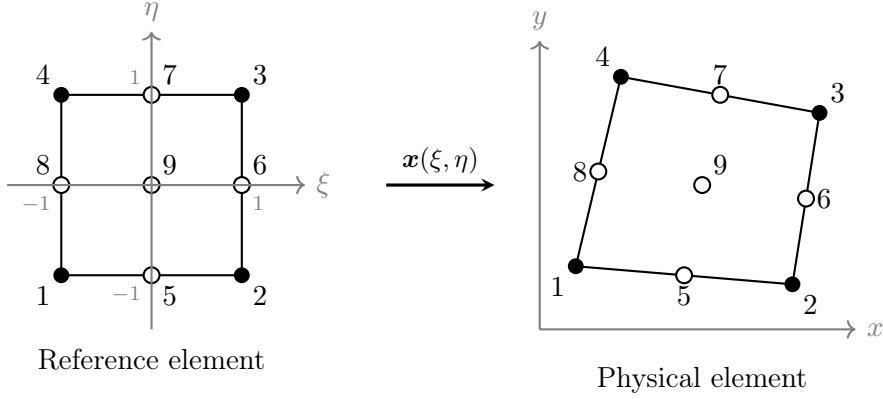

The Q9 element is defined on the reference square $[-1,1]^2$ with 9 nodes:
4 corners, 4 midside nodes, and 1 center node
(Figure~\ref{fig:q9_element}, left).
The reference coordinates $(\xi_a, \eta_a)$ for each node $a$ are also shown in the figure.

\subsubsection{Field interpolation}

Any 2D vector field defined at the 9 element nodes is interpolated using the
full set of Q9 biquadratic shape functions. In the present formulation, the two
fields of interest are the velocity $\bv$ and its time derivative
$\partial_t \bv$; both are interpolated with the same shape functions but with
different nodal coefficients.

The shape functions are constructed as tensor products of 1D quadratic
Lagrange polynomials on $\{-1, 0, 1\}$:
\begin{equation}
  \ell_0(t) = \frac{t(t-1)}{2}, \quad
  \ell_1(t) = 1-t^2, \quad
  \ell_2(t) = \frac{t(t+1)}{2}.
  \label{eq:lagrange_1d}
\end{equation}

The 2D shape functions are
$N_a(\xi, \eta) = \ell_{i_a}(\xi) \, \ell_{j_a}(\eta)$ for $a = 1, \ldots, 9$,
where $(i_a, j_a)$ are the 1D basis indices associated with node $a$.
A generic vector field $\boldsymbol{f} = (f_x, f_y)$ is then represented by its
nodal coefficient vector
\begin{equation}
  \boldsymbol{c} = \begin{pmatrix} c_1, \ldots, c_9, c_{10}, \ldots, c_{18}
  \end{pmatrix}^T \in \Real^{18},
  \label{eq:dof_vector}
\end{equation}
so that
\begin{equation}
  f_x(\xi, \eta) = \sum_{a=1}^{9} N_a \, c_a, \qquad
  f_y(\xi, \eta) = \sum_{a=1}^{9} N_a \, c_{9+a}.
  \label{eq:field_interp}
\end{equation}
For the velocity field, $\boldsymbol{c} = \bd$ with
$\bd = (v_x^1, \ldots, v_x^9, v_y^1, \ldots, v_y^9)^T$; for the time
derivative, $\boldsymbol{c} = \dot{\bd}$ with
$\dot{\bd} = (\dot{v}_x^1, \ldots, \dot{v}_x^9, \dot{v}_y^1, \ldots,
\dot{v}_y^9)^T$.

\subsubsection{Physical gradients}

Computing physical derivatives of the interpolated fields requires the Jacobian
$\bJ$ of the coordinate mapping from the reference element to the physical
element. For a scalar component $f$ with nodal values $(c_1, \ldots, c_9)$,
the physical derivatives are:
\begin{equation}
  \pd{f}{x} = \frac{j_{22} f_{,\xi} - j_{21} f_{,\eta}}{\det(\bJ)}, \qquad
  \pd{f}{y} = \frac{-j_{12} f_{,\xi} + j_{11} f_{,\eta}}{\det(\bJ)},
  \label{eq:phys_derivs}
\end{equation}
where $f_{,\xi} = \bB_\xi \cdot \boldsymbol{c}$
and $f_{,\eta} = \bB_\eta \cdot \boldsymbol{c}$, with $\bB_\xi$ and $\bB_\eta$
the row vectors of parametric shape function derivatives:
\begin{equation}
  \bB_\xi = \begin{pmatrix} \pd{N_1}{\xi} & \cdots & \pd{N_9}{\xi} \end{pmatrix},
  \qquad
  \bB_\eta = \begin{pmatrix} \pd{N_1}{\eta} & \cdots & \pd{N_9}{\eta}
  \end{pmatrix}.
  \label{eq:shape_derivs}
\end{equation}
Applying~\eqref{eq:phys_derivs} to each component of $\bv$ or
$\partial_t \bv$ yields the corresponding physical gradient tensor. The
Jacobian components $j_{ij}$ and $\det(\bJ)$ depend on the specific geometry
mapping, which we detail next.

\subsubsection{Subparametric geometry mapping}
\label{sec:subparametric}

The physical gradient formulas~\eqref{eq:phys_derivs} require the Jacobian of a
coordinate mapping from the reference element to the physical element. In the
standard isoparametric approach, the same biquadratic functions used for velocity
interpolation would also map the geometry, producing a Jacobian that is quadratic
in $(\xi, \eta)$. We instead adopt a \emph{subparametric} mapping that uses
bilinear shape functions constructed from the 4 corner nodes only
(Figure~\ref{fig:q9_element}, right):
\begin{equation}
  \bx(\xi, \eta) = \sum_{a=1}^{4} N_a^{\text{lin}}(\xi, \eta) \, \bx_a, \qquad
  N_a^{\text{lin}} = \frac{(1 + \xi_a \xi)(1 + \eta_a \eta)}{4}.
  \label{eq:geometry_map}
\end{equation}

The Jacobian matrix of this mapping,
\begin{equation}
  \bJ = \begin{pmatrix}
    \pd{x}{\xi} & \pd{x}{\eta} \\[0.3em]
    \pd{y}{\xi} & \pd{y}{\eta}
  \end{pmatrix},
  \label{eq:jacobian_matrix}
\end{equation}
depends at most linearly on $(\xi, \eta)$. It can be parameterized by 6 scalars:
\begin{align}
  j_{11}^0 &= \tfrac{1}{4}(-x_1 + x_2 + x_3 - x_4), &
  j_{12}^0 &= \tfrac{1}{4}(-x_1 - x_2 + x_3 + x_4), \nonumber \\
  j_{21}^0 &= \tfrac{1}{4}(-y_1 + y_2 + y_3 - y_4), &
  j_{22}^0 &= \tfrac{1}{4}(-y_1 - y_2 + y_3 + y_4), \label{eq:jacobian_params} \\
  h_1 &= \tfrac{1}{4}(x_1 - x_2 + x_3 - x_4), &
  h_2 &= \tfrac{1}{4}(y_1 - y_2 + y_3 - y_4), \nonumber
\end{align}
so that the Jacobian components at $(\xi, \eta)$ are:
\begin{equation}
  j_{11} = j_{11}^0 + h_1 \eta, \quad
  j_{12} = j_{12}^0 + h_1 \xi, \quad
  j_{21} = j_{21}^0 + h_2 \eta, \quad
  j_{22} = j_{22}^0 + h_2 \xi.
  \label{eq:jac_components}
\end{equation}

The Jacobian determinant is at most linear in $(\xi, \eta)$:
\begin{equation}
  \det(\bJ) = J_0 + J_\xi \xi + J_\eta \eta,
  \label{eq:detJ}
\end{equation}
where $J_0 = j_{11}^0 j_{22}^0 - j_{12}^0 j_{21}^0$,
$J_\xi = j_{11}^0 h_2 - j_{21}^0 h_1$, and
$J_\eta = h_1 j_{22}^0 - h_2 j_{12}^0$.

\begin{remark}
The subparametric choice is central to the formulation. The physical
derivatives~\eqref{eq:phys_derivs} involve division by $\det(\bJ)$. In
Section~\ref{sec:div_constraint}, the divergence constraint
$\partial v_x / \partial x + \partial v_y / \partial y = 0$ is formed from
these physical derivatives; multiplying through by $\det(\bJ)$ clears the
denominator and yields a constraint row that is a linear combination of the
products $j_{ij} \, \partial N_a / \partial \xi_k$, linear in the nodal
velocities $\bd$. Because the subparametric Jacobian components $j_{ij}$ are
at most bilinear in $(\xi, \eta)$, these products remain low-degree
polynomials, well-suited to standard Gauss quadrature. An isoparametric
formulation with full biquadratic geometry would produce higher-degree Jacobian
entries, complicating the constraint structure without benefit for
straight-sided or gently curved elements.
\end{remark}

\subsection{Divergence-free constraint}
\label{sec:div_constraint}

The incompressibility condition $\nabla \cdot \bv = 0$ is the sole element-level
constraint in the viscous formulation. At a quadrature point
$(\xi_k, \eta_k)$, after multiplying through by $\det(\bJ)$ to avoid division
by the Jacobian determinant, the constraint becomes:
\begin{equation}
  j_{22} v_{x,\xi} - j_{21} v_{x,\eta} - j_{12} v_{y,\xi}
  + j_{11} v_{y,\eta} = 0.
  \label{eq:div_constraint}
\end{equation}

Substituting the parametric derivatives in terms of $\bd$:
\begin{equation}
  \bc_k^{\text{div}} \cdot \bd = 0, \qquad
  \bc_k^{\text{div}} = \begin{pmatrix}
    j_{22}\bB_\xi - j_{21}\bB_\eta & -j_{12}\bB_\xi + j_{11}\bB_\eta
  \end{pmatrix}_k \in \Real^{1 \times 18}.
  \label{eq:div_row}
\end{equation}

This constraint is \emph{linear} in $\bd$ because the Jacobian components
depend only on the (fixed) corner coordinates, not on the velocity DOFs.

We enforce~\eqref{eq:div_row} at $2 \times 2$ Gauss quadrature points, yielding
4 constraint rows per element. The $2 \times 2$ rule is the minimum order that
captures the divergence of the biquadratic velocity field at sufficient points
for accuracy, while leaving $18 - 4 = 14$ unconstrained degrees of freedom per
element, enough for the velocity to satisfy the momentum equation. A
higher-order rule (e.g., $3 \times 3$) would impose 9 constraints per element,
leaving only 9 free DOFs and over-constraining the global system. The 4
constraint rows are always linearly independent (rank 4) for non-degenerate elements, so none are
redundant at the element level.

\begin{remark}\label{rem:orthogonality_main}
	The $2\times 2$ Gauss points have a structural property beyond the
	DOF-count argument above.
	Since $(\nabla\!\cdot\!\bm{v})\,\det(\bm{J})$ is at most degree~$(2,2)$
	in $(\xi,\eta)$ (the biquadratic shape function derivatives contribute
	degree~2 and the subparametric Jacobian components are at most linear),
	its product with any bilinear function $q(\xi,\eta)$ is at most
	degree~$(3,3)$, which the $2\times 2$ Gauss--Legendre rule integrates
	exactly.  Consequently, if $(\nabla\!\cdot\!\bm{v})\,\det(\bm{J}) = 0$
	at the four Gauss points, then for \emph{every} bilinear~$q$,
	\begin{equation}\label{eq:weak_orthog_main}
		\int_{\Omega_e}(\nabla\!\cdot\!\bm{v})\,q\;d\Omega = 0.
	\end{equation}
	Any divergence residual that survives the constraints is therefore
	$L^2$-orthogonal to the bilinear polynomial space $Q_1 =
	\operatorname{span}\{1,\,\xi,\,\eta,\,\xi\eta\}$ on each element: it has
	zero mean, zero first moments, and zero bilinear moment.  Only high-order
	polynomial modes remain, and these oscillate within each element and tend
	to cancel across neighbors rather than accumulate coherently.  This
	``high-pass filtering'' of the divergence residual explains why the
	$2\times 2$ Gauss rule outperforms collocation at points that minimize a
	per-element leakage measure; see the Kovasznay flow comparison in
	Section~3.2 and the detailed analysis in Appendix~A.
\end{remark}

The element constraint rows are assembled into a global constraint matrix
$\bC$ by scattering through the element DOF map, in the same way that element
stiffness matrices are assembled into a global stiffness matrix. Note that the
element-level DOF ordering~\eqref{eq:dof_vector} groups all $v_x$ components
before all $v_y$ components, whereas the global DOF vector
$\bd \in \Real^{n_{\text{dofs}}}$ uses an interleaved ordering: DOF $2i$
corresponds to $v_x$ at global node $i$ and DOF $2i+1$ to $v_y$ at global
node $i$. The element DOF map translates between these two orderings during
assembly. Each row of
$\bC$ enforces $\nabla \cdot \bv = 0$ at a specific quadrature point in a
specific element. Boundary conditions (Section~\ref{sec:boundary_conditions})
are appended as additional rows of $\bC$, and the complete system is solved via
the saddle-point formulation described in Section~\ref{sec:monolithic}.

A key feature of this formulation is that incompressibility is enforced
\emph{pointwise} at the quadrature points, not in a weak or averaged sense.
The resulting velocity field is exactly divergence-free at every Gauss point
in every element. This is a direct consequence of the variational structure:
in the PMPG framework, incompressibility enters as an equality constraint on
the minimization, and the Lagrange multipliers enforce it exactly. There is no
pressure finite element space, no inf-sup condition to satisfy, and no penalty
parameter to tune. The number and placement of the constraints are determined
entirely by the choice of quadrature rule.

This stands in contrast to mixed velocity--pressure formulations, where
incompressibility is enforced weakly: $\int_\Omega q \, \nabla \cdot \bv \,
d\Omega = 0$ for all test functions $q$ in a pressure space $Q_h$. The velocity
field in a mixed method is divergence-free only in the $L^2$ projection onto
$Q_h$, not pointwise. Furthermore, the pressure space must be chosen to satisfy
the Babu\v{s}ka--Brezzi inf-sup condition~\cite{babuska1973,brezzi1974}; if it
is not, the formulation is unstable and requires pressure
stabilization~\cite{hughes1986}. The present formulation avoids velocity--pressure inf--sup requirements because no pressure space is introduced; instead, robustness depends on the rank/consistency of the pointwise constraint operator and the conditioning of the resulting KKT system.

\subsection{Boundary conditions}
\label{sec:boundary_conditions}

All boundary conditions in the present formulation are of Dirichlet type:
prescribed velocity values at boundary nodes. Each prescribed component adds
one row to the global constraint matrix $\bC$ and a corresponding entry to the
right-hand side $\bb$. Specifically, prescribing $v_{x,i} = \bar{v}_{x,i}$
at node $i$ contributes a row with a single unit entry at DOF $2i$ in $\bC$
and the value $\bar{v}_{x,i}$ in $\bb$; an analogous row is added for any
prescribed $v_{y,i}$.

No-slip walls are the special case $\bar{v}_{x,i} = \bar{v}_{y,i} = 0$
(2 constraints per wall node). Inlet profiles are nonzero Dirichlet conditions
with values chosen to match the desired inflow (e.g., a parabolic Poiseuille
profile). In both cases, the constraint enters the saddle-point
system~\eqref{eq:saddle_point} in the same way as the divergence constraints.

At free-outlet boundaries, no constraints are imposed. The velocity at outlet
nodes is determined entirely by the element divergence constraints and the PMPG
minimization; the solver naturally adjusts the outflow to satisfy
incompressibility globally. This is analogous to the ``do-nothing'' or natural
outflow condition in conventional formulations~\cite{gresho1998}.

\subsection{Monolithic saddle-point solver}
\label{sec:monolithic}

\subsubsection{Coupled system}

A common approach in computational fluid dynamics is to
solve the unconstrained momentum equation~\eqref{eq:semi_implicit} first and
then project the result onto the constraint
manifold~\eqref{eq:constraints_abstract} in a separate step. This strategy,
introduced by Chorin~\cite{chorin1968} and widely adopted as the projection or
fractional-step method~\cite{guermond2006}, is straightforward to implement but
introduces a splitting error between the viscous and incompressibility
sub-problems that can compromise steady-state
accuracy~\cite{guermond2006}.
Exact projection methods without splitting error can also be formulated at the discrete level by projecting onto the null space of the discrete divergence operator (not including Dirichlet constraints) when an explicit (preferably sparse) divergence-free basis is available; constructing such a basis is straightforward for some staggered discretizations but can be nontrivial in general FE settings~\cite{chang2002analysis}.

In the present formulation, we instead incorporate the
constraints~\eqref{eq:constraints_abstract} directly into the time step via
Lagrange multipliers. As discussed in Section~\ref{sec:rayleigh_ritz}, the
constraint $\bC \bd^{n+1} = \bb$ is equivalent to
$\bC \dot{\bd} = \bm{0}$ and is therefore linear in the PMPG optimization
variable. Combined with the semi-implicit
system~\eqref{eq:semi_implicit}, this yields a coupled saddle-point system:
\begin{equation}
  \underbrace{
  \begin{pmatrix}
    \bA & \bC^T \\
    \bC & \bm{0}
  \end{pmatrix}
  }_{\text{saddle-point matrix}}
  \begin{pmatrix}
    \bd^{n+1} \\
    \blambda
  \end{pmatrix}
  =
  \begin{pmatrix}
    \bM \bd^n - \Delta t \, \bm{r}_{\text{conv}} \\
    \bb
  \end{pmatrix},
  \label{eq:saddle_point}
\end{equation}
where $\bA = \bM + \nu \Delta t \, \bK$ is the implicit operator,
$\bC \in \Real^{m \times n_{\text{dofs}}}$ is the full constraint matrix
(divergence + boundary conditions), $\bb$ is the constraint right-hand side
(zero for divergence and no-slip, prescribed values for inlet Dirichlet), and
$\blambda \in \Real^m$ is the vector of Lagrange multipliers.

The first block row of~\eqref{eq:saddle_point} is the constrained momentum
equation: the term $\bC^T \blambda$ is the discrete analogue of the constraint
forces that enforce incompressibility and boundary conditions. The second block
row enforces $\bC \bd^{n+1} = \bb$ exactly. This is a standard saddle-point
structure~\cite{benzi2005} that arises naturally from the constrained
minimization of the PMPG functional.

\subsubsection{Implementation}

The saddle-point matrix in~\eqref{eq:saddle_point} is sparse and its sparsity
pattern is fixed throughout the simulation. At each time step:
\begin{enumerate}
  \item The convective residual $\bm{r}_{\text{conv}}(\bd^n)$ is computed from
    the current velocity field.
  \item The right-hand side $\bM \bd^n - \Delta t \, \bm{r}_{\text{conv}}$ is
    assembled.
  \item The saddle-point system~\eqref{eq:saddle_point} is solved either by
    sparse LU factorization (SuperLU) for moderate-size problems, or by
    preconditioned GMRES for larger systems. The GMRES solver uses a
    block-triangular preconditioner based on an incomplete LU factorization
    of $\bA$ and an approximate Schur complement
    $\hat{S} = \bC \, \diag(\bM_L)^{-1} \bC^T$, where $\bM_L$ is the
    lumped mass matrix.
\end{enumerate}

Since the matrix $\bA$ depends on $\Delta t$ (through the term
$\nu \Delta t \bK$), the factorization (or preconditioner) is recomputed only
when $\Delta t$ changes by more than 10\%, which typically occurs only a few
times during the warm-up phase.

\begin{remark}
The saddle-point system~\eqref{eq:saddle_point} resembles the structure of mixed
finite element formulations~\cite{benzi2005}, but differs in two respects.
First, the constraint matrix $\bC$ encodes both incompressibility and boundary
conditions in a single operator; the Lagrange multipliers $\blambda$ are
constraint forces from the optimization, not nodal pressure values from a finite
element pressure space. Second, each row of $\bC$ enforces a pointwise equality
(divergence-free at a specific quadrature point, or a prescribed velocity at
a boundary node) rather than the weighted integral $\int_\Omega q \, \nabla \cdot
\bv \, d\Omega = 0$ of a mixed formulation. The velocity field therefore
satisfies $\nabla \cdot \bv = 0$ exactly at all quadrature points, a property
confirmed numerically by constraint residuals at machine precision ($\sim
10^{-13}$) in all test cases.
An equivalent velocity-only alternative is to parameterize the solution directly in the null space of the discrete divergence operator (with compatibility to the imposed Dirichlet constraints), but constructing a sparse explicit basis is generally the practical bottleneck in non-staggered discretizations~\cite{chang2002analysis}.
\end{remark}

\begin{remark}
The computational complexity per time step is governed by the saddle-point
solve. With $2 \times 2$ divergence quadrature, each element contributes 4
constraint rows, so the constraint matrix has $m \approx 4 n_e + m_{\text{bc}}$
rows, where $n_e$ is the number of elements and $m_{\text{bc}}$ is the number
of boundary DOFs. The total system dimension is $n + m$, where $n = 2 n_{\text{nodes}}$
is the number of velocity DOFs. For Q9 meshes on regular grids, $n_e \approx
n_{\text{nodes}} / 4$ (each interior node is shared by 4 elements), so
$m \approx n$, and the saddle-point system is roughly twice the size of the
velocity system alone. A sparse direct solver (LU factorization) requires
$O((n+m)^{3/2})$ operations in 2D for typical mesh topologies~\cite{benzi2005};
iterative solvers (e.g., preconditioned GMRES) reduce this to $O(n+m)$ per
iteration with an appropriate preconditioner, at the cost of iteration counts
that depend on the condition number. Since $\bA = \bM + \nu \Delta t \, \bK$ is
symmetric positive definite, established block preconditioners for saddle-point
systems~\cite{benzi2005} are directly applicable.
\end{remark}

\subsection{Adaptive CFL time stepping}
\label{sec:cfl}

The time step is determined adaptively by the convective CFL condition:
\begin{equation}
  \Delta t = c_{\text{CFL}} \cdot \frac{h_{\min}}{|\bv|_{\max}},
  \label{eq:cfl}
\end{equation}
where $h_{\min} = \min_e l_{c,e}$ is the global minimum characteristic element
length (computed from the shortest edge of the bilinear mapping),
$|\bv|_{\max}$ is the global maximum velocity magnitude, and
$c_{\text{CFL}} = 0.25$ is a safety factor. A ceiling $\Delta t_{\max}$
prevents unnecessarily large steps during the initial transient when velocities
are small.

\subsection{Wall force extraction from Lagrange multipliers}
\label{sec:wall_force}

The Lagrange multipliers $\blambda$ from the saddle-point
system~\eqref{eq:saddle_point} carry physical information about the constraint
forces. From the first block row of~\eqref{eq:saddle_point}:
\begin{equation}
  \bA \bd^{n+1} + \bC^T \blambda = \bM \bd^n - \Delta t \, \bm{r}_{\text{conv}},
\end{equation}
which can be rearranged as:
\begin{equation}
  \bM \frac{\bd^{n+1} - \bd^n}{\Delta t}
  + \bm{r}_{\text{conv}} + \nu \bK \bd^{n+1}
  = -\frac{\bC^T \blambda}{\Delta t}.
  \label{eq:force_balance}
\end{equation}
The left-hand side is the discrete momentum equation; the right-hand side
$-\bC^T \blambda / \Delta t$ is the total constraint force exerted on the
fluid.

Because the constraint matrix $\bC$ contains rows of different types
(divergence-free, wall boundary conditions, and far-field Dirichlet), the
product $\bC^T \blambda$ at a given DOF mixes contributions from all
constraints that involve that DOF. In particular, divergence rows from elements
sharing a wall node contribute to $[\bC^T \blambda]_i$ at wall DOFs, but these
terms represent volume-averaged pressure-like forces from the incompressibility
constraint, not the surface traction at the wall.

The wall surface traction is obtained by isolating the multipliers associated
with the wall boundary condition rows. For a no-slip wall, each wall DOF~$i$
has a dedicated constraint row $r(i)$ with a single unit entry:
$C_{r(i),\,i} = 1$. The corresponding multiplier $\lambda_{r(i)}$ is the force
needed to hold that DOF at zero, precisely the constraint reaction at the
wall. By Newton's third law, the force the fluid exerts on the wall at DOF~$i$
is:
\begin{equation}
  f_i = \frac{\lambda_{r(i)}}{\Delta t}, \qquad
  i \in \{\text{wall DOFs}\}.
  \label{eq:wall_force}
\end{equation}

For example, if dealing with a lifting body, the total drag and lift on the wall are obtained by summing the $x$- and
$y$-components over all wall nodes:
\begin{equation}
  F_x = \sum_{i \in \text{wall}} f_{2i}, \qquad
  F_y = \sum_{i \in \text{wall}} f_{2i+1}.
  \label{eq:total_force}
\end{equation}
Importantly, these nodal forces include both pressure and viscous shear
contributions: the multiplier $\lambda_{r(i)}$ is the total force needed to
maintain the no-slip condition at node~$i$, encompassing the effect of the full
stress tensor without requiring separate pressure or viscous stress
reconstructions. This provides a natural and accurate mechanism for computing
lift and drag without reconstructing the pressure field. Moreover, the
availability of nodal forces directly from the Lagrange multipliers makes the
formulation particularly attractive for fluid--structure interaction, where
these forces serve as the coupling loads applied to the structural solver.

\section{Verification}
\label{sec:verification}

Before applying the formulation to problems of engineering interest, we verify
it against two exact solutions of the Navier--Stokes equations. The two tests
are chosen to probe complementary aspects of the solver. Poiseuille channel flow
has a quadratic velocity profile that lies exactly in the Q9 biquadratic
interpolation space; the monolithic solver should therefore recover it to
machine precision, providing a stringent test of the constraint enforcement and
time-stepping machinery. The Kovasznay flow, by contrast, has an
exponential-trigonometric solution that the Q9 space cannot represent exactly;
it tests the spatial convergence rate of the discretization on a non-trivial
Navier--Stokes solution.

\subsection{Poiseuille flow}
\label{sec:poiseuille}

\subsubsection{Problem setup}

\begin{figure}[htbp]
\centering
\begin{tikzpicture}[scale=1.0, every node/.style={font=\small},
  hatch/.style={pattern=north east lines, pattern color=gray}]
  \pgfmathsetmacro{\Hd}{1.5}   
  \pgfmathsetmacro{\Ld}{7.5}   
  \fill[hatch] (0,-0.25) rectangle (\Ld,0);
  \draw[thick] (0,0) -- (\Ld,0);
  \fill[hatch] (0,\Hd) rectangle (\Ld,\Hd+0.25);
  \draw[thick] (0,\Hd) -- (\Ld,\Hd);
  \draw[thick, blue] plot[smooth, domain=0:\Hd, samples=30]
    ({0.7*4*\x*(\Hd-\x)/(\Hd*\Hd)}, \x);
  \foreach \f in {0.15, 0.3, 0.5, 0.7, 0.85} {
    \pgfmathsetmacro{\yy}{\f*\Hd}
    \pgfmathsetmacro{\vv}{0.7*4*\yy*(\Hd-\yy)/(\Hd*\Hd)}
    \draw[thick, blue, -{Stealth[length=4pt]}] (0,\yy) -- (\vv,\yy);
  }
  \draw[thick, dashed] (\Ld,0) -- (\Ld,\Hd);
  \foreach \f in {0.2, 0.4, 0.6, 0.8} {
    \pgfmathsetmacro{\yy}{\f*\Hd}
    \draw[thick, gray, -{Stealth[length=3pt]}] (\Ld,\yy) -- ++(0.3,0);
  }
  \draw[<->] (0,-0.6) -- (\Ld,-0.6) node[midway, below] {$L = 5$};
  \draw[<->] (-0.8,0) -- (-0.8,\Hd) node[midway, left] {$H = 1$};
  \draw[->, thick, gray] (-0.8,-1.1) -- (0.7,-1.1) node[right] {$x$};
  \draw[->, thick, gray] (-0.8,-1.1) -- (-0.8,-0.3) node[left] {$y$};
\end{tikzpicture}
\caption{Poiseuille flow: problem setup. Hatched boundaries denote no-slip
  walls ($\bv = \bm{0}$), blue arrows indicate the prescribed parabolic inlet
  profile, and the dashed boundary marks the free outflow.}
\label{fig:poiseuille_setup}
\end{figure}
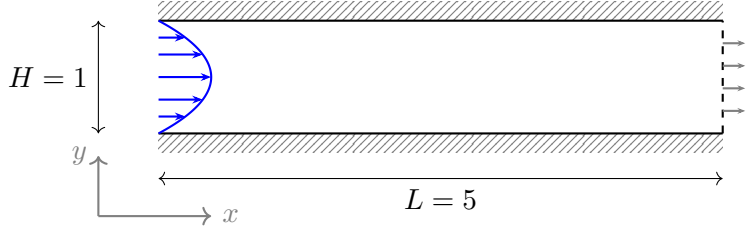

The first verification test is plane Poiseuille flow in a rectangular channel of
height $H = 1$ and length $L = 5$ (Figure~\ref{fig:poiseuille_setup}). The
boundary conditions are: no-slip on both walls ($y = 0$ and $y = H$), parabolic
Poiseuille profile at the inlet ($x = 0$),
\begin{equation}
  v_x(0, y) = u_{\max} \frac{4 y (H - y)}{H^2}, \qquad v_y(0, y) = 0,
  \label{eq:poiseuille_inlet}
\end{equation}
with $u_{\max} = 1.5$ ($U_{\text{mean}} = 1$), and free outflow at $x = L$.
The kinematic viscosity is $\nu = 0.1$, giving $Re = H \cdot U_{\text{mean}} /
\nu = 10$. The domain is discretized with structured Q9 meshes of $n_x \times
n_y$ elements, where $n_x = n_y L / H$ to maintain approximately unit aspect
ratio. We report results for $n_y = 4$, $8$, and $16$ (corresponding to $20
\times 4$, $40 \times 8$, and $80 \times 16$ element grids).

The simulation is initialized with zero velocity everywhere and advanced in
time using the monolithic saddle-point solver (Section~\ref{sec:monolithic})
until a steady state is reached, defined by $\|\partial_t \bv\| / \|\bv\| < 10^{-12}$.

This problem has a closed-form steady-state solution:
$v_x(x, y) = u_{\max} \cdot 4 y (H - y) / H^2$, $v_y = 0$, which is a
quadratic polynomial in $y$. Since the Q9 biquadratic interpolation space
contains all quadratic polynomials, the exact Poiseuille solution lies in the
finite element space. However, membership in the interpolation space is a
necessary but not sufficient condition for exact recovery: the time-stepping
scheme, constraint enforcement, and saddle-point solver must all preserve the
exact solution through the temporal evolution. This makes Poiseuille flow a
particularly stringent test of the monolithic formulation.

\subsubsection{Results}

\begin{figure}[htbp]
\centering
\includegraphics[width=0.8\textwidth]{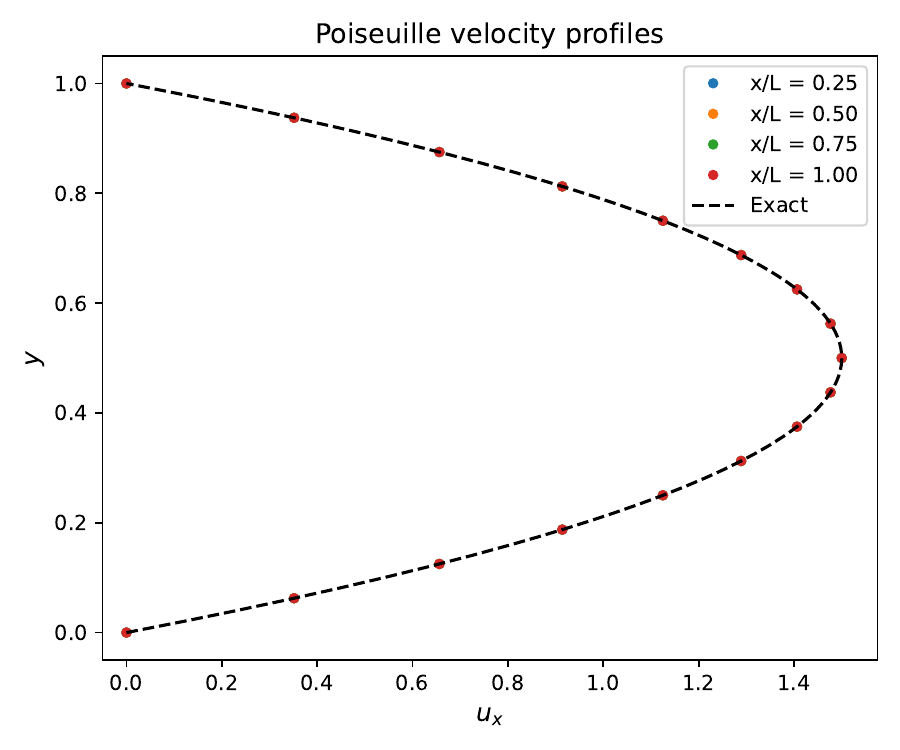}
\caption{Poiseuille flow: computed velocity profiles (symbols) on the $40
  \times 8$ mesh compared with the exact parabolic solution (solid line) at
  multiple $x$-stations.}
\label{fig:poiseuille_profiles}
\end{figure}

Figure~\ref{fig:poiseuille_profiles} shows the computed velocity profiles at
several $x$-stations, compared with the exact parabolic profile on the $40
\times 8$ mesh. The agreement is visually perfect; the results are
indistinguishable on the other meshes.

Quantitatively, the monolithic solver achieves machine-precision accuracy ($L^2$
error $\sim 10^{-13}$) on all three meshes, confirming that the time-stepping
scheme, constraint enforcement, and saddle-point solve introduce no spurious
error.

\begin{figure}[htbp]
\centering
\includegraphics[width=0.8\textwidth]{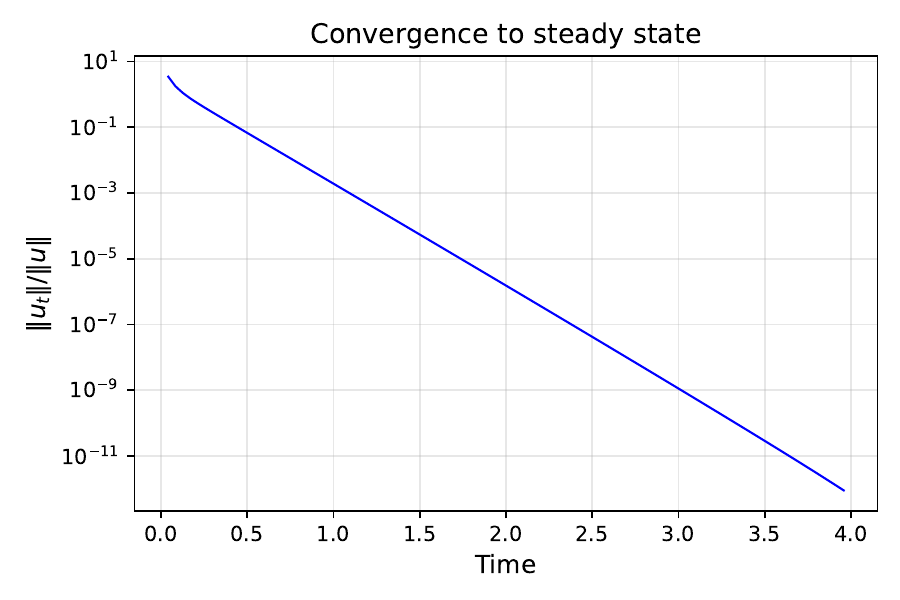}
\caption{Poiseuille flow: convergence history on the $40 \times 8$ mesh. The
  time-derivative norm $\|\partial_t \bv\|$ converges to machine precision,
  confirming exact steady-state recovery.}
\label{fig:poiseuille_temporal}
\end{figure}

Figure~\ref{fig:poiseuille_temporal} shows the convergence history for the $40
\times 8$ mesh. The time-derivative norm $\|\partial_t \bv\|$ decreases
monotonically and reaches machine precision within approximately 100 time steps,
confirming rapid convergence to steady state.

\subsection{Kovasznay flow}
\label{sec:kovasznay}

\subsubsection{Problem setup}

\begin{figure}[htbp]
\centering
\newlength{\kovsquare}\setlength{\kovsquare}{4.5cm}%
\begin{subfigure}[t]{0.35\textwidth}
  \vspace{0pt}%
  \centering
  \begin{tikzpicture}[scale=1.05, every node/.style={font=\small},
    baseline={(0,3.0)}]
    \pgfmathsetmacro{\S}{3.0}
    \draw[thick, blue] (0,0) rectangle (\S,\S);
    \node[below, blue] at (0.5*\S,0) {$\bv = \bv_{\text{exact}}$};
    \node[above, blue] at (0.5*\S,\S) {$\bv = \bv_{\text{exact}}$};
    \node[left, blue, rotate=90, anchor=south] at (0,0.5*\S)
      {$\bv = \bv_{\text{exact}}$};
    \node[right, blue, rotate=-90, anchor=south] at (\S,0.5*\S)
      {$\bv = \bv_{\text{exact}}$};
    \foreach \f in {0.15, 0.35, 0.5, 0.65, 0.85} {
      \pgfmathsetmacro{\yy}{\f*\S}
      \draw[-{Stealth[length=3pt]}, gray!70] (0.2,\yy) -- (0.9,\yy);
      \draw[-{Stealth[length=3pt]}, gray!70] (1.2,\yy) -- (1.9,\yy);
      \draw[-{Stealth[length=3pt]}, gray!70] (2.2,\yy) -- (2.7,\yy);
    }
    \draw[->, thick, gray] (-1,-0.6) -- (0.3,-0.6) node[right] {$x$};
    \draw[->, thick, gray] (-1,-0.6) -- (-1,0.2) node[left] {$y$};
  \end{tikzpicture}
  \caption{Problem setup.}
  \label{fig:kovasznay_setup_schematic}
\end{subfigure}
\hfill
\begin{subfigure}[t]{0.5\textwidth}
  \vspace{0pt}%
  \centering
  \raisebox{-\height}{\includegraphics[height=\kovsquare]{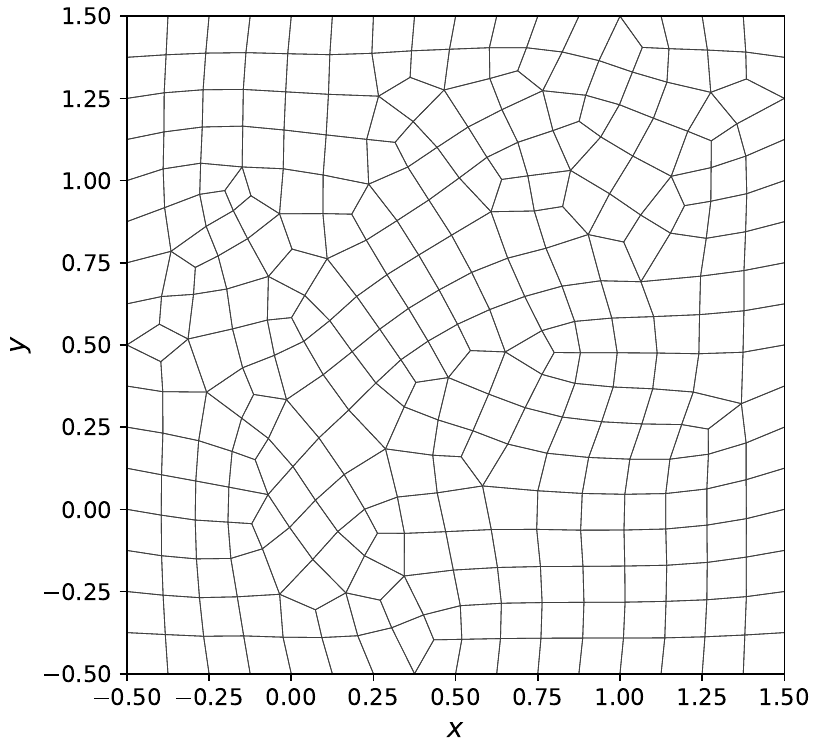}}
  \caption{Unstructured Q9 mesh ($n = 16$, 302 elements).}
  \label{fig:kovasznay_mesh}
\end{subfigure}
\caption{Kovasznay flow: (a) problem setup: square domain $[-0.5, 1.5]^2$
  with the exact Kovasznay solution~\eqref{eq:kovasznay} prescribed as Dirichlet
  conditions (blue) on all sides; gray arrows indicate the predominantly
  rightward mean flow; (b) representative unstructured Q9 mesh ($n = 16$).}
\label{fig:kovasznay_setup}
\end{figure}
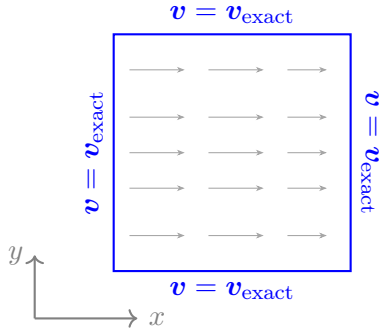
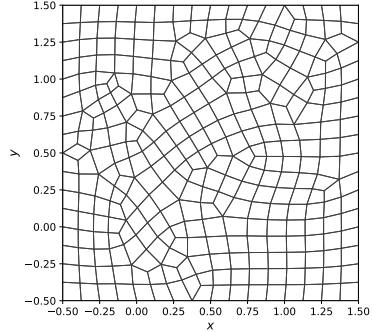

The second verification test uses the exact Kovasznay
solution~\cite{kovasznay1948}, a non-trivial steady-state solution of the
incompressible Navier--Stokes equations that describes the flow downstream of a
two-dimensional grid. The velocity field is given by:
\begin{align}
  v_x &= 1 - e^{\lambda x} \cos(2\pi y), \nonumber \\
  v_y &= \frac{\lambda}{2\pi} e^{\lambda x} \sin(2\pi y),
  \label{eq:kovasznay}
\end{align}
where $\lambda = Re/2 - \sqrt{Re^2/4 + 4\pi^2}$ and $Re = 1/\nu$. The domain
is $[-0.5, 1.5]^2$ (Figure~\ref{fig:kovasznay_setup}), and all boundary
conditions are Dirichlet with the exact solution prescribed on every edge. There
are no solid boundaries; this is a pure spatial accuracy test for the
Navier--Stokes operator. We use $Re = 40$ ($\nu = 0.025$).

Unlike Poiseuille flow, the Kovasznay solution contains exponential and
trigonometric terms that the Q9 interpolation space cannot represent exactly.
The expected convergence rate in the $L^2$ norm is $O(h^3)$ for biquadratic
elements, and the purpose of this test is to verify that the solver attains it.

We perform a spatial convergence study on unstructured Q9 meshes
(Figure~\ref{fig:kovasznay_mesh}) with approximate element size $h = 2/n$ for
$n = 4$, $8$, $16$, and $32$. Using unstructured meshes provides a more
stringent test of the formulation than a regular grid. Each simulation is initialized with zero velocity and advanced in
time using the monolithic solver until the steady-state criterion $\|\partial_t
\bv\| / \|\bv\| < 10^{-12}$ is met. The $L^2$ error is then computed via element-wise
Gauss quadrature comparing the finite element interpolant to the exact
solution~\eqref{eq:kovasznay}.

\subsubsection{Results}

\begin{figure}[htbp]
\centering
\includegraphics[width=0.5\textwidth]{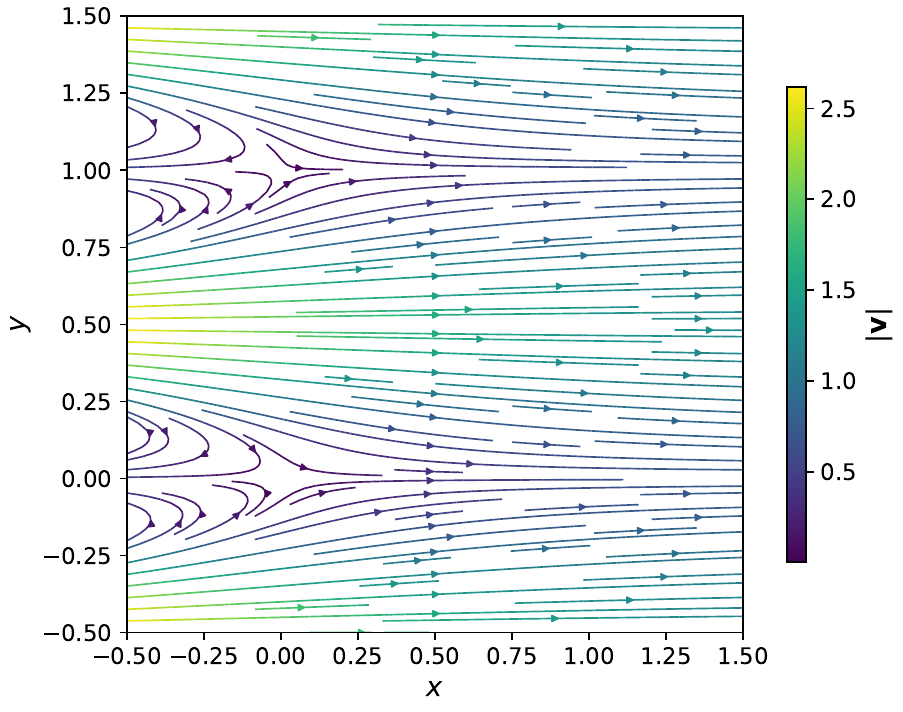}
\caption{Kovasznay flow: computed velocity streamlines at steady state on the
  $n = 16$ unstructured mesh.}
\label{fig:kovasznay_streamlines}
\end{figure}

Figure~~\ref{fig:kovasznay_streamlines} shows the computed velocity streamlines at steady state on the
$n=16$ mesh.  The solution exhibits the characteristic Kovasznay pattern:
a predominantly rightward flow with exponentially decaying sinusoidal
perturbations in the cross-stream direction.

\begin{figure}[htbp]
\centering
\includegraphics[width=1.0\textwidth]{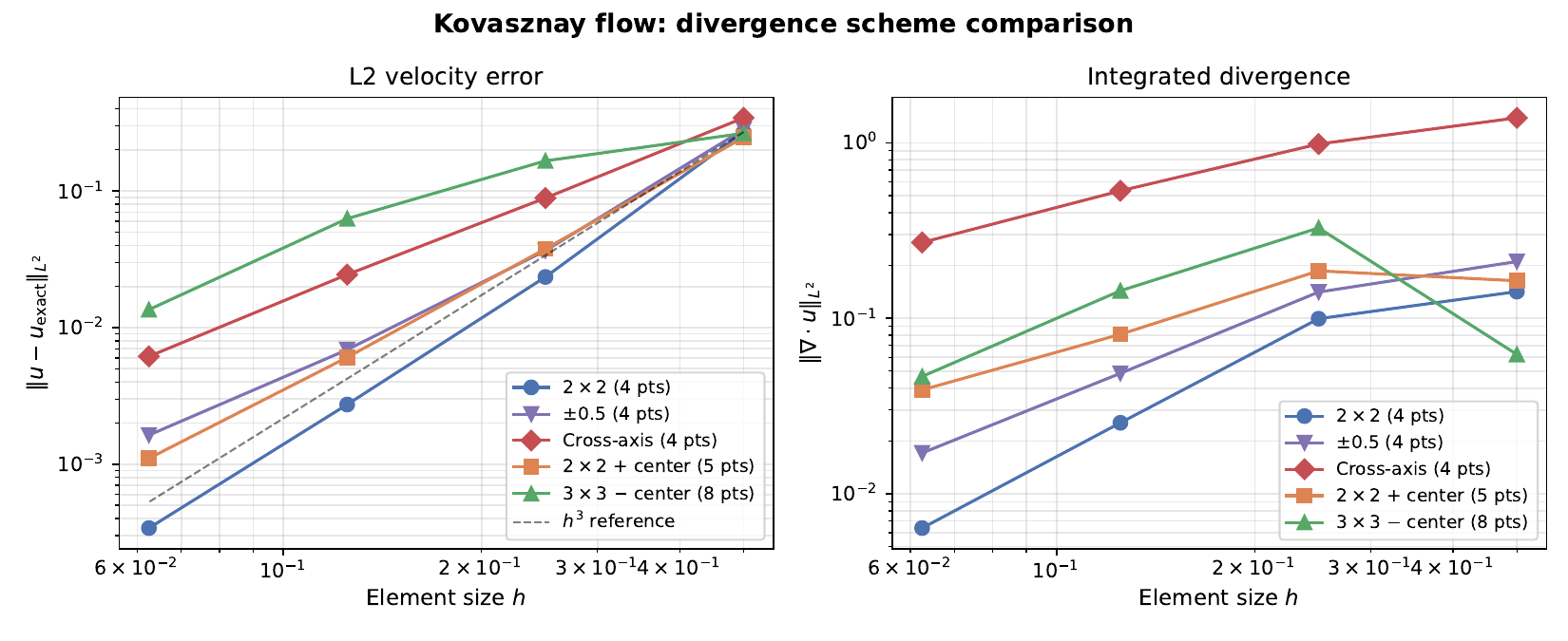}
\caption{Kovasznay flow: comparison of divergence
	collocation schemes on unstructured Q9 meshes.
	\textbf{Left:} $L^2$ velocity error vs.\ element size~$h$.
	\textbf{Right:} integrated divergence
	$\|\nabla\!\cdot\!\bm{v}\|_{L^2}$ vs.~$h$.
	The $2\times 2$ Gauss rule (blue circles) achieves optimal $O(h^3)$
	convergence and the lowest divergence at every refinement level.  The
	polynomial-orthogonality property of the Gauss points (Remark~\ref{rem:orthogonality_main},
	Appendix~A) prevents low-frequency divergence residuals from accumulating
	across elements.}
\label{fig:kovasznay_convergence}
\end{figure}

To assess the spatial accuracy, Figure~\ref{fig:kovasznay_convergence} (left) shows the $L^2$ velocity
error as a function of element size~$h$ on a log--log scale for the baseline $2\times 2$
Gauss collocation rule.  The measured convergence rate ranges from 3.40 on the coarsest
pair to 3.18 on the finest, in good agreement with the theoretical
$O(h^3)$ expectation for biquadratic elements.

Figure~\ref{fig:kovasznay_convergence} also compares the effect of different collocation-point placements
for the divergence constraint~\eqref{eq:div_row}.  Five schemes are tested, all on the
same sequence of unstructured meshes: the $2\times 2$ Gauss rule (4~points),
the diagonal $(\pm\frac{1}{2},\, \pm\frac{1}{2})$ rule (4~points), a
cross-axis placement (4~points on the coordinate axes), $2\times 2$ Gauss
plus the element center (5~points), and $3\times 3$ Gauss minus the center
(8~points).  The $2\times 2$ Gauss rule achieves the lowest velocity error
and the lowest integrated divergence $\|\nabla\!\cdot\!\bm{v}\|_{L^2}$ at
every refinement level, with clean $O(h^3)$ convergence.  The diagonal
$(\pm\frac{1}{2},\, \pm\frac{1}{2})$ scheme, which minimizes a
per-element divergence-leakage measure (Appendix~A), produces
noticeably higher divergence and degraded velocity accuracy in the
assembled system.  Over-constrained schemes (5 or 8~points) show no
improvement, consistent with the observation that excessive constraints
reduce the available DOFs for the momentum equation.

The superiority of the $2\times 2$ Gauss rule in the assembled computation,
despite its slightly worse single-element leakage, is explained by the
polynomial orthogonality property described in Remark~\ref{rem:orthogonality_main} and analyzed in
detail in Appendix~A: the Gauss-point constraints strip the divergence
residual of all low-frequency content, so that neighboring-element
residuals cancel rather than accumulate.

\section{Demonstrations}
\label{sec:demonstrations}

Having verified the formulation against exact solutions, we now apply it to
benchmark problems of increasing complexity that lack closed-form solutions and
must be validated against published experimental or computational data. The
lid-driven cavity tests the solver on a confined recirculating flow with
velocity singularities at the corners, comparing against the widely used
reference data of Ghia et al.~\cite{ghia1982}. The backward-facing step
introduces flow separation and reattachment, exercising the wall force
extraction mechanism (Section~\ref{sec:wall_force}) and enabling comparison
with the experimental measurements of Armaly et al.~\cite{armaly1983}. Flow past a circular cylinder validates
the drag and lift computation (Section~\ref{sec:wall_force}) against the
classical data of Dennis \& Chang~\cite{dennis1970},
Tritton~\cite{tritton1959}, and Williamson~\cite{williamson1996}, spanning steady and unsteady regimes.

\subsection{Lid-driven cavity}
\label{sec:cavity}

\subsubsection{Problem setup}

\begin{figure}[htbp]
\centering
\newlength{\cavsetupheight}\setlength{\cavsetupheight}{6.5cm}%
\begin{subfigure}[t]{0.46\textwidth}
  \centering
  \begin{tikzpicture}[scale=1.5, every node/.style={font=\small},
    hatch/.style={pattern=north east lines, pattern color=gray}]
    \pgfmathsetmacro{\S}{2.5}
    \pgfmathsetmacro{\hw}{0.2}
    \fill[hatch] (0,-\hw) rectangle (\S,0);
    \draw[thick] (0,0) -- (\S,0);
    \fill[hatch] (-\hw,0) rectangle (0,\S);
    \draw[thick] (0,0) -- (0,\S);
    \fill[hatch] (\S,0) rectangle (\S+\hw,\S);
    \draw[thick] (\S,0) -- (\S,\S);
    \draw[thick] (0,\S) -- (\S,\S);
    \draw[very thick, blue, -{Stealth[length=5pt]}] (0.15,\S+0.15)
      -- (\S-0.15,\S+0.15);
    \node[above, blue] at (0.5*\S,\S+0.15) {$v_x = 1$, \; $v_y = 0$};
    \draw[<->] (0,-0.55) -- (\S,-0.55) node[midway, below] {$1$};
    \draw[<->] (\S+0.55,0) -- (\S+0.55,\S) node[midway, right] {$1$};
    \draw[->, thick, gray] (-0.55,-1.0) -- (0.6,-1.0) node[right] {$x$};
    \draw[->, thick, gray] (-0.55,-1.0) -- (-0.55,-0.15) node[left] {$y$};
  \end{tikzpicture}
  \caption{Problem setup.}
  \label{fig:cavity_setup_schematic}
\end{subfigure}
\hfill
\begin{subfigure}[t]{0.50\textwidth}
  \centering
  \includegraphics[height=\cavsetupheight]{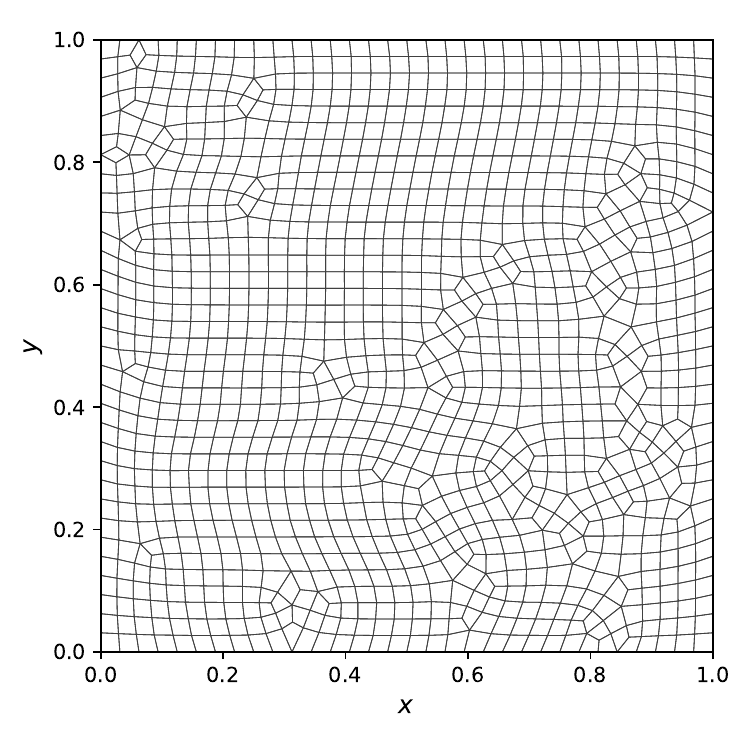}
  \caption{Unstructured Q9 mesh (1\,183 elements).}
  \label{fig:cavity_mesh}
\end{subfigure}
\caption{Lid-driven cavity: (a) problem setup: the top wall moves with
  prescribed velocity (blue arrow) while hatched boundaries denote stationary
  no-slip walls ($\bv = \bm{0}$); (b) unstructured Q9 mesh used for both
  $Re = 100$ and $Re = 400$.}
\label{fig:cavity_setup}
\end{figure}
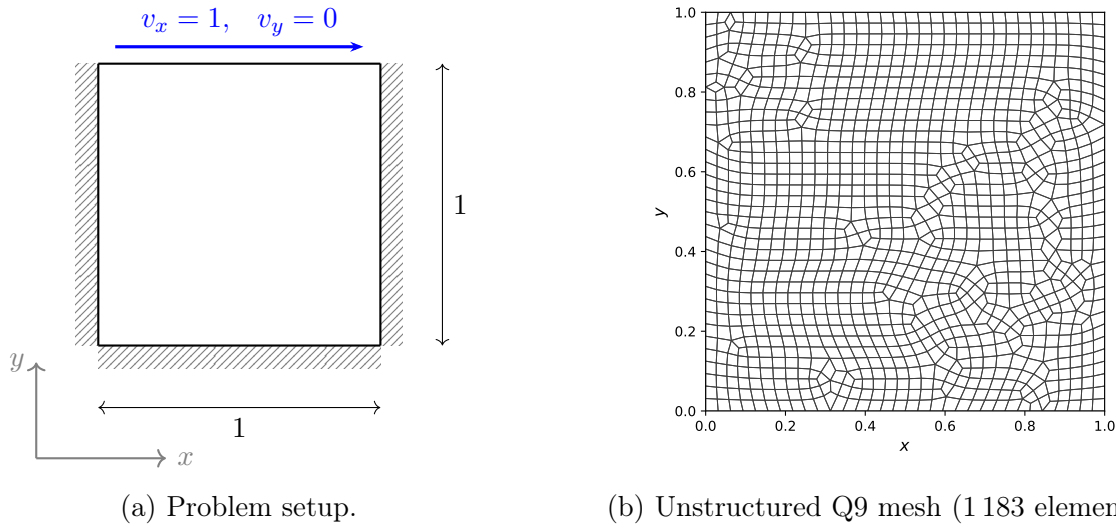

The lid-driven cavity is a standard benchmark for incompressible flow
solvers~\cite{ghia1982,erturk2005}. The domain is $[0,1]^2$
(Figure~\ref{fig:cavity_setup}) with no-slip boundary conditions on all
walls. The top wall (lid) moves with unit velocity $v_x = 1$, $v_y = 0$,
driving a recirculating flow. The bottom, left, and right walls are stationary
($\bv = \bm{0}$). At the top corners, where the lid meets the stationary
walls, the no-slip condition $\bv = \bm{0}$ is imposed (wall wins over lid),
creating velocity discontinuities that the finite element solution regularizes
over nearby elements. The Reynolds number is $Re = U_{\text{lid}} L / \nu$,
where $L = 1$ is the cavity side length and $U_{\text{lid}} = 1$ is the lid
speed.

For $Re = 100$ ($\nu = 0.01$) and $Re = 400$ ($\nu = 0.0025$), the domain is
discretized with an unstructured Q9 mesh of approximately $32 \times 32$
elements (Figure~\ref{fig:cavity_mesh}; 1\,183 elements, 4\,861 nodes,
9\,722 DOFs). For $Re = 1000$ ($\nu = 0.001$), we use a wall-graded
unstructured Q9 mesh with element size $h_{\text{wall}} = 1/32$ at the walls
and $h_{\text{center}} = 3/32$ in the interior (446~elements, 1\,913~nodes,
3\,826~DOFs). Each simulation is initialized with zero velocity and advanced in
time using the monolithic solver until the steady-state criterion
$\|\partial_t \bv\| / \|\bv\| < 10^{-10}$ is met. Results are compared
against the tabulated reference data of Ghia et al.~\cite{ghia1982}.

\subsubsection{Results}

\begin{figure}[p]
\centering
\begin{subfigure}[t]{0.32\textwidth}
  \centering
  \includegraphics[width=\textwidth]{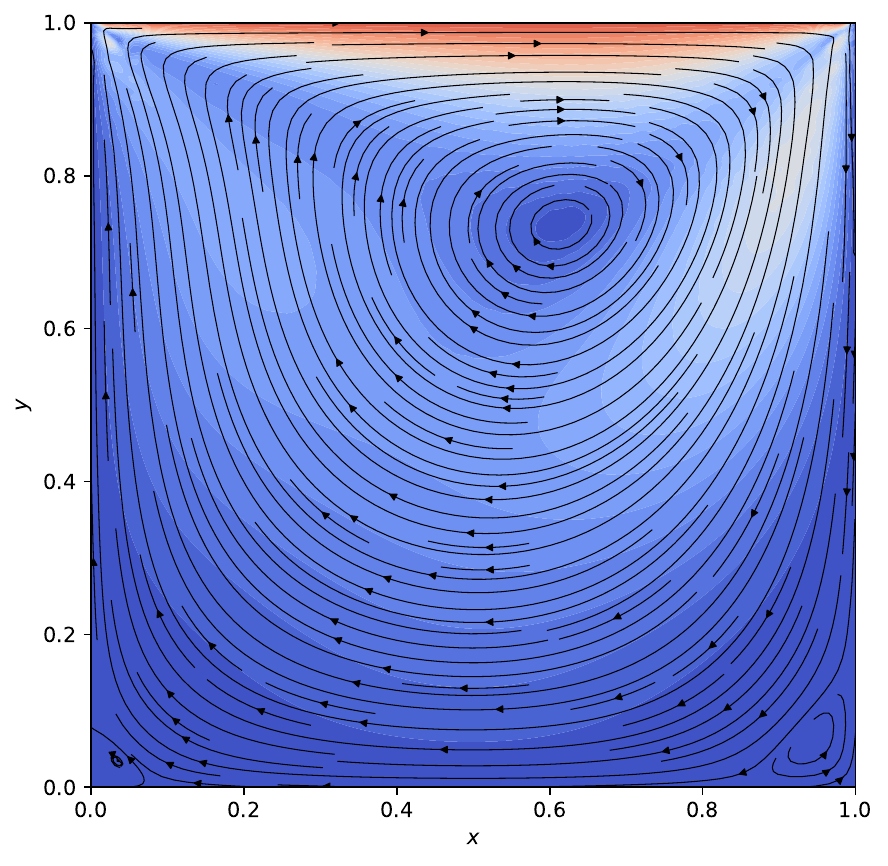}
  \caption{$Re = 100$.}
  \label{fig:cavity_re100}
\end{subfigure}
\hfill
\begin{subfigure}[t]{0.65\textwidth}
  \centering
  \includegraphics[width=\textwidth]{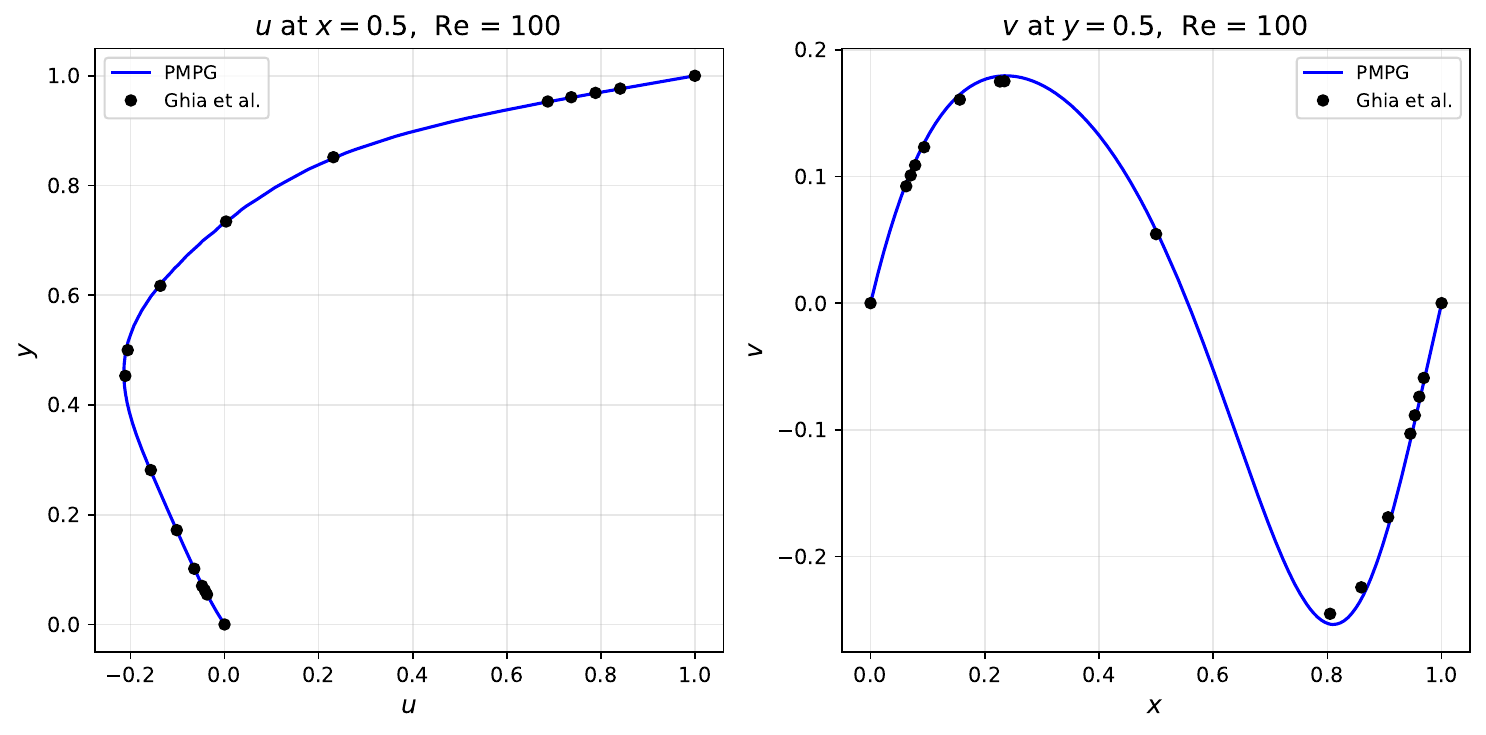}
\end{subfigure}
\\[2pt]
\begin{subfigure}[t]{0.32\textwidth}
  \centering
  \includegraphics[width=\textwidth]{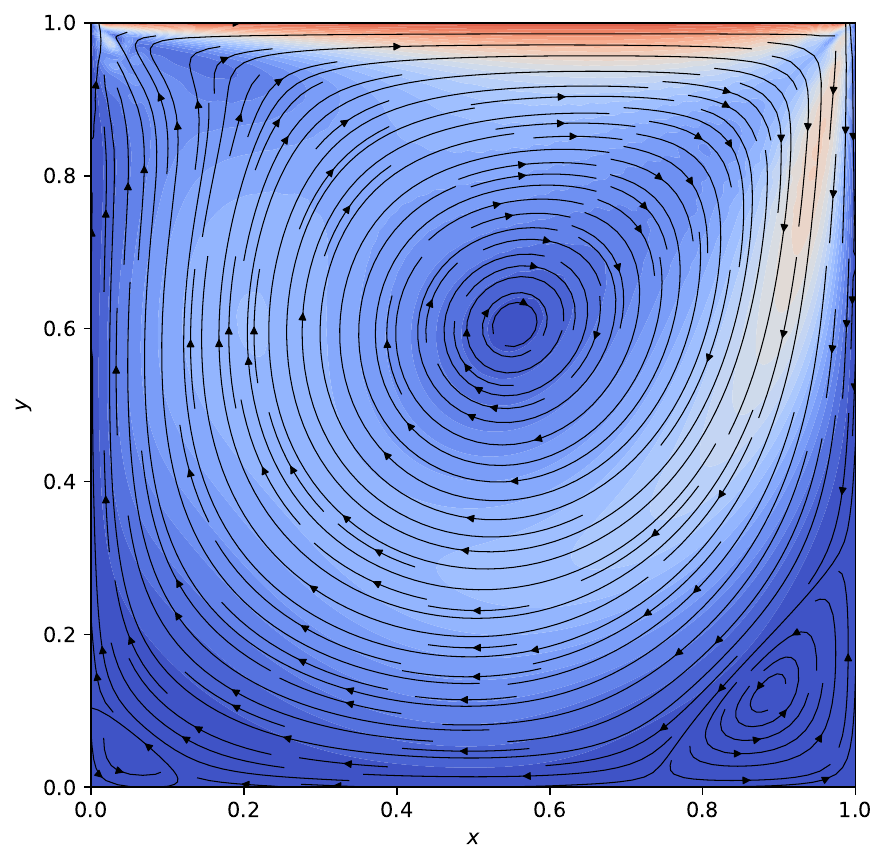}
  \caption{$Re = 400$.}
  \label{fig:cavity_re400}
\end{subfigure}
\hfill
\begin{subfigure}[t]{0.65\textwidth}
  \centering
  \includegraphics[width=\textwidth]{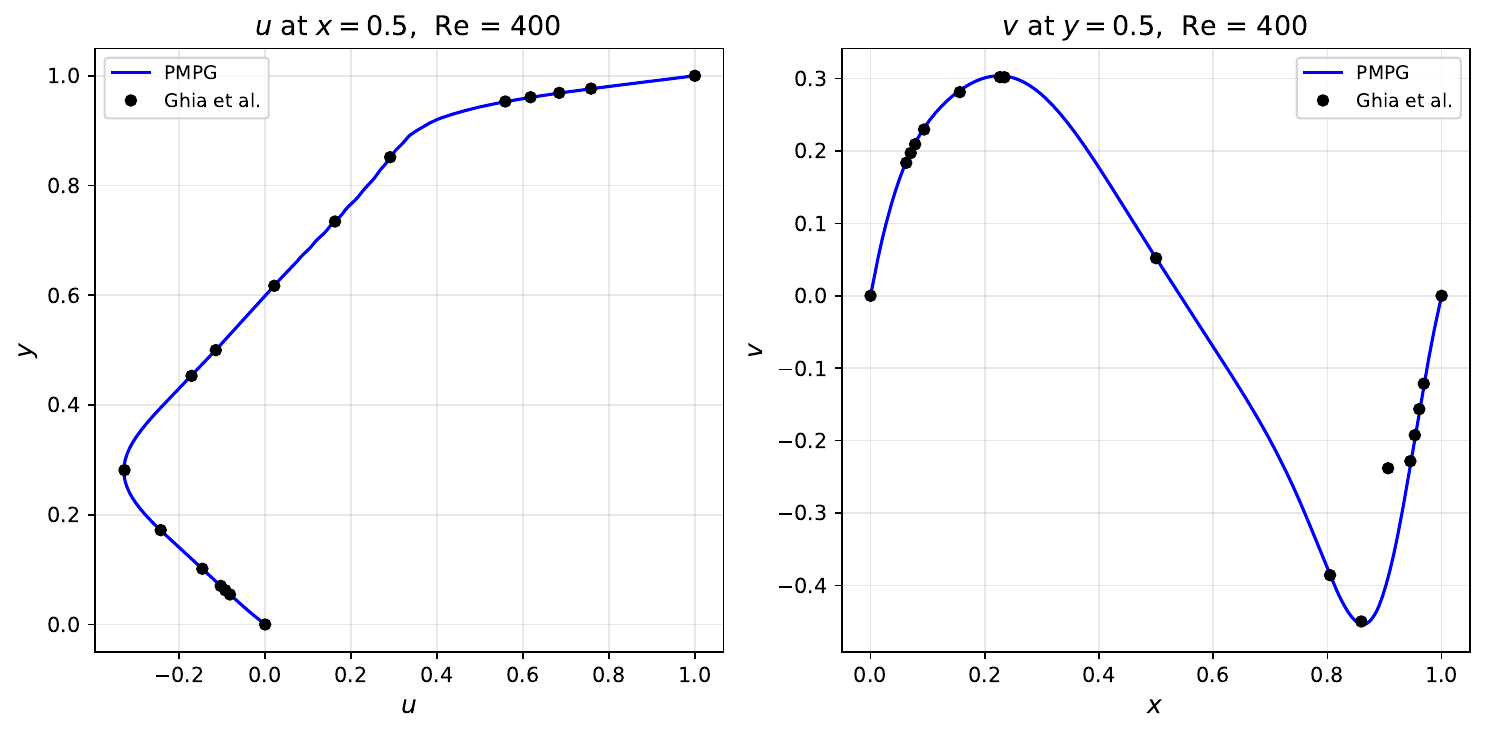}
\end{subfigure}
\\[2pt]
\begin{subfigure}[t]{0.32\textwidth}
  \centering
  \includegraphics[width=\textwidth]{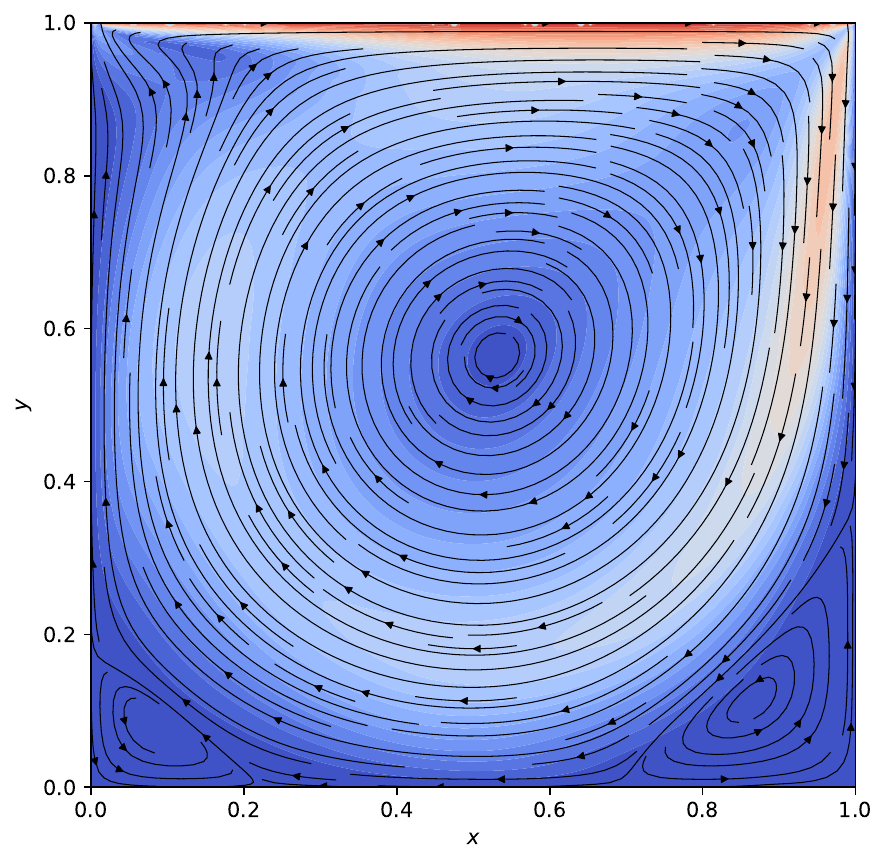}
  \caption{$Re = 1000$.}
  \label{fig:cavity_re1000}
\end{subfigure}
\hfill
\begin{subfigure}[t]{0.65\textwidth}
  \centering
  \includegraphics[width=\textwidth]{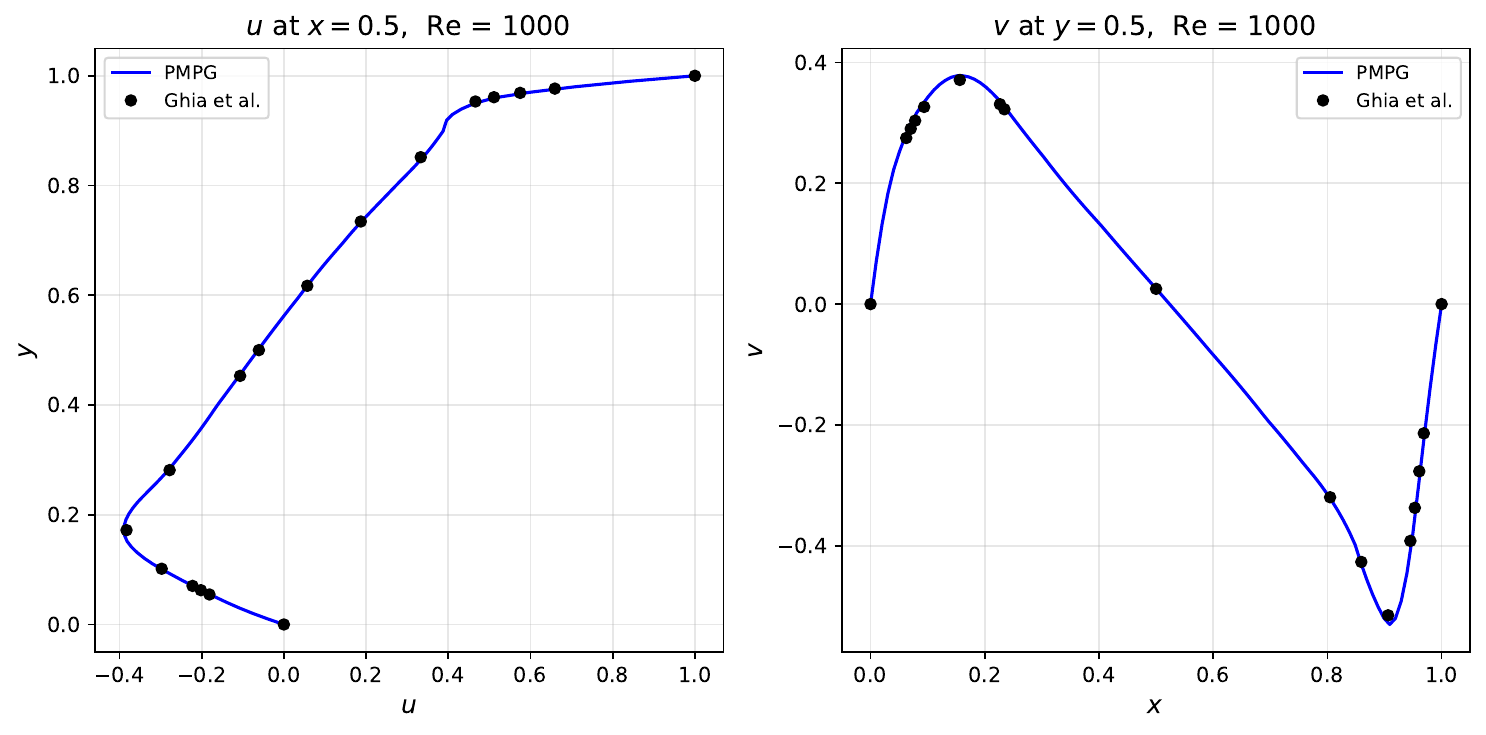}
\end{subfigure}
\caption{Lid-driven cavity: velocity streamlines colored by speed $|\bv|$
  (left) and centerline velocity profiles compared with Ghia et
  al.~\cite{ghia1982} reference data (symbols, right). As $Re$ increases,
  the primary vortex migrates toward the center, the boundary layers thin,
  and secondary corner vortices become more prominent. The PMPG solution
  agrees well with the reference data at all three Reynolds numbers.}
\label{fig:cavity_results}
\end{figure}

Figure~\ref{fig:cavity_results} presents the computed velocity streamlines
(left column, colored by speed $|\bv|$) and centerline velocity profiles
(right columns, compared with Ghia et al.~\cite{ghia1982}) at all three
Reynolds numbers. At $Re = 100$, a single primary vortex occupies the cavity;
at $Re = 400$ the boundary layers thin and secondary corner vortices emerge;
at $Re = 1000$ the primary vortex is nearly centered and the corner vortices
are more prominent. The centerline profiles show good agreement with the
reference data at all three Reynolds numbers, with the velocity gradients near
the walls steepening as $Re$ increases.

\subsubsection{Stability in the convection-dominated regime}
\label{sec:cavity_re1000}

To test the stability of the PMPG formulation at high element P\'{e}clet
numbers, we solve the cavity at $Re = 1000$ on a sequence of deliberately
coarse wall-graded Q9 meshes. Each mesh uses a wall element size
$h_{\text{wall}} = 1/n$ and an interior size $h_{\text{center}} = 3/n$, with
$n = 12, 16, 24, 32$ elements along each boundary, producing genuinely
unstructured meshes ranging from 96 to 446 elements.
The element P\'{e}clet number at the wall is
$\text{Pe}_h = U h_{\text{wall}} / (2\nu) = n/2$, ranging from~6 ($n = 12$)
to~16 ($n = 32$); in the interior, where $h_{\text{center}} = 3/n$,
$\text{Pe}_h$ reaches~42 on the coarsest mesh, well into the regime where
standard Galerkin methods exhibit spurious oscillations.
Moreover, the laminar boundary layer thickness scales as
$\delta \sim L / \sqrt{Re} \approx 0.032$, so that
$h_{\text{wall}} / \delta$ ranges from~2.6 ($n = 12$) to~1.0 ($n = 32$):
on the coarsest meshes the boundary layer is not even resolved by a single
element.

\begin{figure}[htbp]
\centering
\includegraphics[width=0.85\textwidth]{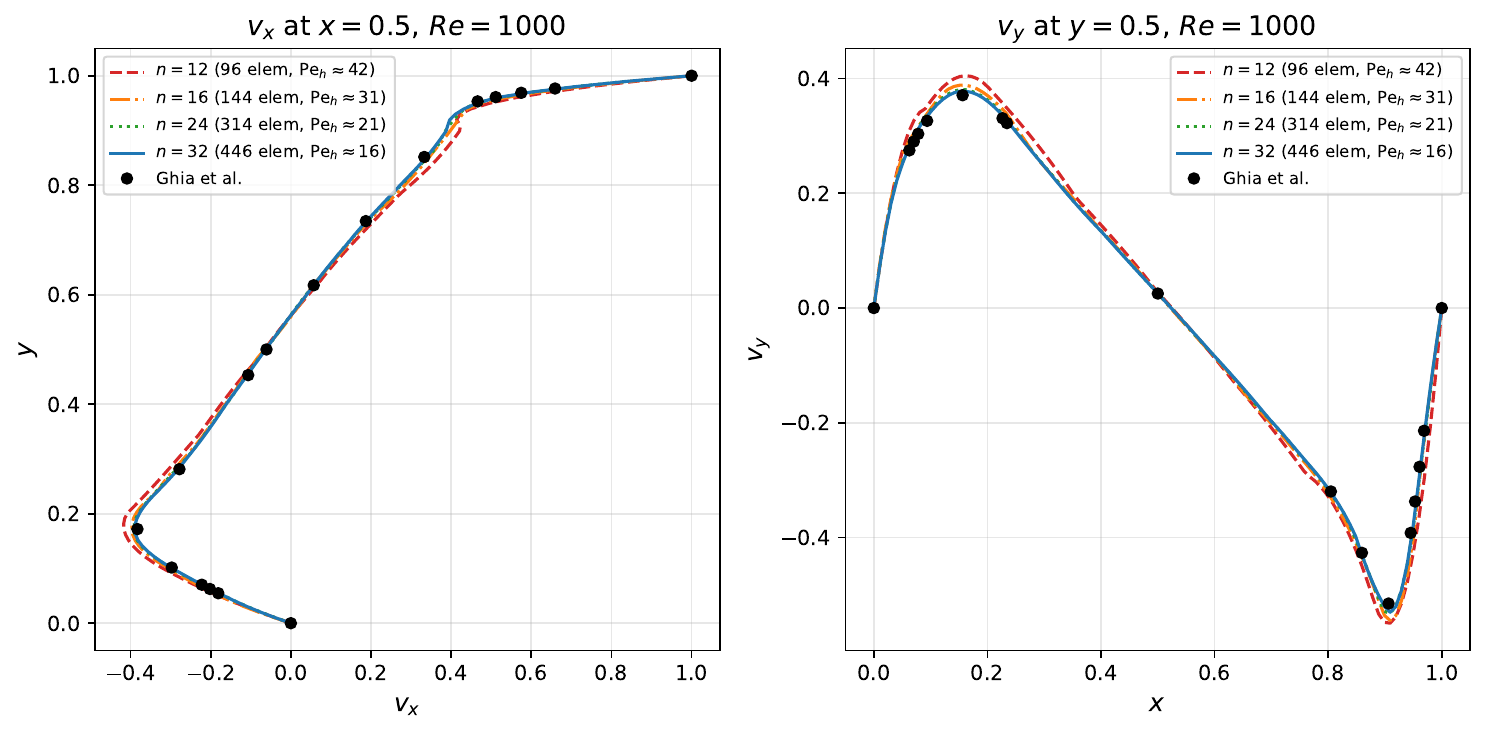}
\caption{Lid-driven cavity at $Re = 1000$: centerline velocity profiles on
  coarse wall-graded meshes with $n = 12, 16, 24, 32$ elements per side.
  Element P\'{e}clet numbers range from 16 to 42 in the interior, and the
  boundary layer is not resolved by a single element on the coarsest meshes.
  All solutions are smooth and oscillation-free despite the absence of any
  stabilization, converging toward the Ghia et al.~\cite{ghia1982} reference
  data with mesh refinement.}
\label{fig:cavity_re1000_coarse}
\end{figure}

Despite this severe under-resolution, all four meshes produce smooth,
oscillation-free steady-state solutions without any stabilization.
Figure~\ref{fig:cavity_re1000_coarse} overlays the centerline profiles from
all four meshes, demonstrating that the solutions converge toward
the reference data as the mesh is refined, with no sign of oscillation even on
the coarsest grids. This behavior confirms the structural stability property
discussed in Remark~\ref{rem:stability}: since convection enters only through
the explicit right-hand side, the PMPG time-stepping scheme does not suffer
from the coercivity loss that triggers Galerkin instabilities at high
P\'{e}clet numbers.

\subsection{Backward-facing step}
\label{sec:backstep}

\subsubsection{Problem setup}

\begin{figure}[htbp]
\centering
\begin{subfigure}[t]{\textwidth}
  \centering
  \begin{tikzpicture}[scale=0.65, every node/.style={font=\small},
    hatch/.style={pattern=north east lines, pattern color=gray}]
    \pgfmathsetmacro{\Hd}{3.5}     
    \pgfmathsetmacro{\sd}{1.75}    
    \pgfmathsetmacro{\hd}{1.75}    
    \pgfmathsetmacro{\Lin}{2.5}    
    \pgfmathsetmacro{\Lout}{9.0}   
    \pgfmathsetmacro{\hw}{0.2}     
    \fill[hatch] (-\Lin,\Hd) rectangle (\Lout,\Hd+\hw);
    \draw[thick] (-\Lin,\Hd) -- (\Lout,\Hd);
    \fill[hatch] (-\Lin,\sd-\hw) rectangle (0,\sd);
    \draw[thick] (-\Lin,\sd) -- (0,\sd);
    \fill[hatch] (-\hw,0) rectangle (0,\sd);
    \draw[thick] (0,0) -- (0,\sd);
    \fill[hatch] (0,-\hw) rectangle (\Lout,0);
    \draw[thick] (0,0) -- (\Lout,0);
    \draw[thick, blue] plot[smooth, domain=\sd:\Hd, samples=20]
      ({-\Lin + 0.8*4*(\x-\sd)*(\Hd-\x)/(\hd*\hd)}, \x);
    \foreach \f in {0.15, 0.35, 0.5, 0.65, 0.85} {
      \pgfmathsetmacro{\yy}{\sd + \f*\hd}
      \pgfmathsetmacro{\dy}{\yy - \sd}
      \pgfmathsetmacro{\vv}{0.8*4*\dy*(\hd-\dy)/(\hd*\hd)}
      \draw[thick, blue, -{Stealth[length=3pt]}] (-\Lin,\yy) -- +(\vv,0);
    }
    \draw[thick, dashed] (\Lout,0) -- (\Lout,\Hd);
    \foreach \f in {0.2, 0.4, 0.6, 0.8} {
      \pgfmathsetmacro{\yy}{\f*\Hd}
      \draw[thick, gray, -{Stealth[length=3pt]}] (\Lout,\yy) -- ++(0.6,0);
    }
    \draw[dashed, gray!70, thick] (0.08,\sd) .. controls (1.5,1.4)
      and (3.5,0.8) .. (4.5,0.08);
    \draw[<->] (\Lout+1.3,0) -- (\Lout+1.3,\Hd)
      node[midway, right] {$H = 1$};
    \draw[<->] (-\Lin-0.5,\sd) -- (-\Lin-0.5,\Hd)
      node[midway, left] {$h = 0.5$};
    \draw[<->] (0.6,0) -- (0.6,\sd)
      node[pos=0.3, right] {$s = 0.5$};
    \draw[<->] (-\Lin,\sd-0.8) -- (0,\sd-0.8)
      node[midway, below] {$L_{\text{in}} = 2$};
    \draw[<->] (0,-0.6) -- (4.5,-0.6)
      node[midway, below] {$X_r$};
    \draw[<->] (0,-1.6) -- (\Lout,-1.6)
      node[midway, below] {$L_{\text{out}} = 15$};
    \draw[->, thick, gray] (-\Lin,-2.6) -- (-\Lin+1.5,-2.6)
      node[right] {$x$};
    \draw[->, thick, gray] (-\Lin,-2.6) -- (-\Lin,-1.3)
      node[left] {$y$};
  \end{tikzpicture}
  \caption{Problem setup.}
  \label{fig:backstep_setup_schematic}
\end{subfigure}\\[6pt]
\begin{subfigure}[t]{\textwidth}
  \centering
  \includegraphics[width=\textwidth]{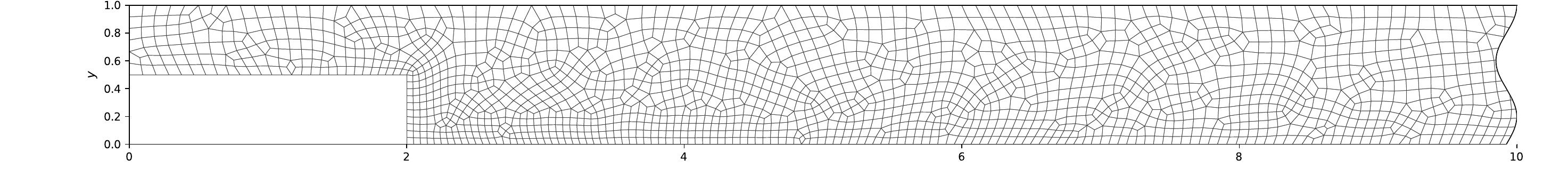}
  \caption{Unstructured Q9 mesh (2\,757 elements, $l_c = 0.1$).}
  \label{fig:backstep_mesh}
\end{subfigure}
\caption{Backward-facing step: (a) problem setup: an inlet channel of height
  $h = 0.5$ feeds into an expanded channel of height $H = 1$ past a step of
  height $s = 0.5$; blue arrows indicate the parabolic inlet profile, hatched
  boundaries denote no-slip walls, and the dashed boundary marks the free
  outflow; the dashed curve indicates the recirculation region with
  reattachment length $X_r$; (b) unstructured Q9 mesh used for $Re = 141$.}
\label{fig:backstep_setup}
\end{figure}

The backward-facing step is a canonical benchmark for separated
flows~\cite{armaly1983}. The geometry
(Figure~\ref{fig:backstep_setup}) consists of an L-shaped channel: an inlet
section of height $h = H - s$ feeds into an expanded section of height $H$,
with a sharp step of height $s$ at the junction. The expansion ratio is $ER =
H/h$. We use $H = 1$, $s = 0.5$ ($ER = 2$, $h = 0.5$), inlet length
$L_{\text{in}} = 2$, and outlet length $L_{\text{out}} = 15$ (sufficiently long
to avoid outlet effects on the recirculation region).

Boundary conditions are: Poiseuille inlet profile
$v_x = \frac{3}{2} U_{\text{mean}} \cdot 4(y - s)(H - y)/h^2$, $v_y = 0$
with $U_{\text{mean}} = 1$; no-slip on all walls (top, bottom, and step face);
free outflow at the outlet. The Reynolds number is defined as
$Re = U_{\text{mean}} \cdot h / \nu$.

The domain is discretized with unstructured Q9 meshes generated by gmsh
(Figure~\ref{fig:backstep_mesh}), with the mesh refined near the step corner
where the shear layer originates. The characteristic element size is $l_c =
0.15$ for $Re \leq 100$ and $l_c = 0.1$ for $Re = 141$, yielding approximately
1\,260 and 2\,760 elements, respectively. Each simulation is initialized with zero velocity and advanced in
time using the monolithic solver until the steady-state criterion $\|\partial_t
\bv\| / \|\bv\| < 10^{-10}$ is met.

\subsubsection{Results}

We compute solutions for several Reynolds numbers in the range
$Re = 34$--$141$ and compare against the experimental data of Armaly et
al.~\cite{armaly1983} ($ER = 1.94$).

\begin{figure}[htbp]
\centering
\begin{subfigure}[t]{\textwidth}
  \centering
  \includegraphics[width=\textwidth]{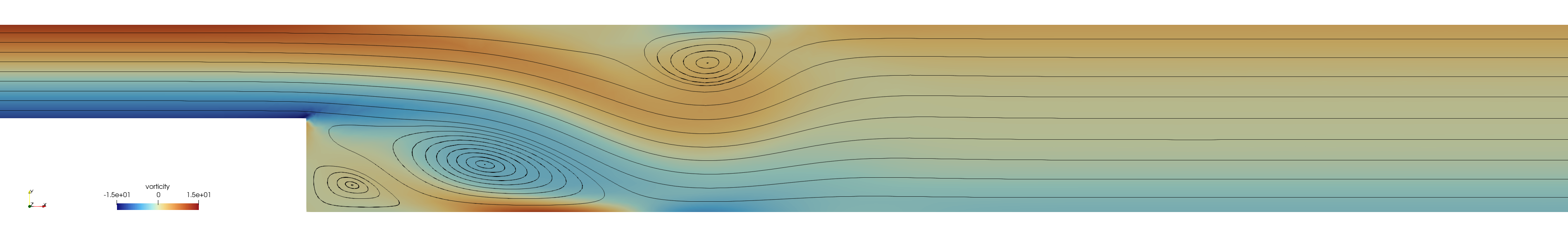}
  \caption{Transient streamlines (step 1000): two corner bubbles and an
    upper-wall vortex.}
  \label{fig:backstep_t1000}
\end{subfigure}\\[4pt]
\begin{subfigure}[t]{\textwidth}
  \centering
  \includegraphics[width=\textwidth]{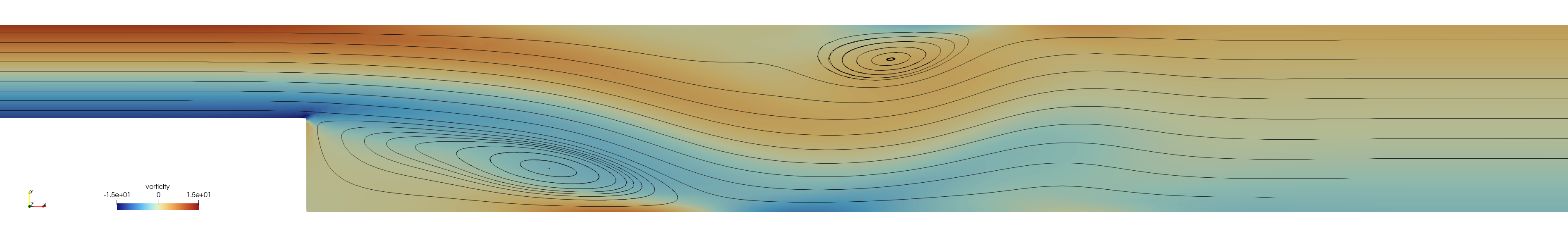}
  \caption{Transient streamlines (step 1600): corner bubbles merged, upper
    vortex convecting downstream.}
  \label{fig:backstep_t1600}
\end{subfigure}\\[4pt]
\begin{subfigure}[t]{\textwidth}
  \centering
  \includegraphics[width=\textwidth]{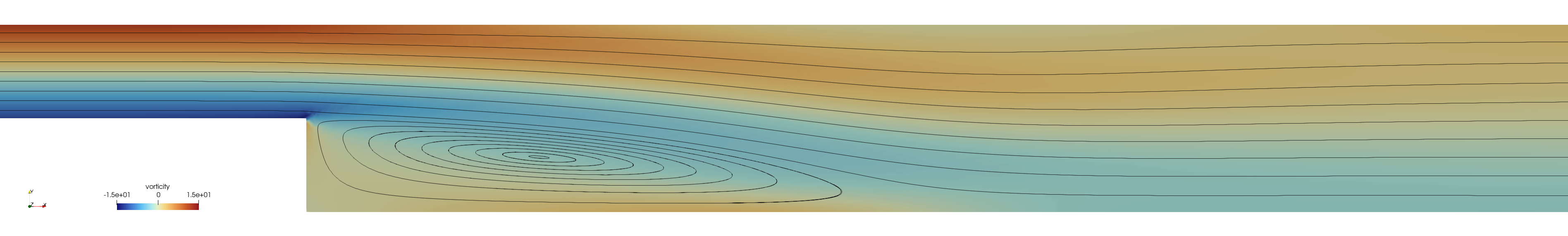}
  \caption{Steady-state streamlines.}
  \label{fig:backstep_steady}
\end{subfigure}
\caption{Backward-facing step at $Re = 141$: transient evolution of the
  velocity streamlines showing the development and merging of recirculation
  zones. Only the region $x \leq 10$ is shown; the full domain extends to
  $x = 17$ ($L_{\text{out}} = 15$).}
\label{fig:backstep_streamlines}
\end{figure}

Figure~\ref{fig:backstep_streamlines} illustrates the transient flow
development at $Re = 141$. At
step~1000 (panel~a), three distinct recirculation zones are visible: two
separate bubbles behind the step corner and a third vortex along the upper
wall, driven by the separated shear layer. By step~1600 (panel~b), the two
corner bubbles have merged into a single larger recirculation zone while the
upper-wall vortex is convected downstream. Panel~(c) shows the final steady
state, featuring the characteristic primary recirculation region extending from
the step corner. This transient evolution is representative of the rich
separated flow dynamics captured by the PMPG formulation.

At steady state, flow separates at the step corner and reattaches downstream,
forming a recirculation bubble. The primary quantity of interest is the
nondimensional reattachment length $X_r/s$, defined as the distance from the
step corner to the reattachment point, normalized by the step height $s$.

The reattachment point is determined from the wall force extraction mechanism
of Section~\ref{sec:wall_force}. When bottom-wall nodes are treated as no-slip
constraints, the Lagrange multipliers from the saddle-point
system~\eqref{eq:saddle_point} provide the nodal reaction forces
via~\eqref{eq:wall_force}. The tangential (streamwise) component at each wall
node is directly related to the local wall shear stress: it is negative in the
recirculation region, where the flow is reversed, and positive downstream of the
reattachment point. The sign change locates the reattachment.

\begin{figure}[htbp]
\centering
\includegraphics[width=0.85\textwidth]{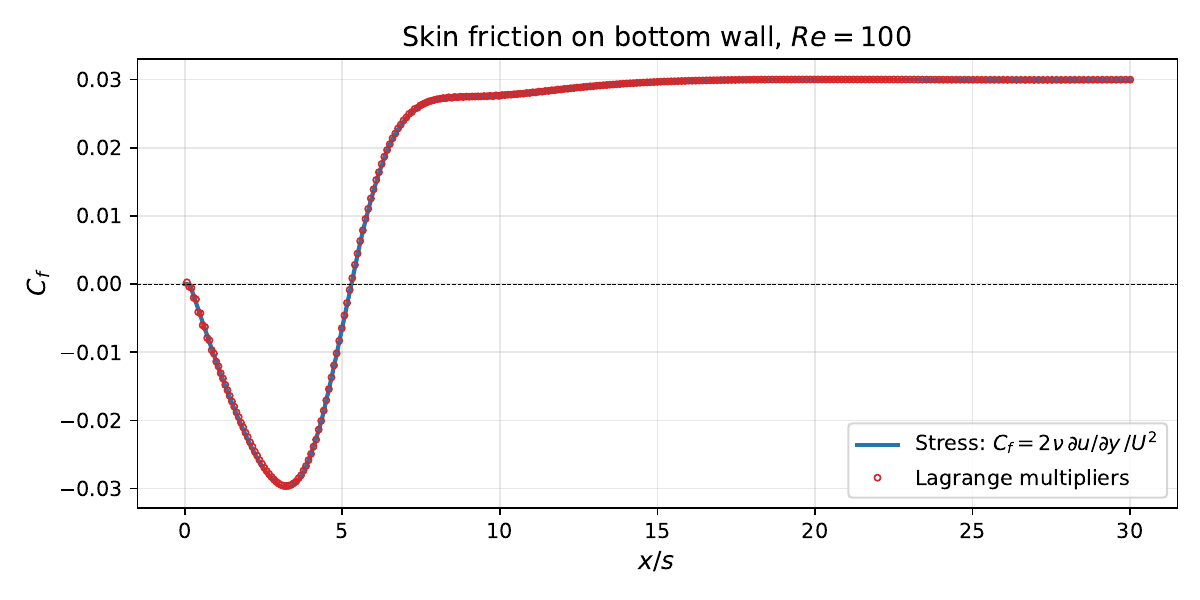}
\caption{Backward-facing step at $Re = 100$: skin friction coefficient along
  the bottom wall. Solid line: classical stress-based computation from velocity
  gradients; circles: Lagrange multiplier nodal forces converted to $C_f$.
  The zero crossing indicates the reattachment point.}
\label{fig:backstep_skinfriction}
\end{figure}

Figure~\ref{fig:backstep_skinfriction} illustrates this for $Re = 100$ by
converting the nodal tangential forces to a skin friction coefficient
$C_f = 2 f_x / (\Delta x_i \, U_{\text{mean}}^2)$,
where $f_x$ is the tangential nodal force from~\eqref{eq:wall_force} and
$\Delta x_i$ is the tributary length (the Gauss--Lobatto integration weight
along the element edge). For validation, the figure also shows the classical
stress-based estimate
$C_f = 2\nu\,(\partial v_x / \partial y)|_{y=0} / U_{\text{mean}}^2$,
computed from the velocity gradients using the Q9 shape function derivatives.
The two distributions are virtually indistinguishable. The relative $L^2$ error
over the full wall, defined as
$\|\,C_f^{\lambda} - C_f^{\text{stress}}\|_2 \, / \, \|C_f^{\text{stress}}\|_2$,
is 0.5\%; restricting to the region $x/s > 1$ (away from the step-corner
singularity where both methods give near-zero values), the median pointwise
relative error is 0.03\%. This confirms that the Lagrange multiplier wall
forces faithfully reproduce the viscous shear stress without requiring explicit
velocity gradient evaluation, and that the nodal forces can be used to detect
flow features such as separation and reattachment.

\begin{figure}[htbp]
\centering
\includegraphics[width=0.85\textwidth]{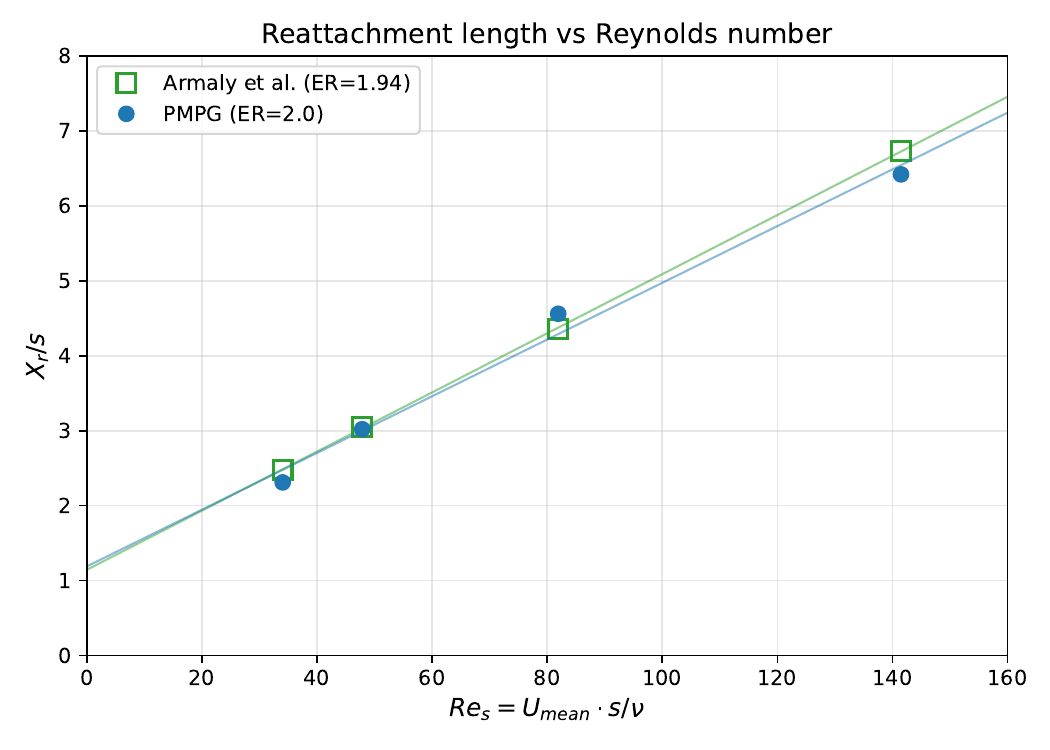}
\caption{Backward-facing step: nondimensional reattachment length $X_r/s$ as a
  function of Reynolds number. PMPG results (circles with least-squares fit) compared with
  experimental data of Armaly et al.~\cite{armaly1983}.}
\label{fig:backstep_reattachment}
\end{figure}

Figure~\ref{fig:backstep_reattachment} collects
the nondimensional reattachment lengths for all Reynolds numbers considered. The PMPG results
follow the expected linear trend $X_r/s \propto Re$ in good agreement with both
experimental datasets despite the slight differences in expansion ratio. At
lower Reynolds numbers ($Re \leq 100$), the error relative to the nearest
experimental value is below 7\%, with the best agreement at $Re \approx 48$
(1.1\% error). At $Re = 141$ the error is 4.6\%, consistent with the increased
complexity of the flow at higher Reynolds numbers.

\subsection{Flow past a circular cylinder}
\label{sec:cylinder}

\subsubsection{Problem setup}

\begin{figure}[htbp]
\centering
\begin{tikzpicture}[scale=1.0, every node/.style={font=\small},
  hatch/.style={pattern=north east lines, pattern color=gray}]
  \pgfmathsetmacro{\W}{10.0}     
  \pgfmathsetmacro{\Hd}{3.5}     
  \pgfmathsetmacro{\hw}{0.2}     
  \pgfmathsetmacro{\cx}{3.75}    
  \pgfmathsetmacro{\cy}{\Hd/2}   
  \pgfmathsetmacro{\cr}{0.35}    
  \fill[hatch] (0,-\hw) rectangle (\W,0);
  \draw[thick] (0,0) -- (\W,0);
  \fill[hatch] (0,\Hd) rectangle (\W,\Hd+\hw);
  \draw[thick] (0,\Hd) -- (\W,\Hd);
  \draw[thick] (0,0) -- (0,\Hd);
  \foreach \f in {0.12, 0.30, 0.50, 0.70, 0.88} {
    \pgfmathsetmacro{\yy}{\f*\Hd}
    \draw[thick, blue, -{Stealth[length=4pt]}] (-0.5,\yy) -- (0,\yy);
  }
  \node[blue, left] at (-0.5,\cy) {$U_\infty$};
  \draw[thick, dashed] (\W,0) -- (\W,\Hd);
  \foreach \f in {0.2, 0.4, 0.6, 0.8} {
    \pgfmathsetmacro{\yy}{\f*\Hd}
    \draw[thick, gray, -{Stealth[length=3pt]}] (\W,\yy) -- ++(0.3,0);
  }
  \fill[white] (\cx,\cy) circle (\cr);
  \fill[hatch] (\cx,\cy) circle (\cr);
  \draw[thick] (\cx,\cy) circle (\cr);
  \node[below right] at (\cx+\cr+0.05, \cy-\cr-0.05) {$(0,0)$};
  \node[above] at (0.5*\W, \Hd+\hw+0.05) {$\bv = (U_\infty, 0)$};
  \node[below] at (0.5*\W, -\hw-0.05) {$\bv = (U_\infty, 0)$};
  \node[right] at (\W+1.3, 0.85*\Hd) {free outflow};
  \draw[<->] (0,-1.2) -- (\cx,-1.2) node[midway, below] {$15D$};
  \draw[<->] (\cx,-1.2) -- (\W,-1.2) node[midway, below] {$25D$};
  \draw[<->] (\W+0.8,0) -- (\W+0.8,\Hd) node[midway, right] {$30D$};
  \draw[<->] (\cx-\cr, \cy+\cr+0.25) -- (\cx+\cr, \cy+\cr+0.25)
    node[midway, above] {$D$};
  \draw[->, thick, gray] (-0.8,-2.2) -- (0.7,-2.2) node[right] {$x$};
  \draw[->, thick, gray] (-0.8,-2.2) -- (-0.8,-1.0) node[left] {$y$};
\end{tikzpicture}
\caption{Flow past a circular cylinder: computational domain (not to scale).
  Uniform inflow $U_\infty = 1$ on the left, freestream velocity
  $\bv = (U_\infty, 0)$ on the top and bottom, no-slip on the cylinder surface,
  and free outflow on the right. The cylinder of diameter $D = 1$ is centered
  at the origin.}
\label{fig:cylinder_setup}
\end{figure}
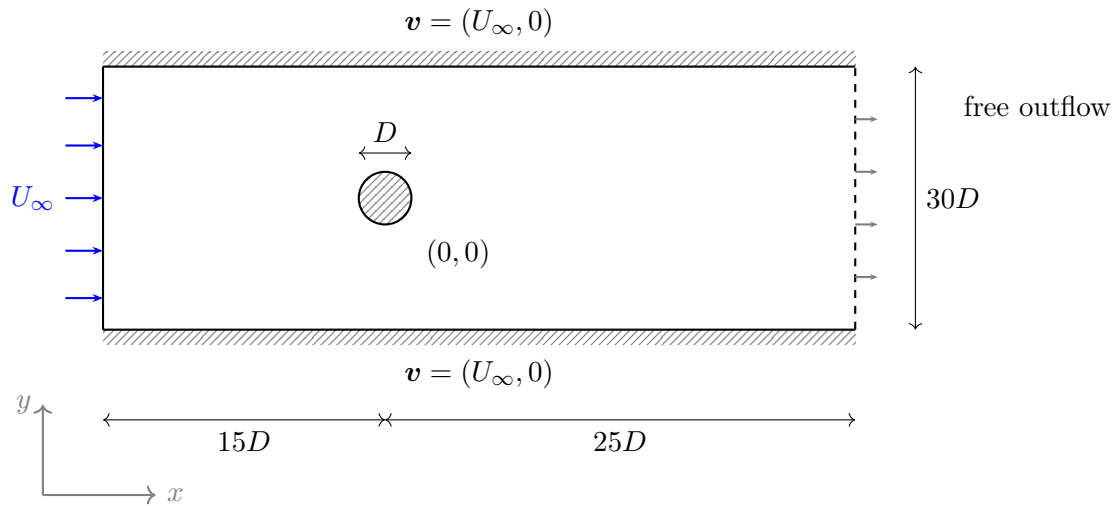

Flow past a circular cylinder is a classical benchmark that spans qualitatively
different flow regimes as the Reynolds number increases: steady symmetric flow
($Re \lesssim 5$), steady recirculation with twin vortices
($5 \lesssim Re \lesssim 47$), and periodic vortex shedding
($Re \gtrsim 47$)~\cite{williamson1996}. We consider three Reynolds numbers
that sample these regimes: $Re = 20$ and $Re = 40$ (steady, twin
vortices), and $Re = 100$ (unsteady, vortex shedding).

The computational domain (Figure~\ref{fig:cylinder_setup}) is
$[-15D,\, 25D] \times [-15D,\, 15D]$, with the cylinder of diameter $D = 1$
centered at the origin. Boundary conditions are: uniform inflow $\bv =
(U_\infty, 0)$ with $U_\infty = 1$ on the left boundary; freestream velocity
$\bv = (U_\infty, 0)$ on the top and bottom boundaries; no-slip ($\bv =
\bm{0}$) on the cylinder surface; and free outflow on the right boundary. The Reynolds number is defined as $Re = U_\infty D / \nu$.

The domain is discretized with unstructured Q9 meshes generated by gmsh, with
far-field characteristic element size $l_c = 0.3$, refined to $l_c/4 \approx
0.075$ at the cylinder surface. This yields approximately 26\,000 elements and
106\,000 nodes (212\,000 velocity DOFs). Each simulation is initialized with
uniform flow $\bv = (U_\infty, 0)$ everywhere (with $\bv = \bm{0}$ on the
cylinder) and advanced in time using the monolithic solver with adaptive CFL
time stepping ($c_{\text{CFL}} = 0.25$).

Drag and lift coefficients are defined as:
\begin{equation}
  C_D = \frac{F_x}{\frac{1}{2}\rho U_\infty^2 D}, \qquad
  C_L = \frac{F_y}{\frac{1}{2}\rho U_\infty^2 D},
  \label{eq:cd_cl}
\end{equation}
where $F_x$ and $F_y$ are the total wall forces from~\eqref{eq:total_force}
and the reference length is the cylinder diameter $D$ (standard 2D convention,
per unit span).

\subsubsection{Results}


At $Re = 20$ and $Re = 40$, the flow reaches a steady state.
Table~\ref{tab:cylinder_cd} compares the computed drag coefficients with the
numerical results of Dennis \& Chang~\cite{dennis1970} and the experimental
measurements of Tritton~\cite{tritton1959}. At $Re = 20$ the present result
($C_D = 2.13$) is within 4\% of Dennis \& Chang and 2\% of Tritton. At
$Re = 40$ the computed value ($C_D = 1.57$) is within 3\% of Dennis \& Chang
and 1.3\% of Tritton. At $Re = 100$, where the flow is unsteady, the
time-averaged drag coefficient ($C_D = 1.36$) is within 3\% of Tritton's
experimental measurement.

\begin{figure}[htbp]
\centering
\begin{subfigure}[t]{\textwidth}
  \centering
  \includegraphics[width=0.85\textwidth]{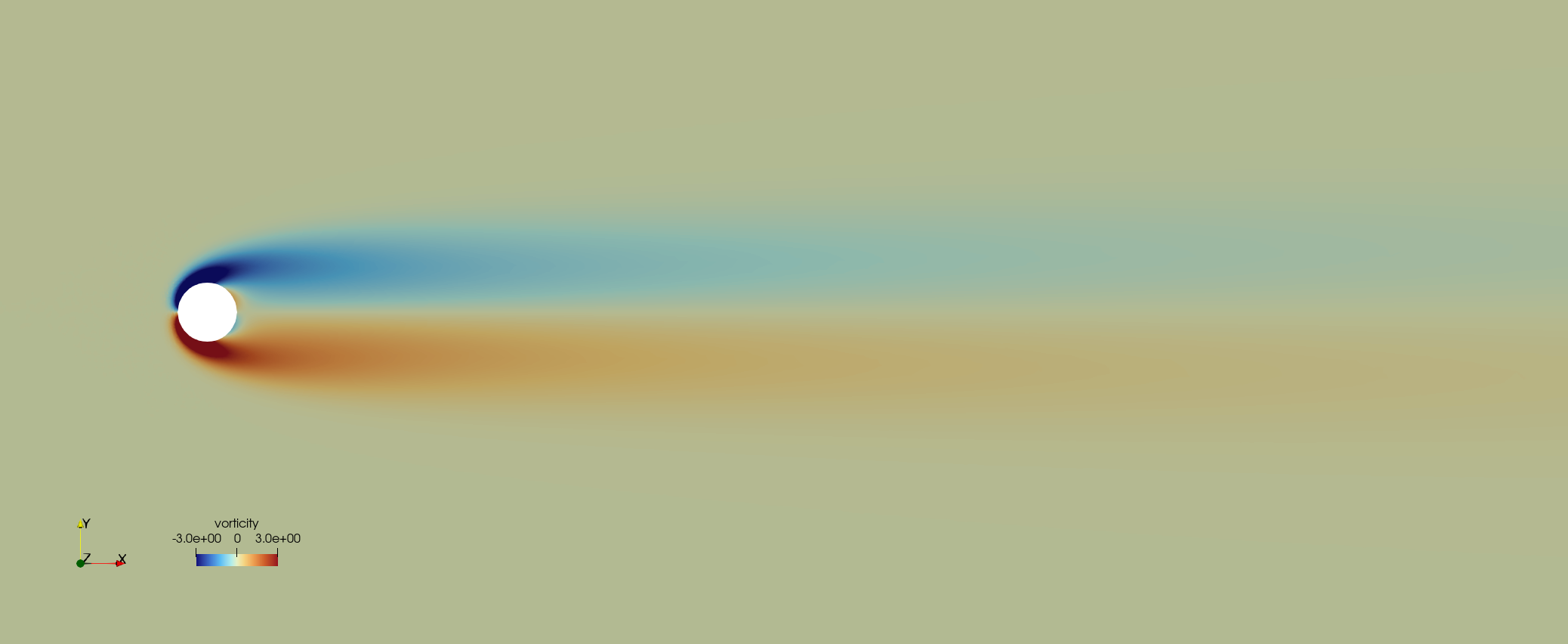}
  \caption{$Re = 40$: steady-state symmetric twin-vortex wake.}
  \label{fig:cylinder_wake_Re40}
\end{subfigure}
\par\medskip
\begin{subfigure}[t]{\textwidth}
  \centering
  \includegraphics[width=0.85\textwidth]{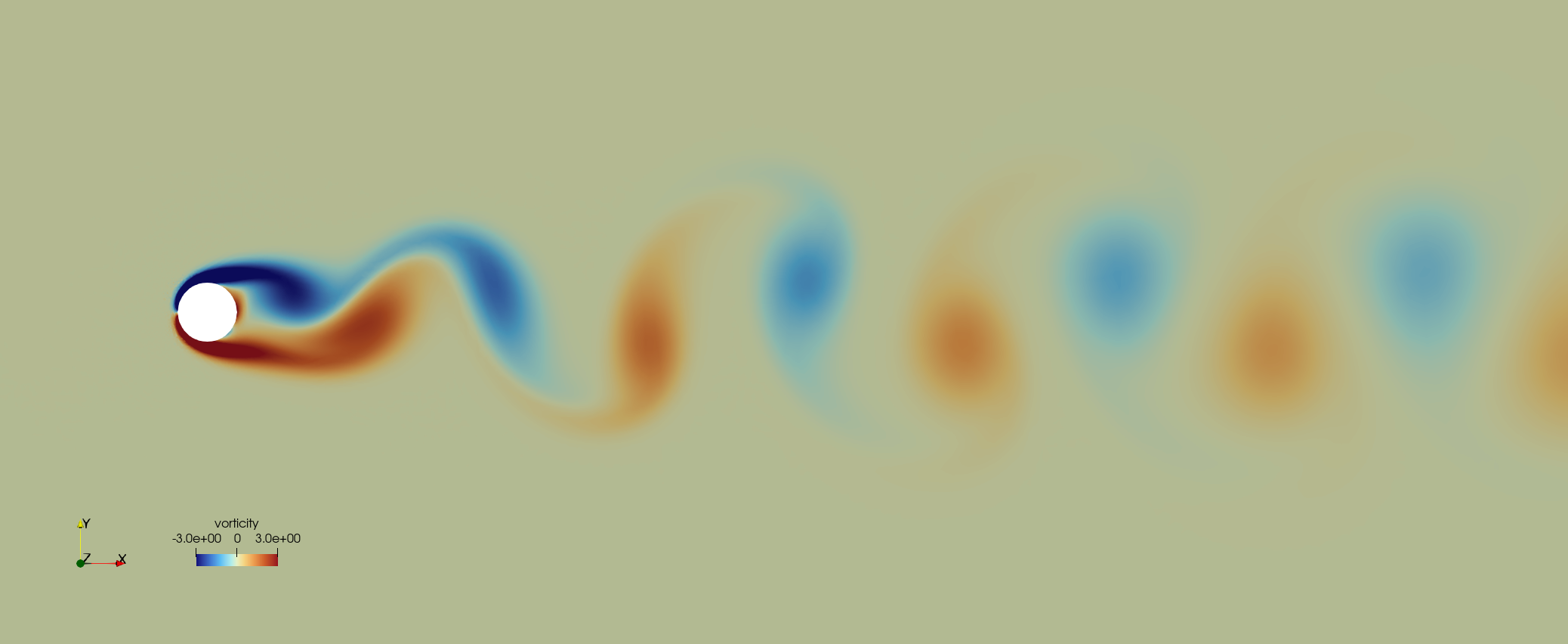}
  \caption{$Re = 100$: instantaneous von K\'{a}rm\'{a}n vortex street.}
  \label{fig:cylinder_vorticity_Re100}
\end{subfigure}
\caption{Flow past a circular cylinder: vorticity contours showing the
  transition from the steady regime to periodic vortex shedding.}
\label{fig:cylinder_vorticity}
\end{figure}

At $Re = 100$, the flow develops periodic vortex shedding (von K\'{a}rm\'{a}n
street). Figure~\ref{fig:cylinder_vorticity} compares the steady twin-vortex
wake at $Re = 40$ with the fully developed von K\'{a}rm\'{a}n vortex street at
$Re = 100$.

\begin{figure}[htbp]
\centering
\includegraphics[width=0.85\textwidth]{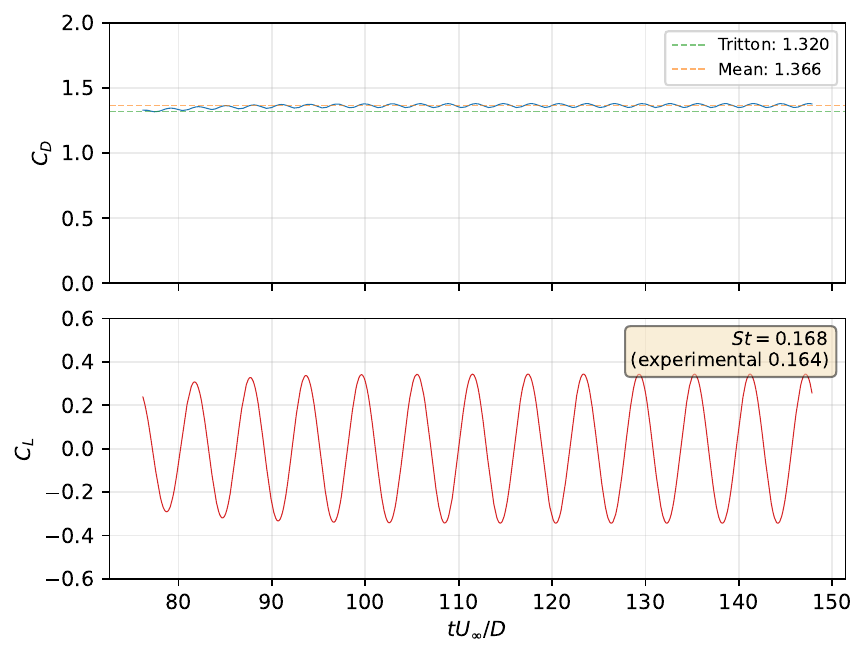}
\caption{Flow past a circular cylinder at $Re = 100$: time history of
  (a)~drag coefficient $C_D$ and (b)~lift coefficient $C_L$. The periodic
  oscillations correspond to vortex shedding.}
\label{fig:cylinder_forces_Re100}
\end{figure}

After an initial transient, the lift coefficient $C_L$ oscillates symmetrically
about zero, indicating alternating vortex detachment from the upper and lower
surfaces of the cylinder. Figure~\ref{fig:cylinder_forces_Re100} shows the time
history of $C_D$ and $C_L$. The drag oscillates at twice the shedding frequency
(as expected from symmetry), while the lift oscillates at the fundamental
frequency.

The Strouhal number characterizing the shedding frequency is:
\begin{equation}
  St = \frac{f D}{U_\infty},
  \label{eq:strouhal}
\end{equation}
where $f$ is determined from the zero crossings of the periodic $C_L$ signal,
and the reported $C_D$ is the time average over the fully developed oscillatory
regime. The computed Strouhal number ($St = 0.168$) agrees to within 2.5\% of
the experimental correlation of Williamson~\cite{williamson1996} ($St = 0.164$),
as reported in Table~\ref{tab:cylinder_cd}.


\begin{table}[htbp]
\centering
\caption{Flow past a circular cylinder: drag coefficient $C_D$ and Strouhal
  number $St$ compared with published data. Dashes indicate quantities not
  applicable to the flow regime.}
\label{tab:cylinder_cd}
\begin{tabular}{cccccc}
\toprule
& \multicolumn{3}{c}{$C_D$} & \multicolumn{2}{c}{$St$} \\
\cmidrule(lr){2-4} \cmidrule(lr){5-6}
$Re$ & PMPG & Dennis \& Chang & Tritton & PMPG & Williamson \\
\midrule
 20 & 2.13 & 2.045 & 2.09 & ---   & --- \\
 40 & 1.57 & 1.522 & 1.59 & ---   & --- \\
100 & 1.36 & ---   & 1.32 & 0.168 & 0.164 \\
\bottomrule
\end{tabular}
\end{table}

The cylinder benchmark demonstrates the wall force extraction mechanism
(Section~\ref{sec:wall_force}) on a curved geometry with quantitative
validation. The drag coefficients computed from the Lagrange multipliers agree
with published values without requiring pressure reconstruction or stress
integration, confirming that the constraint reactions provide a reliable
mechanism for computing aerodynamic forces.

\section{Error estimation and adaptive refinement}
\label{sec:error_estimation}

A distinctive feature of the PMPG formulation is that the objective function
itself provides a natural measure of solution quality.  The PMPG
functional~\eqref{eq:pmpg_continuous} evaluates the squared $L^2$ norm of the
momentum residual,
$\mathcal{J}_h = \frac{1}{2}\int_\Omega |\bm{R}_h|^2 \, d\Omega$,
where $\bm{R}_h = (\bv_h \cdot \nabla)\bv_h - \nu\nabla^2\bv_h$ is the
residual~\eqref{eq:residual} evaluated on the discrete velocity field.
The exact Navier--Stokes velocity $\bv_{\text{exact}}$ has its own residual
$\bm{R}_{\text{exact}} = (\bv_{\text{exact}} \cdot \nabla)\bv_{\text{exact}}
- \nu\nabla^2\bv_{\text{exact}}$
and the corresponding functional value
\begin{equation}
  \mathcal{J}_{\text{exact}}
  = \tfrac{1}{2}\int_\Omega |\bm{R}_{\text{exact}}|^2 \, d\Omega.
  \label{eq:J_exact}
\end{equation}
The element-wise functional density $\mathcal{J}_e / A_e$, where $A_e$ is the
element area, therefore measures the local momentum imbalance and serves as a
built-in error indicator: regions where the residual is large relative to its
neighbors signal under-resolution.  Unlike residual-based \emph{a~posteriori}
estimators for Galerkin methods~\cite{ainsworth2000,verfurth2013}, which
require reliability and efficiency constants that depend on the local mesh
geometry and the PDE coefficients, the PMPG indicator emerges directly from
the variational principle (it is the quantity being minimized) and
requires no additional computation beyond what the solver already performs.

\subsection{Element-wise error indicator}
\label{sec:error_indicator}

To assess the indicator quantitatively, we return to the Kovasznay test
problem (Section~\ref{sec:kovasznay}), for which the exact solution is
available.  Writing $\bm{R}_h = \bm{R}_{\text{exact}} + \delta\bm{R}$, where
$\delta\bm{R} = \bm{R}_h - \bm{R}_{\text{exact}}$ is the residual error, the
gap between the discrete and exact functionals decomposes as
\begin{equation}
  \mathcal{J}_h - \mathcal{J}_{\text{exact}}
  = \underbrace{\int_\Omega \bm{R}_{\text{exact}} \cdot \delta\bm{R}
    \, d\Omega}_{\text{cross term}}
  \;+\; \underbrace{\tfrac{1}{2}\int_\Omega |\delta\bm{R}|^2 \, d\Omega}%
    _{\text{quadratic term} \;\geq\; 0}.
  \label{eq:gap_decomposition}
\end{equation}
The quadratic term is non-negative by construction; the cross term can have
either sign.  For a discrete solution that satisfies the boundary conditions
and the divergence constraint exactly, the functional is an upper bound on
$\mathcal{J}_{\text{exact}}$.  When these constraints are only approximately
satisfied, the gap can become negative, signaling that the constraints
themselves are under-resolved.  This non-monotone behavior is itself a useful
diagnostic.

\begin{figure}[htbp]
\centering
\includegraphics[width=1.0\textwidth]{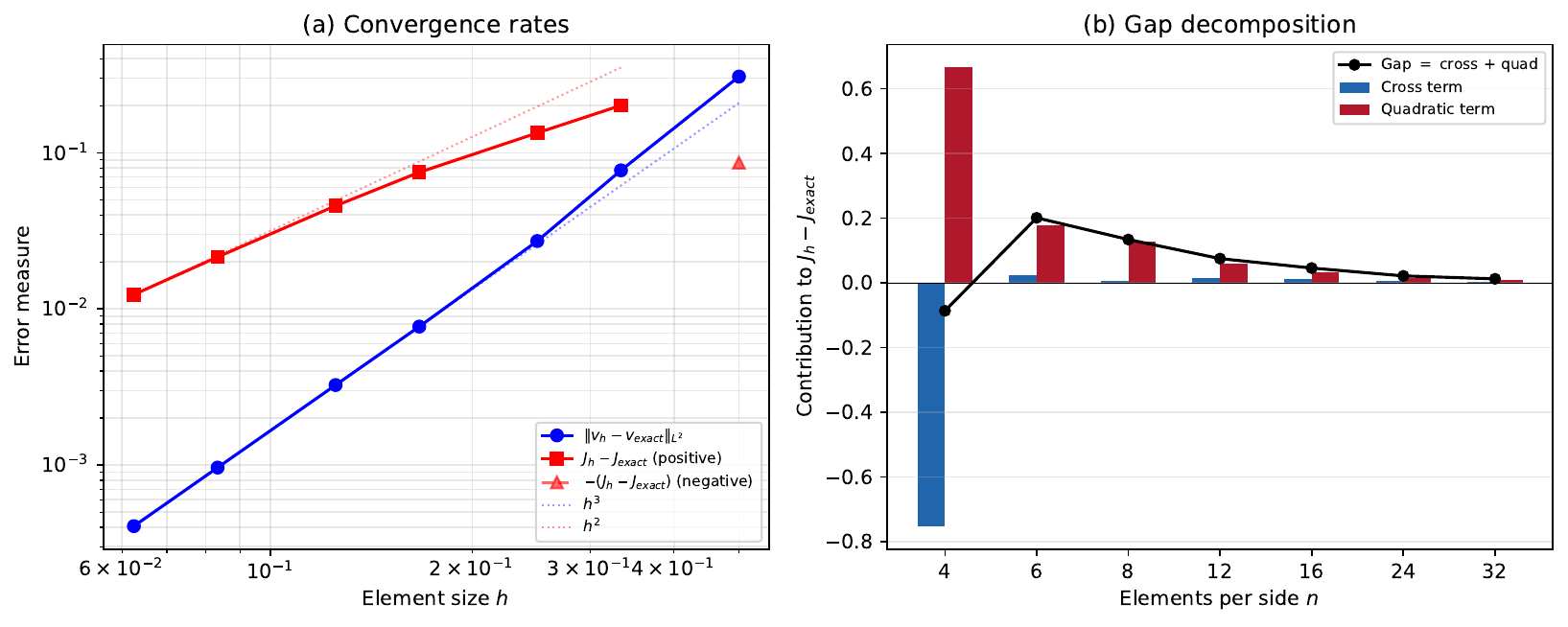}
\caption{Kovasznay flow: convergence of the PMPG functional.
  \textbf{(a)}~Velocity $L^2$ error and functional gap
  $\mathcal{J}_h - \mathcal{J}_{\text{exact}}$ versus element size~$h$.
  The gap changes sign between $n = 4$ and $n = 6$; filled squares denote
  positive gap, open triangles denote negative gap (plotted as absolute
  value).  Reference slopes $h^3$ and $h^2$ are shown.
  \textbf{(b)}~Decomposition of the gap into cross and quadratic
  terms~\eqref{eq:gap_decomposition}.  The negative gap on the coarsest mesh
  is due to the cross term; the quadratic term is always non-negative.}
\label{fig:functional_convergence}
\end{figure}

Figure~\ref{fig:functional_convergence} shows the convergence of
$\mathcal{J}_h - \mathcal{J}_{\text{exact}}$ and the velocity $L^2$ error
on the Kovasznay problem for mesh sizes $n = 4, 6, 8, 12, 16, 24, 32$.
The gap changes sign between $n = 4$ (negative, due to constraint violation
on the coarsest mesh) and $n = 6$ (positive), and thereafter converges at a
rate $\sim h^2$.  The gap decomposition~\eqref{eq:gap_decomposition}
confirms that the cross term drives the sign change on coarse meshes, while
the quadratic term, which is always non-negative, dominates on finer meshes.

\begin{figure}[htbp]
\centering
\includegraphics[width=1.0\textwidth]{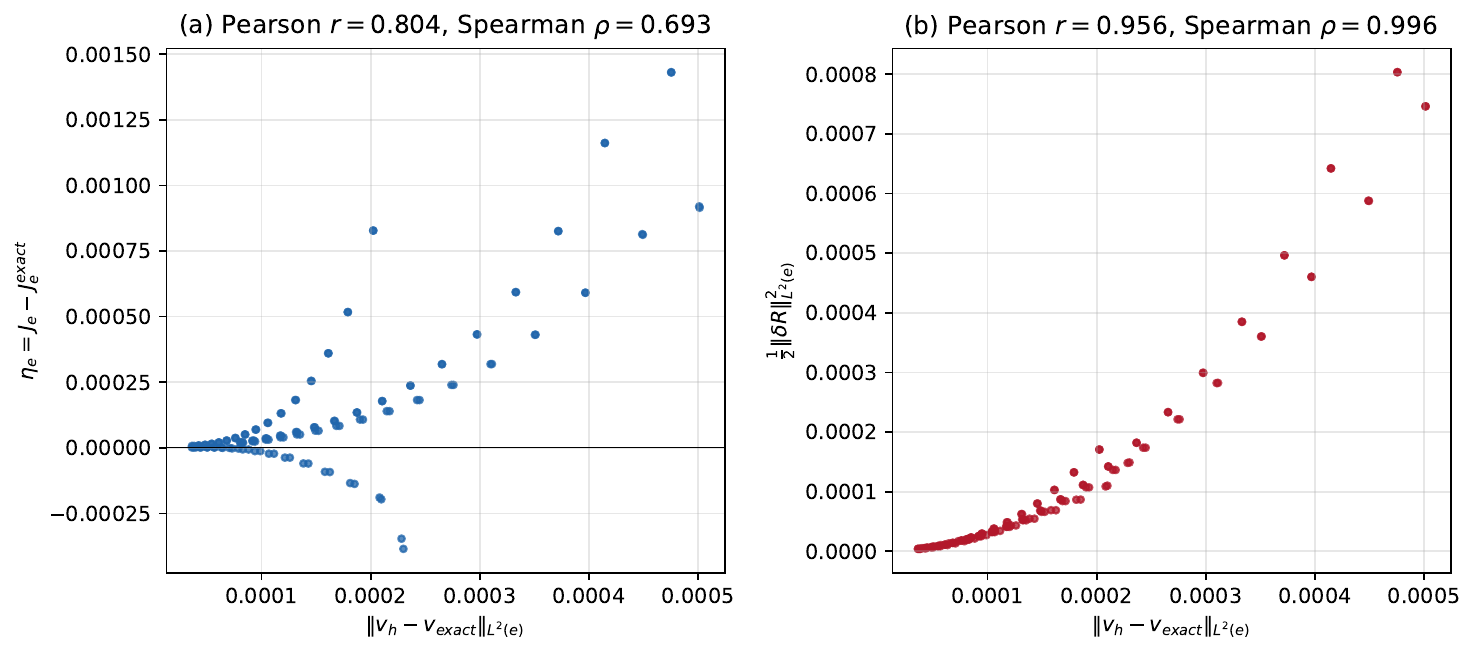}
\caption{Kovasznay flow ($n = 16$): element-wise error indicator versus true
  velocity $L^2$ error.  \textbf{Left:}~Full indicator
  $\eta_e = \mathcal{J}_e - \mathcal{J}_e^{\text{exact}}$ (Spearman
  $\rho = 0.69$).  \textbf{Right:}~Quadratic term
  $\frac{1}{2}\int_{\Omega_e}|\delta\bm{R}|^2\,d\Omega$ (Spearman
  $\rho = 0.996$).  The quadratic term provides near-perfect element-error
  ranking.}
\label{fig:elementwise_correlation}
\end{figure}

At the element level, the quadratic component of the gap
$\frac{1}{2}\int_{\Omega_e} |\delta\bm{R}|^2 \, d\Omega$ correlates
near-perfectly with the true element-wise velocity error.
Figure~\ref{fig:elementwise_correlation} compares, on the $n = 16$ mesh,
the full element-wise indicator
$\eta_e = \mathcal{J}_e - \mathcal{J}_e^{\text{exact}}$ with the
element-wise $L^2$ velocity error, and shows separately the correlation of
the quadratic term alone.  While the full indicator $\eta_e$ shows moderate
rank correlation (Spearman $\rho = 0.69$) because of sign-indefinite cross
terms, the quadratic term achieves $\rho = 0.996$: it is an almost perfect
ranker of elements by error magnitude.

\begin{remark}
The Kovasznay test is an unusual case: all boundary conditions involve
transcendental functions that the Q9 space cannot represent exactly, leading
to constraint violation on coarse meshes and a negative gap at $n = 4$.
In typical applications (channel flows, cavity flows, cylinder flows),
the boundary conditions are polynomial or zero (no-slip walls), and the Q9
space represents them exactly.  In such cases the functional is a true upper
bound on $\mathcal{J}_{\text{exact}}$, and monotone decrease with mesh
refinement is expected.
\end{remark}

\subsection{Spatial distribution on the backward-facing step}
\label{sec:backstep_functional}

\begin{figure}[htbp]
\centering
\includegraphics[width=1.0\textwidth]{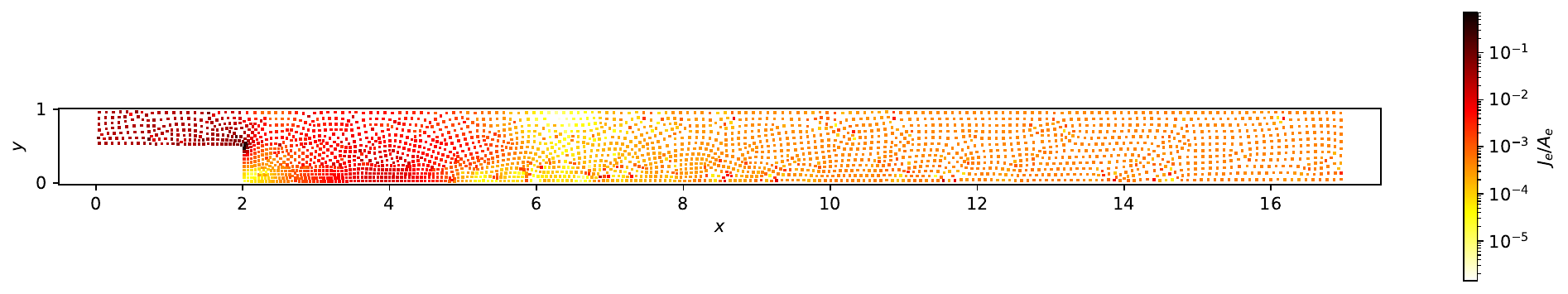}
\caption{Backward-facing step (Re~$= 100$): element-wise PMPG functional
  density $\mathcal{J}_e / A_e$ on a logarithmic color scale.
  The indicator spans four orders of magnitude and concentrates at the step
  corner, the free shear layer, and the reattachment zone.}
\label{fig:backstep_functional_map}
\end{figure}

When no exact solution is available, the functional density
$\mathcal{J}_e / A_e$ can still be used as a spatial diagnostic: elements
with high density relative to the mesh average are candidates for
refinement.  Figure~\ref{fig:backstep_functional_map} shows the element-wise
functional density on the backward-facing step (Re $= 100$,
Section~\ref{sec:backstep}) computed on an unstructured Q9 mesh with
characteristic element size $l_c = 0.10$.

The density naturally concentrates at the step corner, a geometric
singularity where the velocity gradients are strongest, and along the
free shear layer emanating from the step edge.  In the fully developed
downstream region, the density drops by four orders of magnitude.  The
indicator therefore identifies, without any additional computation, precisely
the flow features that require resolution: the separation point, the shear
layer, and the reattachment zone.

\subsection{Adaptive mesh refinement}
\label{sec:adaptive}

The functional density can drive mesh adaptation through equidistribution:
elements with high density are refined, and elements with low density are
coarsened, so as to equalize the density across the mesh.  Specifically, the
target element size is computed as
\begin{equation}
  h_e^{\text{new}}
  = h_e \left( \frac{\bar{\mathcal{J}}}{\mathcal{J}_e / A_e} \right)^{\!\alpha},
  \label{eq:equidistribution}
\end{equation}
where $h_e = \sqrt{A_e}$ is the current element size,
$\bar{\mathcal{J}}$ is the mean functional density, and $\alpha = 0.25$.
The target sizes are clipped to $[h_{\min}, h_{\max}]$, projected from
element centers to mesh nodes by averaging, and passed to
Gmsh~\cite{geuzaine2009gmsh} as a background size field to generate a new
Q9 mesh.

\begin{figure}[htbp]
\centering
\includegraphics[width=0.85\textwidth]{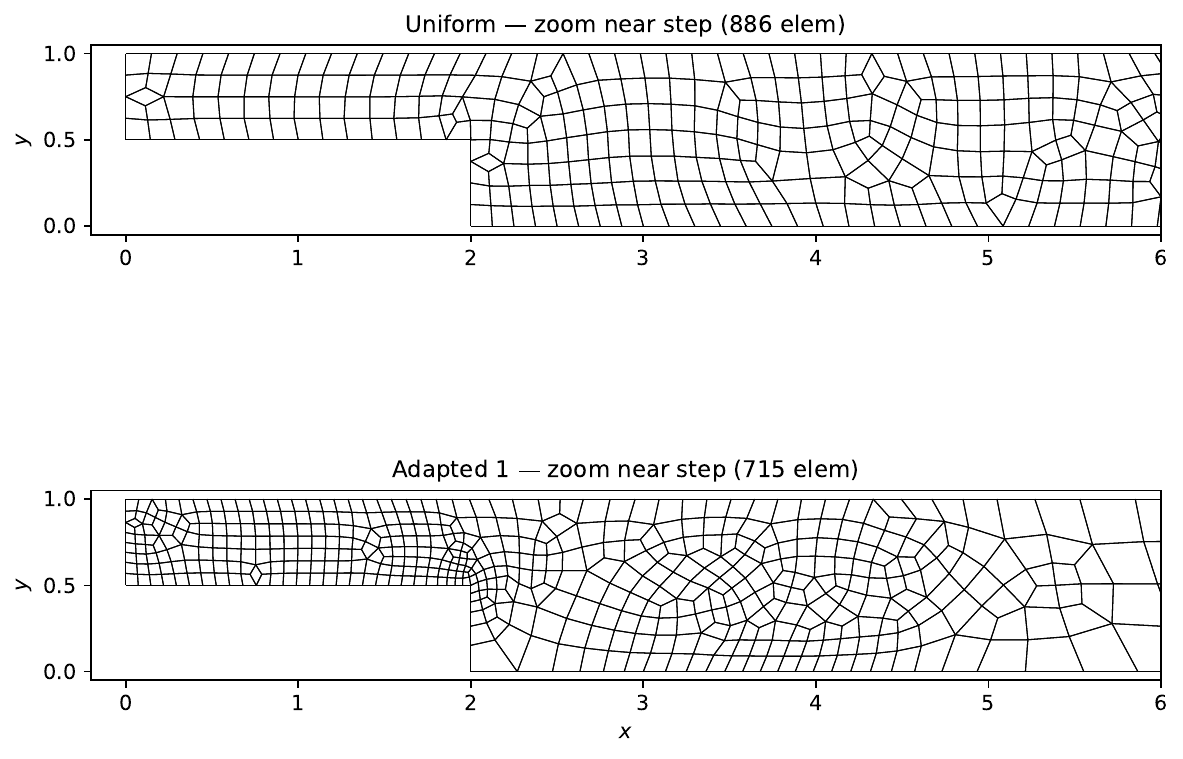}
\caption{Backward-facing step: mesh topology near the step corner.
  \textbf{Top:}~Uniform mesh (886~elements).
  \textbf{Bottom:}~After one adaptation cycle (715~elements, $-19\%$).
  Elements concentrate at the step corner and shear layer, coarsening
  downstream where the flow is well resolved.}
\label{fig:backstep_adaptive_zoom}
\end{figure}

Figure~\ref{fig:backstep_adaptive_zoom} compares the mesh topology near the
step corner before and after adaptation.  The initial uniform
mesh ($l_c = 0.15$) contains 886 elements.  After a single adaptation
cycle, the adapted mesh contains 715~elements (a 19\% reduction) while
maintaining the same global functional value.  The elements concentrate at
the step corner and along the shear layer, precisely where the functional
density is largest, and coarsen in the developed downstream region where the
flow is well resolved.

The purpose of this demonstration is not to develop a full adaptive mesh
refinement algorithm (issues such as convergence criteria, oscillation
control, and optimal refinement strategies are beyond the present scope)
but rather to illustrate that the PMPG functional provides a natural,
zero-cost error indicator that can drive standard mesh adaptation tools
without any additional machinery.

\section{Viscosity estimation from velocity field data}
\label{sec:inverse}

Estimating fluid properties from velocity field measurements is an active
area of research.  Physics-informed neural networks
(PINNs)~\cite{raissi2019physics,raissi2020hidden} co-learn velocity,
pressure, and Navier--Stokes parameters by minimizing PDE residuals through
iterative training.  Bayesian inverse Navier--Stokes
methods~\cite{kontogiannis2025learning} reconstruct the full flow field and
learn rheological parameters via adjoint-based optimization with
uncertainty quantification.  These approaches are powerful and general, but
they involve iterative optimization loops and operate on the
Navier--Stokes equations in a form that includes pressure as an
additional unknown.

In experimental solid mechanics, the Virtual Fields Method
(VFM)~\cite{pierron2012virtual} takes a different approach: it tests the
weak form of equilibrium against judiciously chosen virtual displacement
fields to extract constitutive parameters directly from full-field strain
measurements, without solving any forward problem.  No iterative
optimization is needed; the unknown parameters appear linearly in the
weak form, and the identification reduces to a direct linear solve.
Crucially, the weak form involves only first spatial derivatives (strains),
not second derivatives, making the method inherently more robust to
measurement noise.

The PMPG stationarity condition~\eqref{eq:pmpg_stationarity} naturally
suggests an analogous approach for fluid viscosity.  In the forward solver,
this condition determines~$\dot{\bd}$ for a known~$\nu$; reading it
backwards, with the velocity and its time derivative known from experiment,
it becomes a constraint on~$\nu$.  Since the formulation is velocity-only
(no pressure field is ever introduced), the measured velocity enters
directly into the same finite element operators ($\bM$, $\bK$,
$\bm{r}_{\text{conv}}$) used in the forward solver.  The constraint
forces~$\bC^T\blambda$ are removed by projecting onto the null space
of~$\bC$, and the viscosity is recovered from a single ratio of inner
products.  Like the VFM, only first spatial derivatives of the measured
velocity are needed (through the stiffness matrix~$\bK$), not the
Laplacian~$\nabla^2\bv$ that appears in the strong-form Navier--Stokes
equations.  The entire identification is a direct algebraic computation:
no forward solve, no adjoint, no iterative optimization.

While the underlying idea (testing a weak momentum balance against
divergence-free fields to isolate~$\nu$) could in principle be pursued
with any consistent Galerkin discretization, the velocity-only PMPG
formulation makes the implementation particularly clean: there is no
pressure field to reconstruct or eliminate, and the collocation-point
divergence constraint (Section~\ref{sec:div_constraint}) provides the
sparse matrix~$\bC_{\text{div}}$ from which the projector is built via a
single sparse factorization.

In this section we develop this idea into a concrete method for identifying
the kinematic viscosity from particle image velocimetry (PIV) data.  The
method is demonstrated on synthetic data (converged numerical solutions
used as surrogate PIV measurements, with and without added noise),
leaving validation on experimental PIV data for future work.

\subsection{Formulation}
\label{sec:inverse_formulation}

Consider an experimentalist who has measured the velocity field in a
rectangular window at a regular array of points, the typical output of a
PIV system.  The measured velocities are assigned to a Q9 finite element
mesh: each $3 \times 3$ block of measurement points defines one biquadratic
element.  From this mesh, the mass matrix~$\bM$, the stiffness
matrix~$\bK$, the convective residual~$\bm{r}_{\text{conv}}(\bd)$, and the
divergence constraint matrix~$\bC$ (Section~\ref{sec:div_constraint}) are
assembled in the usual way.  No boundary conditions are imposed;
the only constraint is incompressibility,
$\bC\bd = \bm{0}$.

The PMPG stationarity
condition~\eqref{eq:pmpg_stationarity} extended with the constraint force
from incompressibility reads
\begin{equation}\label{eq:inverse_balance}
  \bM \dot{\bd} + \bm{r}_{\text{conv}}(\bd) + \nu\,\bK\bd
  + \bC^T\blambda = \bm{0}.
\end{equation}
In the forward solver, $\nu$ is known and~$\dot{\bd}$ is the
unknown; here, the velocity and its time derivative are data and~$\nu$ is
the unknown.  Defining
\begin{equation}\label{eq:ah_bh}
  \ba_h = \bM\dot{\bd} + \bm{r}_{\text{conv}}(\bd),
  \qquad
  \bb_h = \bK\bd,
\end{equation}
\eqref{eq:inverse_balance} becomes
$\ba_h + \nu\,\bb_h + \bC^T\blambda = \bm{0}$.
The constraint force~$\bC^T\blambda$ is unknown, but it can be eliminated as follows.
Let $\mathbb{P}$ denote the orthogonal projector onto~$\Null(\bC)$,
defined for any vector~$\bff$ by
\begin{equation}\label{eq:projector_def}
  \mathbb{P}\,\bff = \bff - \bC^T(\bC\bC^T)^{-1}\bC\bff.
\end{equation}
Applying $\mathbb{P}$ to~\eqref{eq:inverse_balance} eliminates the
constraint force and gives
$\mathbb{P}\ba_h + \nu\,\mathbb{P}\bb_h = \bm{0}$, an overdetermined
system in the single scalar~$\nu$.  The least-squares solution in the
$\bM$-weighted inner product is
\begin{equation}\label{eq:nu_estimate}
  \boxed{
  \nu
  = -\frac{(\mathbb{P}\bb_h)^T\bM\,(\mathbb{P}\ba_h)}
          {(\mathbb{P}\bb_h)^T\bM\,(\mathbb{P}\bb_h)}.
  }
\end{equation}
Each projection requires only a sparse triangular solve with a
pre-factored~$\bC\bC^T$; no dense null-space basis is formed.

\begin{remark}
Every quantity in~\eqref{eq:nu_estimate} (the matrices $\bM$, $\bK$,
$\bC$, the convective residual~$\bm{r}_{\text{conv}}$, and the velocity
$\bd$ and its time derivative~$\dot{\bd}$) is computed entirely from the
measured velocities and the Q9 shape functions.  No boundary conditions,
pressure reconstruction, or forward simulation enter the computation.
\end{remark}

\paragraph{Interior nodes.}
At the boundary of the measurement window, the outermost elements have nodes
whose basis function support extends to elements that lie outside the field
of view.  The momentum balance~\eqref{eq:inverse_balance} at such a node is
incomplete because contributions from the missing elements are absent.
We therefore restrict the
estimate to \emph{interior} nodes: those whose complete element neighborhood
lies within the measurement window.  Let $\mathcal{I}$ denote the set of
interior DOFs.  The divergence constraint is restricted to these DOFs,
$\bC_{\mathcal{I}} = \bC(:, \mathcal{I})$, and the
projector~\eqref{eq:projector_def} and
estimate~\eqref{eq:nu_estimate} are evaluated on this reduced set with
$\bM_{\mathcal{I}} = \bM(\mathcal{I}, \mathcal{I})$.

Two features of this formulation deserve emphasis.  First, the
vectors~$\ba_h$ and~$\bb_h$ involve at most first spatial derivatives of the
measured velocity; no second spatial derivatives are needed, unlike
methods based on the strong-form Navier--Stokes equations.
Second, the entire identification is a direct algebraic computation:
no forward solve, no adjoint, no iterative optimization.  The procedure is
summarized in Algorithm~\ref{alg:inverse}.

\begin{algorithm}[htbp]
\caption{Viscosity estimation from velocity field data.}
\label{alg:inverse}
\begin{algorithmic}[1]
\REQUIRE Velocity $\bd$ and its time derivative $\dot{\bd}$ at grid points
\ENSURE  Kinematic viscosity $\nu$
\medskip
\STATE Assign Q9 elements to the measurement grid.
\STATE Assemble $\bM$, $\bK$, $\bC$ on these elements.
\STATE Identify interior DOFs~$\mathcal{I}$ (nodes with complete element support).
\STATE Restrict: $\bC_{\mathcal{I}} \leftarrow \bC(:, \mathcal{I})$.
\STATE Factor $\bC_{\mathcal{I}}\bC_{\mathcal{I}}^T$ (sparse, once).
\STATE Compute $\ba_h = \bM\dot{\bd} + \bm{r}_{\text{conv}}(\bd)$
       and $\bb_h = \bK\bd$.
\STATE Restrict to interior DOFs:
       $\ba_{\mathcal{I}} \leftarrow \ba_h(\mathcal{I})$, \;
       $\bb_{\mathcal{I}} \leftarrow \bb_h(\mathcal{I})$.
\STATE Project: $\mathbb{P}\ba_{\mathcal{I}}
       \leftarrow \ba_{\mathcal{I}}
       - \bC_{\mathcal{I}}^T
         (\bC_{\mathcal{I}}\bC_{\mathcal{I}}^T)^{-1}
         \bC_{\mathcal{I}}\ba_{\mathcal{I}}$
       \; (likewise for $\bb_{\mathcal{I}}$).
\STATE $\displaystyle \nu \leftarrow
       -\frac{(\mathbb{P}\bb_{\mathcal{I}})^T
              \bM_{\mathcal{I}}\,(\mathbb{P}\ba_{\mathcal{I}})}
             {(\mathbb{P}\bb_{\mathcal{I}})^T
              \bM_{\mathcal{I}}\,(\mathbb{P}\bb_{\mathcal{I}})}$.
\end{algorithmic}
\end{algorithm}

\subsection{Temporal polynomial regression}
\label{sec:inverse_regression}

Algorithm~\ref{alg:inverse} requires the velocity~$\bd$ and its time
derivative~$\dot{\bd}$ at the measurement grid points.  In a PIV experiment,
what is available is a sequence of~$N$ velocity snapshots at
times $t_0, t_1, \ldots, t_{N-1}$, so both~$\bd$ and~$\dot{\bd}$ must be
estimated from the discrete time series.  A simple finite difference $\dot{\bd} \approx (\bd^{n+1} - \bd^n) /
\Delta t$ amplifies measurement noise by a factor~$1/\Delta t$, degrading
the viscosity estimate.

We address this with a sliding-window polynomial regression, following the
Savitzky--Golay approach~\cite{savitzky1964}.  Within a window of~$W$
consecutive snapshots at times $t_0, \ldots, t_{W-1}$, a polynomial
of degree~$p < W$ is fit to each DOF's time series by least squares.
The fitted polynomial is then evaluated, along with its time derivative,
at the window midpoint $t_{\text{mid}} = (t_0 + t_{W-1})/2$.

The regression weights are independent of the data.  Defining the centered
Vandermonde matrix $V_{ij} = (t_i - t_{\text{mid}})^j$ and its
pseudoinverse $V^+ = (V^T V)^{-1} V^T$, the fitted value and derivative at
$t_{\text{mid}}$ are
\begin{equation}\label{eq:regression}
  d_k(t_{\text{mid}}) = \bm{w}_{\text{val}}^T \bm{d}_k,
  \qquad
  \dot{d}_k(t_{\text{mid}}) = \bm{w}_{\text{der}}^T \bm{d}_k,
\end{equation}
where $\bm{w}_{\text{val}} = V^+_{0,:}$ and
$\bm{w}_{\text{der}} = V^+_{1,:}$ are the zeroth- and first-order weight
vectors, and $\bm{d}_k = (d_k(t_0), \ldots, d_k(t_{W-1}))^T$ is the vector
of samples for DOF~$k$.  The weights are computed once and applied to all
DOFs by a single matrix--vector product.

When $p < W - 1$, the polynomial fit has $W - p - 1$ residual degrees of
freedom that average out noise.  The noise in the smoothed velocity and its
derivative scales as $\sim \sigma / \sqrt{W - p - 1}$ compared to the raw
measurement noise~$\sigma$.  The polynomial degree~$p$ controls the
bias--variance tradeoff: higher~$p$ captures more temporal curvature (relevant
for oscillating flows) but leaves fewer degrees of freedom for averaging.

Sliding the window one step at a time through the~$N$ snapshots
produces $N - W + 1$ viscosity estimates.  For noisy data, these are
averaged to reduce variance.  When multiple independent noise realizations
are available (e.g., from different PIV acquisition runs), the procedure is
applied to each realization, and the results are pooled.

\subsubsection{Pre-filtering of noisy snapshots}

Measurement noise violates the incompressibility constraint
$\bC \bd = \bm{0}$, introducing a compressible velocity
component that biases the viscosity estimate.  Before applying the temporal
regression, each noisy snapshot is projected onto the divergence-free
subspace using
$\mathbb{P}\,\bd = \bd - \bC^T
(\bC\bC^T)^{-1}
\bC\bd$, where the projection uses all DOFs in the measurement window
(not just the interior ones).  For isotropic noise in two dimensions, this
removes roughly 30--50\% of the noise variance.

\subsection{Numerical demonstration}
\label{sec:inverse_demo}

We demonstrate the method on the backward-facing step
(Section~\ref{sec:backstep}), using its converged velocity field as
synthetic PIV data.  The backward-facing step at $Re = 100$
($\nu_{\text{true}} = 0.005$) provides a nearly steady test case: the
converged flow changes by~$O(10^{-4})$ between consecutive time steps, so
the time derivative~$\dot{\bd}$ is negligible compared to the spatial terms.
Nevertheless, we use the full transient formulation with $N = 65$ snapshots
separated by $\Delta t = 0.007$.

\begin{figure}[htbp]
  \centering
  \includegraphics[width=0.85\textwidth]{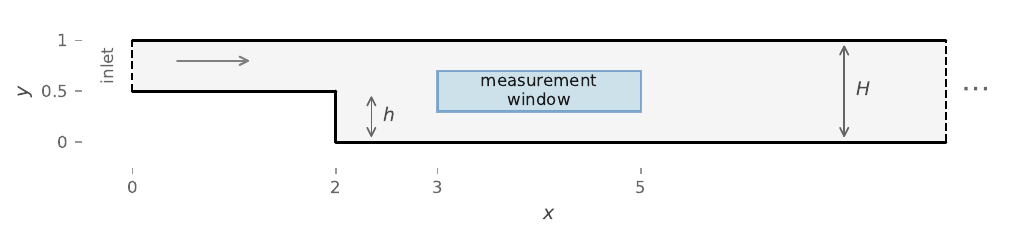}
  \caption{Backward-facing step geometry with the measurement window used for
    viscosity estimation.  The window is placed in the recirculation region
    downstream of the step, away from all domain boundaries.}
  \label{fig:backstep_patch_layout}
\end{figure}

A rectangular measurement window of $75$~elements ($498$~interior DOFs) is
placed just downstream of the step ($x \in [3, 5]$, $y \in [0.3, 0.7]$,
Figure~\ref{fig:backstep_patch_layout}),
capturing the recirculation zone where the velocity gradients, and
hence the viscous signal, are strongest.
With clean data, the sliding
window ($W = 33$, $p = 1$) recovers
$\nu_{\text{est}} / \nu_{\text{true}} = 0.9998$ (averaged over all window
positions).  Four additional windows placed elsewhere in the channel
(developed flow, far downstream, outlet region) give comparably accurate
results, confirming that the method is not sensitive to the choice of window
location provided the velocity gradients are sufficiently strong.

\begin{figure}[htbp]
  \centering
  \includegraphics[width=0.65\textwidth]{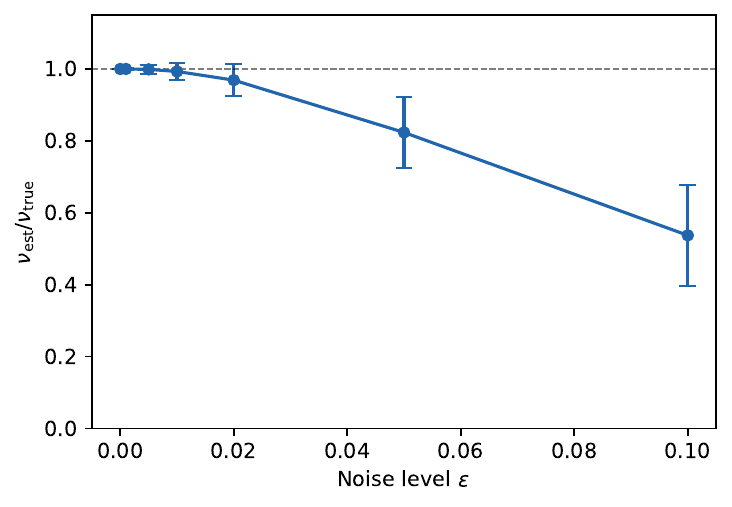}
  \caption{Viscosity estimate vs.\ noise level for the near-step measurement
    window ($W = 33$, $p = 1$).  Error bars show $\pm 1$ standard deviation
    over 20 noise realizations.  Pre-filtering onto the divergence-free
    subspace is applied to each noisy snapshot before regression.}
  \label{fig:backstep_noise}
\end{figure}

To simulate measurement uncertainty, isotropic Gaussian noise of standard
deviation $\sigma = \varepsilon\,V_{\text{rms}}$ is added to each velocity
component, where $V_{\text{rms}} = \sqrt{\langle |\bv|^2 \rangle}$ is the
spatial root-mean-square velocity.  Each noisy snapshot is pre-filtered onto
the divergence-free subspace before regression.
Figure~\ref{fig:backstep_noise} shows the resulting viscosity estimates as a
function of noise level~$\varepsilon$.  At~$1\%$ noise, the method recovers
$\nu_{\text{est}} / \nu_{\text{true}} = 0.99 \pm 0.02$ (mean $\pm$ one
standard deviation over 20 independent noise realizations and all
sliding-window positions).

%

The dominant noise degradation mechanism is an \emph{errors-in-variables}
attenuation bias: the viscous term $\bb_h = \bK\bd$ appears in the
denominator of~\eqref{eq:nu_estimate}, and noise in~$\bd$ biases it toward
zero.  This effect can be mitigated by increasing the temporal window
size~$W$, which averages the noise over more snapshots.  Among the
polynomial degrees tested ($p = 1, 2, 3$), $p = 1$ (linear regression)
consistently provides the best results because it leaves the most degrees of
freedom ($W - 2$) for noise averaging.  In a typical PIV experiment with
frame rates of hundreds of Hz, windows of $W \geq 30$ frames are easily
achievable.

\subsection{Discussion}
\label{sec:inverse_discussion}

The numerical demonstration shows that the PMPG stationarity condition can
be read backwards to recover the kinematic viscosity directly from velocity
field data, with near-exact results on clean data and graceful degradation
under measurement noise.  The most important practical consideration is the
placement of the measurement window: the accuracy depends strongly on the
local signal-to-noise ratio of the viscous term~$\bK\bd$, so windows should
target regions with strong velocity gradients (shear layers,
recirculation zones, boundary layers) where the viscous contribution to
the momentum balance is largest relative to measurement noise.

The demonstration above assumed a spatially uniform Newtonian viscosity, but
the formulation extends naturally to spatially varying viscosity
$\nu(\bx) = \sum_a N_a(\bx)\,\nu_a$.  The viscous term becomes
$\sum_a \nu_a\,\bK_a\bd$ with
$(K_a)_{ij} = \int_\Omega N_a \nabla N_i \cdot \nabla N_j \, d\Omega$,
and the projected system becomes a linear system in the
vector~$\bm{\nu} = (\nu_1, \ldots, \nu_{n_{\text{nodes}}})^T$.  This
opens the possibility of mapping the spatial distribution of effective
viscosity (including turbulent eddy viscosity) directly from PIV
measurements.  For generalized Newtonian fluids, where the viscosity depends
on the local strain rate, $\nu = \nu(\dot{\gamma})$, the same framework
applies: the rheological parameters enter the projected momentum balance
algebraically and can be identified without pressure reconstruction or
forward simulation.

\section{Summary and future work}
\label{sec:conclusions}

We have presented the first finite element implementation of the principle of
minimum pressure gradient (PMPG) for incompressible viscous flow. Unlike
Galerkin formulations, which discretize a weak form of the governing equations,
the present approach applies the Rayleigh--Ritz method directly to the PMPG
variational principle: at each instant, the nodal velocity
rates~$\dot{\bd}$ are determined by minimizing the $L^2$ norm of the implied
pressure gradient, subject to incompressibility and boundary constraints.
This variational structure has three important consequences:

\begin{enumerate}
  \item \textbf{Velocity-rate-only formulation.}
    No pressure finite element space is introduced and no inf--sup condition
    must be satisfied. Pressure-related quantities enter solely through the
    Lagrange multipliers of the constraint system, which provide wall forces
    (pressure and viscous shear) as a direct byproduct of the solve.

  \item \textbf{Symmetric positive definite system matrix.}
    Because convection appears only in the explicit right-hand side, the system
    matrix $\bM + \nu \Delta t\, \bK$ is SPD for any positive viscosity and time
    step, regardless of the Reynolds number. The formulation therefore requires
    no stabilization, even on coarse meshes in the convection-dominated regime.

  \item \textbf{Built-in error indicator.}
    The element-wise PMPG functional density measures the local momentum
    residual at zero additional cost, providing a natural criterion for adaptive
    mesh refinement without adjoint problems or reference solutions.
\end{enumerate}

The formulation was verified against two exact solutions: Poiseuille channel
flow (machine-precision recovery) and the Kovasznay solution (spatial
convergence rate $\sim 3.3$, consistent with $O(h^3)$ for biquadratic
elements). It was validated against published benchmark data for the lid-driven
cavity (Ghia et al.~\cite{ghia1982}), the backward-facing step (Armaly et
al.~\cite{armaly1983}), and flow past a circular cylinder (Dennis \&
Chang~\cite{dennis1970}, Tritton~\cite{tritton1959},
Williamson~\cite{williamson1996}), demonstrating good agreement across a range
of Reynolds numbers. In particular, the lid-driven cavity at $Re = 1000$ on
coarse wall-graded meshes with element P\'{e}clet numbers up to~42 confirmed
the oscillation-free behavior predicted by the SPD structure.

The Lagrange multipliers from the saddle-point system were used to extract wall
forces without pressure reconstruction, yielding reattachment lengths on the
backward-facing step and drag and lift coefficients on the cylinder that agreed
with published values to within a few percent.

The error indicator was validated on the Kovasznay problem (Spearman
$\rho = 0.996$ correlation with the true element-wise velocity error) and
applied to the backward-facing step, where a single round of
equidistribution-based adaptation reduced the element count by 19\% while
maintaining the same functional value.

Beyond forward simulation, we showed that the PMPG stationarity condition can
be read backwards to estimate the kinematic viscosity directly from velocity
field data. The constraint forces are eliminated by projecting onto the
divergence-free subspace, and the viscosity is recovered from a single ratio of
inner products, requiring no boundary conditions, pressure reconstruction, or
iterative optimization. On a backward-facing step test case, the estimator
recovers the true viscosity to within $0.02\%$ from clean data, and to within
$1\%$ at $1\%$ measurement noise. The formulation extends naturally to
spatially varying and non-Newtonian viscosities.

The present implementation is a first step toward realizing the full potential
of the PMPG framework. Several extensions are envisioned:

\begin{itemize}
  \item \textbf{Higher Reynolds numbers.}
    The cavity results at $Re = 1000$ (Section~\ref{sec:cavity_re1000})
    show that the PMPG formulation is structurally stable at high element
    P\'{e}clet numbers, producing smooth solutions even when the boundary
    layer is not resolved by a single element. At still higher Reynolds
    numbers, boundary and shear layers become thinner and require finer
    meshes for \emph{accuracy}. The element-wise PMPG functional density
    (Section~\ref{sec:error_estimation}) already identifies under-resolved
    regions, and the adaptive refinement strategy demonstrated on the
    backward-facing step (Section~\ref{sec:adaptive}) provides a natural
    mechanism for concentrating elements where they are needed, without
    requiring any stabilization machinery.

  \item \textbf{Three-dimensional extension.}
    The formulation extends naturally to 3D using 27-node hexahedral elements
    (Q27) with trilinear geometry mapping. The divergence constraint remains
    linear, and the saddle-point structure is unchanged.

  \item \textbf{Turbulence modeling.}
    The PMPG framework may offer new perspectives on turbulence modeling, as the
    variational principle provides a natural energy-based criterion for
    subgrid-scale closures.

  \item \textbf{Moving boundaries and fluid--structure interaction.}
    The constraint-based architecture is well-suited for incorporating moving
    boundaries, as the constraints can be updated at each time step without
    modifying the underlying element formulation. Furthermore, the Lagrange
    multipliers provide the total nodal forces (pressure and viscous shear)
    at wall nodes as a direct byproduct of the solve, eliminating the need for
    separate force reconstruction. These nodal forces are precisely the
    coupling loads required by a structural solver, making the PMPG formulation
    a natural candidate for partitioned or monolithic fluid--structure
    interaction schemes.

  \item \textbf{Connections to the variational framework.}
    The PMPG principle is part of a broader variational framework for fluid
    mechanics~\cite{gonzalez2022variational,taha2023minimization} that includes
    Gauss's principle and the Appell function. Exploring these connections may
    yield new computational strategies and physical insights.
\end{itemize}

\section*{Acknowledgments}
The author thanks Professors Haithem Taha and Perry Johnson (Department of Mechanical and Aerospace Engineering, University of California, Irvine) for valuable discussions during the development of this work.

\appendix
\section*{Appendices}
\section{Optimality of the $2\times 2$ Gauss collocation points}
\label{app:collocation}

This appendix analyzes the choice of collocation points for the
divergence-free constraint~\eqref{eq:div_row}.  We introduce a rigorous measure of
divergence leakage, derive optimal collocation-point placements in closed
form, and explain why the $2\times 2$ Gauss points (which are not
per-element optimal) produce the best results in assembled finite
element computations.  All notation follows the main text; in particular,
$\bm{d}$, $\bm{B}_\xi$, $\bm{B}_\eta$, $\bm{J}$, and $\bm{c}_k^{\text{div}}$
are defined in equations~\eqref{eq:dof_vector}, \eqref{eq:shape_derivs}, \eqref{eq:jacobian_matrix}, and~\eqref{eq:div_row}.

\subsection{Divergence leakage operator}
\label{app:leakage}

At any point $(\xi,\eta)$ in the reference element, the scaled divergence
$(\nabla\!\cdot\!\bm{v})\,\det(\bm{J}) = \bm{c}^{\text{div}}\cdot\bm{d}$
is a linear function of the element DOF vector
$\bm{d}\in\mathbb{R}^{18}$.  To measure the integrated divergence
residual over the element we define the \emph{divergence mass matrix}
\begin{equation}\label{eq:app_Mdef}
	\bm{M}_{\text{div}} = \int_{-1}^{1}\!\!\int_{-1}^{1}
	(\bm{c}^{\text{div}})^T \bm{c}^{\text{div}}\;d\xi\,d\eta
	\;\in\;\mathbb{R}^{18\times 18},
\end{equation}
which is symmetric positive semi-definite and satisfies
$\bm{d}^T\bm{M}_{\text{div}}\,\bm{d}
= \int_{-1}^{1}\!\int_{-1}^{1}
[(\nabla\!\cdot\!\bm{v})\det(\bm{J})]^2\;d\xi\,d\eta$.
For a square element where $\det(\bm{J})$ is constant, this is
proportional to $\int(\nabla\!\cdot\!\bm{v})^2\,d\Omega$.

Evaluating the constraint~\eqref{eq:div_row} at four collocation points yields a
$4\times 18$ constraint matrix~$\bm{A}$ whose null space
$\Null(\bm{A})\subset\mathbb{R}^{18}$ has dimension~14.  Let
$\bm{V}\in\mathbb{R}^{18\times 14}$ be an orthonormal basis for this null
space.  Every admissible velocity field satisfies
$\bm{d} = \bm{V}\bm{z}$ for some $\bm{z}\in\mathbb{R}^{14}$, and its
divergence residual is
\begin{equation}\label{eq:app_leakage}
	\bm{d}^T\bm{M}_{\text{div}}\,\bm{d}
	= \bm{z}^T
	\underbrace{\bm{V}^T\bm{M}_{\text{div}}\,\bm{V}}_{\displaystyle\bm{L}}
	\;\bm{z},
\end{equation}
where $\bm{L}\in\mathbb{R}^{14\times 14}$ is the \emph{leakage operator}.
Its trace $\tr(\bm{L})$ measures the total divergence leakage and admits
the basis-free formula
\begin{equation}\label{eq:app_trace}
	\tr(\bm{L}) = \tr(\bm{M}_{\text{div}})
	- \tr\!\bigl(\bm{A}\,\bm{M}_{\text{div}}\,\bm{A}^T
	(\bm{A}\bm{A}^T)^{-1}\bigr).
\end{equation}

Since the 14~eigenvalues of $\bm{L}$ interlace with those of
$\bm{M}_{\text{div}}$, the leakage satisfies the lower bound
\begin{equation}\label{eq:app_bound}
	\tr(\bm{L}) \;\geq\; \sum_{i=1}^{14}\mu_i,
\end{equation}
where $\mu_1\leq\cdots\leq\mu_{18}$ are the eigenvalues of
$\bm{M}_{\text{div}}$.  This bound is the best any four linear constraints
can achieve.

\subsection{Analysis on the unit square}
\label{app:square}

On the reference square $[-1,1]^2$, the Jacobian is the identity and
$\bm{c}^{\text{div}} = (\bm{B}_\xi\;\;\bm{B}_\eta)$.  The tensor-product
structure of the Q9 shape functions allows $\bm{M}_{\text{div}}$ to be
assembled from the $3\times 3$ one-dimensional mass, stiffness, and mixed
matrices of the 1D Lagrange basis~\eqref{eq:lagrange_1d}.

The matrix has $\rank(\bm{M}_{\text{div}}) = 8$: the 10-dimensional null
space consists of the biquadratic velocity fields that are exactly
divergence-free.  The 8~nonzero eigenvalues, reflecting the $D_4$ symmetry
of the square, are:
\begin{equation}\label{eq:app_eigvals}
	\renewcommand{\arraystretch}{1.2}
	\begin{array}{llll}
		\mu_{1} = \tfrac{17-\sqrt{209}}{15}\approx 0.170, &
		\mu_{2} = \tfrac{4}{15}\approx 0.267, &
		\mu_{3,4} = \tfrac{27-\sqrt{409}}{10}\approx 0.678\;\;(\times 2), \\[6pt]
		\mu_{5} = \tfrac{17+\sqrt{209}}{15}\approx 2.097, &
		\mu_{6} = \tfrac{8}{3}\approx 2.667, &
		\mu_{7,8} = \tfrac{27+\sqrt{409}}{10}\approx 4.722\;\;(\times 2),
	\end{array}
\end{equation}
with $\sum_{i=1}^{8}\mu_i = 16 = \tr(\bm{M}_{\text{div}})$.
The theoretical minimum leakage~\eqref{eq:app_bound} equals the sum of the
four smallest nonzero eigenvalues:
\begin{equation}\label{eq:app_best}
	\tr(\bm{L})_{\min}
	= \frac{34}{5} - \frac{\sqrt{209}}{15} - \frac{\sqrt{409}}{5}
	\approx 1.791.
\end{equation}

\subsection{Optimal collocation points}
\label{app:optimal}

The $D_4$ symmetry of the square constrains the optimal configuration.  We
consider two natural one-parameter families that respect this symmetry.

\medskip\noindent\textbf{Diagonal placement: $(\pm s,\,\pm s)$.}\quad
Evaluating~\eqref{eq:app_trace} symbolically yields a closed-form leakage:
\begin{equation}\label{eq:app_leakage_diag}
	\mathcal{L}_{\text{diag}}(s)
	= \frac{4(9234s^8 - 18684s^6 + 16632s^4 - 6858s^2 + 1297)}
	{45(3s^4 - 4s^2 + 2)(36s^4 - 48s^2 + 25)}.
\end{equation}
The minimum occurs at $s^*\approx 0.4959$, with the closed-form value at
$s=1/2$ essentially matching it:
$\mathcal{L}_{\text{diag}}(1/2)= 18746/10431 \approx 1.797$.  The
$2\times 2$ Gauss points correspond to $s = 1/\sqrt{3}$, giving
$\mathcal{L}_{\text{diag}}(1/\sqrt{3}) = 1124/585 \approx 1.921$.

\medskip\noindent\textbf{Axis placement: $(0,\pm s)$ and $(\pm s,0)$.}\quad
The leakage is
$\mathcal{L}_{\text{axis}}(s) = (19332s^4-45552s^2+42352)/(3375s^4-9000s^2+9000)$,
with optimum $\approx 4.564$, more than 2.5 times the theoretical bound.
Axis points probe the divergence along the coordinate directions and miss
the dominant diagonal stretching modes ($\mu_{7,8}\approx 4.72$).

Table~\ref{tab:app_comparison} collects the key results.

\begin{table}[ht]
	\centering
	\renewcommand{\arraystretch}{1.4}
	\caption{Leakage $\tr(\bm{L})$ for four-point collocation schemes on the
		unit square.  The ratio is relative to the theoretical
		bound~\eqref{eq:app_best}.}
	\label{tab:app_comparison}
	\begin{tabular}{lccc}
		\toprule
		\textbf{Collocation points} & $s$ & $\tr(\bm{L})$ &
		\textbf{Ratio to bound} \\
		\midrule
		Theoretical bound (any 4 constraints)
		& --- & $1.791$ & $1.000$ \\[2pt]
		Optimal diagonal
		& $0.496$ & $1.797$ & $1.003$ \\[2pt]
		$(\pm\tfrac{1}{2},\,\pm\tfrac{1}{2})$
		& $0.500$ & $1.797$ & $1.003$ \\[2pt]
		$2\times 2$ Gauss
		& $1/\sqrt{3}$ & $1.921$ & $1.073$ \\[2pt]
		Optimal axis
		& $0.755$ & $4.564$ & $2.548$ \\[2pt]
		\bottomrule
	\end{tabular}
\end{table}

By the per-element measure, the diagonal $(\pm\frac{1}{2},\,
\pm\frac{1}{2})$ points are nearly optimal (within~$0.3\%$ of the bound),
the $2\times 2$ Gauss points are $7.3\%$ above it, and axis points are
dramatically inferior.

\medskip\noindent\textbf{General quadrilaterals.}\quad
To verify robustness, the analysis was repeated on 200 random convex
quadrilaterals (corner perturbations with standard deviation~$0.3$).  The
diagonal $(\pm\frac{1}{2},\,\pm\frac{1}{2})$ points maintain a mean
leakage ratio of~$1.02$ and worst case below~$1.19$; the $2\times 2$ Gauss
points give a mean ratio of~$1.09$.  Axis-aligned points exceed a ratio
of~$2.5$ on average.

\subsection{Polynomial orthogonality of the $2\times 2$ Gauss rule}
\label{app:orthogonality}

Despite their worse per-element leakage, the $2\times 2$ Gauss points
produce lower integrated divergence and better velocity convergence in
assembled computations (Figure~\ref{fig:kovasznay_convergence}).  The explanation is a
polynomial-degree argument.

In the subparametric formulation, each Jacobian component $j_{ij}$ is at
most linear in $(\xi,\eta)$; see~\eqref{eq:jac_components}.  The shape function derivatives
$\partial N_a/\partial\xi$ are degree~$(1,2)$ and
$\partial N_a/\partial\eta$ are degree~$(2,1)$.  The products in the
divergence constraint~\eqref{eq:div_row} therefore satisfy
\begin{equation}\label{eq:app_degree}
	(\nabla\!\cdot\!\bm{v})\,\det(\bm{J})
	\;\;\text{is at most degree $(2,2)$ in $(\xi,\eta)$}.
\end{equation}
For any bilinear function $q(\xi,\eta)$ (degree $\leq 1$ in each
variable), the product $[(\nabla\!\cdot\!\bm{v})\det(\bm{J})]\cdot q$ is
at most degree~$(3,3)$, which the $2\times 2$ Gauss rule integrates
exactly.  If the divergence constraint vanishes at the four Gauss points,
then
\begin{equation}\label{eq:app_weak}
	\int_{\Omega_e}(\nabla\!\cdot\!\bm{v})\,q\;d\Omega
	\;=\; \sum_{k=1}^{4}w_k
	\underbrace{[(\nabla\!\cdot\!\bm{v})\det(\bm{J})]_{P_k}}_{=\,0}\,q(P_k)
	\;=\; 0
	\qquad\forall\, q \in Q_1,
\end{equation}
where $w_k=1$ are the Gauss weights and the quadrature is exact.

In words: \emph{pointwise collocation at the $2\times 2$ Gauss points
	implies that the divergence residual is $L^2$-orthogonal to the entire
	bilinear polynomial space $Q_1 = \operatorname{span}\{1,\,\xi,\,\eta,\,
	\xi\eta\}$ on each element.}  The residual divergence can contain only
high-order modes (quadratic and above in $\xi$ or $\eta$).

\begin{remark}\label{rem:app_half_no_gauss}
	The $(\pm\frac{1}{2},\,\pm\frac{1}{2})$ points are not Gauss points for
	any standard rule.  A four-point rule at these locations has
	insufficient polynomial exactness: the product
	$[(\nabla\!\cdot\!\bm{v})\det(\bm{J})]\cdot q$ is degree~$(3,3)$, but
	four points at $(\pm\frac{1}{2},\,\pm\frac{1}{2})$ integrate exactly only
	up to degree~$(1,1)$.  The orthogonality
	condition~\eqref{eq:app_weak} therefore does not hold, and the divergence
	residual retains low-frequency content (including, potentially, a
	nonzero element-average) that can accumulate coherently across the
	mesh.
\end{remark}

\subsection{Consequences for the assembled system}
\label{app:consequences}

The orthogonality property~\eqref{eq:app_weak} explains the three
features observed in Figure~\ref{fig:kovasznay_convergence}:

\medskip\noindent
\textbf{1.\ Inter-element cancellation.}\;
With the $2\times 2$ Gauss rule, each element's divergence residual
is purely high-order: it has zero mean, zero first moments, and zero
bilinear moment.  High-order residuals oscillate in sign and cancel
between neighboring elements whose DOFs are correlated through
inter-element continuity.  Without this orthogonality, low-frequency
residuals (particularly a nonzero element-average divergence) add
constructively across the mesh, explaining the higher global divergence
produced by the per-element-optimal $(\pm\frac{1}{2},\,\pm\frac{1}{2})$
points.

\medskip\noindent
\textbf{2.\ Velocity accuracy.}\;
By confining the divergence residual to high-order modes, the $2\times 2$
Gauss rule avoids introducing spurious low-order errors into the velocity
field.  The Lagrange multipliers~$\bm{\lambda}$ in the saddle-point
system~\eqref{eq:saddle_point} remain well-determined, and the constrained solution retains
the full $O(h^3)$ approximation power of the biquadratic space.

\medskip\noindent
\textbf{3.\ Over-constrained schemes.}\;
Increasing the number of collocation points (5 or 8~Gauss points) leaves
too few unconstrained DOFs for the momentum equation and can degrade
performance. The $2\times 2$ rule is the lowest-order Gauss rule that
provides the 4~constraints needed to leave $18-4=14$ DOFs per element, and
its polynomial exactness is precisely what is needed for the orthogonality
property~\eqref{eq:app_weak}.

\subsection{Reconciling per-element and assembled analyses}
\label{app:reconcile}

The single-element leakage analysis and the assembled computations measure
different quantities:
\begin{itemize}
	\item \textbf{Per-element leakage} (Sections~A.1--A.3) considers the
	worst-case $\int(\nabla\!\cdot\!\bm{v})^2\,d\Omega$ over
	\emph{all} velocity fields in $\Null(\bm{A})$, on a single
	isolated element with arbitrary DOFs.  By this metric, the
	$(\pm\frac{1}{2},\,\pm\frac{1}{2})$ points are optimal.
	\item \textbf{Assembled divergence} (Figure~\ref{fig:kovasznay_convergence}) measures the integrated
	divergence of the actual numerical solution to a specific PDE, on
	coupled elements with constraints and momentum enforced
	simultaneously.  By this metric, the $2\times 2$ Gauss rule is
	superior.
\end{itemize}
The gap arises because the assembled solution inhabits a highly constrained
subspace of the per-element admissible set: inter-element continuity,
the momentum equation, and boundary conditions all restrict which velocity
fields are realized.  Within this restricted space, the polynomial
structure of the divergence residual (whether its low-frequency content
has been eliminated) matters more than the worst-case amplitude bound.
The $2\times 2$ Gauss rule is not the per-element optimum, but it filters
the residual in precisely the way that matters for the assembled system.

\bibliographystyle{unsrt}
\bibliography{pmpg_manuscript}

@article{taha2023minimization,
  author  = {Taha, Haithem E. and Gonzalez, Cody and Shorbagy, Mohamed},
  title   = {A minimization principle for incompressible fluid mechanics},
  journal = {Physics of Fluids},
  volume  = {35},
  number  = {12},
  pages   = {123110},
  year    = {2023},
  doi     = {10.1063/5.0175959},
}

@article{gonzalez2022variational,
  author  = {Gonzalez, Cody and Taha, Haithem E.},
  title   = {A variational theory of lift},
  journal = {Journal of Fluid Mechanics},
  volume  = {941},
  pages   = {A58},
  year    = {2022},
  doi     = {10.1017/jfm.2022.348},
}

@article{appell1900,
  author  = {Appell, Paul},
  title   = {Sur une forme g\'{e}n\'{e}rale des \'{e}quations de la dynamique},
  journal = {Journal f\"{u}r die reine und angewandte Mathematik},
  volume  = {121},
  pages   = {310--319},
  year    = {1900},
}

@article{gauss1829,
  author  = {Gauss, Carl Friedrich},
  title   = {\"{U}ber ein neues allgemeines {G}rundgesetz der {M}echanik},
  journal = {Journal f\"{u}r die reine und angewandte Mathematik},
  volume  = {4},
  pages   = {232--235},
  year    = {1829},
}

@article{brooks1982,
  author  = {Brooks, Alexander N. and Hughes, Thomas J. R.},
  title   = {Streamline upwind/{P}etrov--{G}alerkin formulations for
             convection dominated flows with particular emphasis on
             the incompressible {N}avier--{S}tokes equations},
  journal = {Computer Methods in Applied Mechanics and Engineering},
  volume  = {32},
  number  = {1--3},
  pages   = {199--259},
  year    = {1982},
  doi     = {10.1016/0045-7825(82)90071-8},
}

@article{chorin1968,
  author  = {Chorin, Alexandre Joel},
  title   = {Numerical solution of the {N}avier--{S}tokes equations},
  journal = {Mathematics of Computation},
  volume  = {22},
  number  = {104},
  pages   = {745--762},
  year    = {1968},
  doi     = {10.1090/S0025-5718-1968-0242392-2},
}

@book{gresho1998,
  author    = {Gresho, Philip M. and Sani, Robert L.},
  title     = {Incompressible Flow and the Finite Element Method},
  publisher = {Wiley},
  address   = {Chichester},
  year      = {1998},
}

@article{babuska1973,
  author  = {Babu\v{s}ka, Ivo},
  title   = {The finite element method with {L}agrangian multipliers},
  journal = {Numerische Mathematik},
  volume  = {20},
  number  = {3},
  pages   = {179--192},
  year    = {1973},
  doi     = {10.1007/BF01436561},
}

@article{brezzi1974,
  author  = {Brezzi, Franco},
  title   = {On the existence, uniqueness and approximation of
             saddle-point problems arising from {L}agrangian multipliers},
  journal = {Revue fran\c{c}aise d'automatique, informatique, recherche
             op\'{e}rationnelle. Analyse num\'{e}rique},
  volume  = {8},
  number  = {R2},
  pages   = {129--151},
  year    = {1974},
}

@article{benzi2005,
  author  = {Benzi, Michele and Golub, Gene H. and Liesen, J\"{o}rg},
  title   = {Numerical solution of saddle point problems},
  journal = {Acta Numerica},
  volume  = {14},
  pages   = {1--137},
  year    = {2005},
  doi     = {10.1017/S0962492904000212},
}

@article{guermond2006,
  author  = {Guermond, Jean-Luc and Minev, Peter and Shen, Jie},
  title   = {An overview of projection methods for incompressible flows},
  journal = {Computer Methods in Applied Mechanics and Engineering},
  volume  = {195},
  number  = {44--47},
  pages   = {6011--6045},
  year    = {2006},
  doi     = {10.1016/j.cma.2005.10.010},
}

@article{chang2002analysis,
	title={Analysis of an exact fractional step method},
	author={Chang, Wang and Giraldo, Francis and Perot, Blair},
	journal={Journal of Computational Physics},
	volume={180},
	number={1},
	pages={183--199},
	year={2002},
	publisher={Elsevier},
	doi = {10.1006/jcph.2002.7087},
}

@article{ghia1982,
  author  = {Ghia, Urmila and Ghia, Kirti N. and Shin, C. T.},
  title   = {High-{R}e solutions for incompressible flow using the
             {N}avier--{S}tokes equations and a multigrid method},
  journal = {Journal of Computational Physics},
  volume  = {48},
  number  = {3},
  pages   = {387--411},
  year    = {1982},
  doi     = {10.1016/0021-9991(82)90058-4},
}

@article{armaly1983,
  author  = {Armaly, Bassem F. and Durst, Franz and Pereira, J. C. F.
             and Sch\"{o}nung, Bernhard},
  title   = {Experimental and theoretical investigation of
             backward-facing step flow},
  journal = {Journal of Fluid Mechanics},
  volume  = {127},
  pages   = {473--496},
  year    = {1983},
  doi     = {10.1017/S0022112083002839},
}

@article{kovasznay1948,
  author  = {Kovasznay, Leslie S. G.},
  title   = {Laminar flow behind a two-dimensional grid},
  journal = {Mathematical Proceedings of the Cambridge Philosophical Society},
  volume  = {44},
  number  = {1},
  pages   = {58--62},
  year    = {1948},
  doi     = {10.1017/S0305004100023999},
}

@article{erturk2005,
  author  = {Erturk, Ercan and Corke, Thomas C. and G\"{o}k\c{c}\"{o}l, Cihan},
  title   = {Numerical solutions of 2-{D} steady incompressible driven
             cavity flow at high {R}eynolds numbers},
  journal = {International Journal for Numerical Methods in Fluids},
  volume  = {48},
  number  = {7},
  pages   = {747--774},
  year    = {2005},
  doi     = {10.1002/fld.953},
}

@article{dennis1970,
  author  = {Dennis, S. C. R. and Chang, Gau-Zu},
  title   = {Numerical solutions for steady flow past a circular cylinder
             at {R}eynolds numbers up to 100},
  journal = {Journal of Fluid Mechanics},
  volume  = {42},
  number  = {3},
  pages   = {471--489},
  year    = {1970},
  doi     = {10.1017/S0022112070001428},
}

@article{tritton1959,
  author  = {Tritton, David J.},
  title   = {Experiments on the flow past a circular cylinder at low
             {R}eynolds numbers},
  journal = {Journal of Fluid Mechanics},
  volume  = {6},
  number  = {4},
  pages   = {547--567},
  year    = {1959},
  doi     = {10.1017/S0022112059000829},
}

@article{williamson1996,
  author  = {Williamson, C. H. K.},
  title   = {Vortex dynamics in the cylinder wake},
  journal = {Annual Review of Fluid Mechanics},
  volume  = {28},
  pages   = {477--539},
  year    = {1996},
  doi     = {10.1146/annurev.fl.28.010196.002401},
}

@article{hughes1986,
  author  = {Hughes, Thomas J. R. and Franca, Leopoldo P. and Balestra, Marc},
  title   = {A new finite element formulation for computational fluid
             dynamics: {V}. {C}ircumventing the {B}abu\v{s}ka--{B}rezzi
             condition: a stable {P}etrov--{G}alerkin formulation of the
             {S}tokes problem accommodating equal-order interpolations},
  journal = {Computer Methods in Applied Mechanics and Engineering},
  volume  = {59},
  number  = {1},
  pages   = {85--99},
  year    = {1986},
  doi     = {10.1016/0045-7825(86)90025-3},
}

@article{raissi2019physics,
  author  = {Raissi, Maziar and Perdikaris, Paris and Karniadakis, George Em},
  title   = {Physics-informed neural networks: A deep learning framework for
             solving forward and inverse problems involving nonlinear partial
             differential equations},
  journal = {Journal of Computational Physics},
  volume  = {378},
  pages   = {686--707},
  year    = {2019},
  doi     = {10.1016/j.jcp.2018.10.045},
}

@article{raissi2020hidden,
  author  = {Raissi, Maziar and Yazdani, Alireza and Karniadakis, George Em},
  title   = {Hidden fluid mechanics: Learning velocity and pressure fields
             from flow visualizations},
  journal = {Science},
  volume  = {367},
  number  = {6481},
  pages   = {1026--1030},
  year    = {2020},
  doi     = {10.1126/science.aaw4741},
}

@article{kontogiannis2025learning,
  author  = {Kontogiannis, Alexandros and Elgersma, Scott V. and
             Sederman, Andrew J. and Juniper, Matthew P.},
  title   = {Learning rheological parameters of non-{N}ewtonian fluids from
             velocimetry data},
  journal = {Journal of Fluid Mechanics},
  volume  = {1011},
  pages   = {R3},
  year    = {2025},
  doi     = {10.1017/jfm.2025.92},
}

@book{pierron2012virtual,
  author    = {Pierron, Fabrice and Gr\'{e}dias, Michel},
  title     = {The Virtual Fields Method: Extracting Constitutive Mechanical
               Parameters from Full-field Deformation Measurements},
  publisher = {Springer},
  address   = {New York},
  year      = {2012},
  doi       = {10.1007/978-1-4614-1824-5},
}

@article{savitzky1964,
  author  = {Savitzky, Abraham and Golay, Marcel J. E.},
  title   = {Smoothing and differentiation of data by simplified least
             squares procedures},
  journal = {Analytical Chemistry},
  volume  = {36},
  number  = {8},
  pages   = {1627--1639},
  year    = {1964},
  doi     = {10.1021/ac60214a047},
}

@book{ainsworth2000,
  author    = {Ainsworth, Mark and Oden, J. Tinsley},
  title     = {A Posteriori Error Estimation in Finite Element Analysis},
  publisher = {Wiley},
  year      = {2000},
  doi       = {10.1002/9781118032824},
}

@book{verfurth2013,
  author    = {Verf{\"u}rth, R{\"u}diger},
  title     = {A Posteriori Error Estimation Techniques for Finite Element
               Methods},
  publisher = {Oxford University Press},
  year      = {2013},
  doi       = {10.1093/acprof:oso/9780199679423.001.0001},
}

@article{geuzaine2009gmsh,
  author  = {Geuzaine, Christophe and Remacle, Jean-Fran{\c{c}}ois},
  title   = {{Gmsh}: A 3-{D} finite element mesh generator with built-in
             pre- and post-processing facilities},
  journal = {International Journal for Numerical Methods in Engineering},
  volume  = {79},
  number  = {11},
  pages   = {1309--1331},
  year    = {2009},
  doi     = {10.1002/nme.2579},
}

\end{document}